\documentclass[aps,prd,showpacs,eqsecnum,twocolumn,superscriptaddress]{revtex4-1}
\usepackage{amsmath,amssymb,graphicx,color,ulem,multirow}

\begin{document}

\title{Efficient magnetic-field amplification due to the Kelvin-Helmholtz instability 
in binary neutron star mergers}

\author{Kenta Kiuchi}
\affiliation{Yukawa Institute for Theoretical Physics, 
Kyoto University, Kyoto, 606-8502, Japan~} 

\author{Pablo Cerd\'{a}-Dur\'{a}n}
\affiliation{Departamento de Astronom\'{i}a y Astrof\'{i}sica, Universitat de Val\'{e}ncia, 46100 Burjassot (Valencia), Spain}

\author{Koutarou Kyutoku} 
\affiliation{Interdisciplinary Theoretical Science (iTHES) Research Group, RIKEN, Wako, Saitama 351-0198, Japan}

\author{Yuichiro Sekiguchi}
\affiliation{Department of Physics, Toho University, Funabashi, Chiba 274-8510, Japan}

\author{Masaru Shibata}
\affiliation{Yukawa Institute for Theoretical Physics, 
Kyoto University, Kyoto, 606-8502, Japan} 

\date{\today}

\newcommand{\pcd}[1]{{\color{red}#1}}

\begin{abstract}

We explore magnetic-field amplification due to the Kelvin-Helmholtz instability during binary neutron star 
mergers. By performing high-resolution general relativistic magnetohydrodynamics simulations with a resolution of $17.5$ m 
for $4$--$5$ ms after the onset of the merger on the Japanese supercomputer ``K'', 
we find that an initial magnetic field of moderate maximum strength $10^{13}$ G 
is amplified at least by a factor of $\approx 10^3$. We also explore the saturation of the magnetic-field energy and our result shows that 
it is likely to be $\gtrsim 4 \times 10^{50}$ erg, which is $\gtrsim 0.1\%$ of the bulk kinetic energy of the merging binary neutron stars.

\end{abstract}

\pacs{04.25.D-, 04.30.-w, 04.40.Dg}

\maketitle


\section{Introduction.}

The merger of binary neutron stars (BNS) is one of the most promising sources for the ground-based 
gravitational wave detectors such as advanced LIGO, advanced VIRGO, and 
KAGRA~\cite{LIGO,VIRGO,KAGRA}. If gravitational waves from them are detected, 
we will be able to assess the validity of general relativity in a strong gravitational field 
and explore the equation of state (EOS) of neutron star (NS) matter. 
Furthermore, the merger of BNSs could be a central engine of short-hard gamma-ray bursts (sGRB) and 
the simultaneous detection of gravitational waves and sGRB will give a constraint on this 
merger hypothesis~\cite{Narayan:1992iy}. 

During the merger, the elements heavier than the iron peak elements could be synthesized via the so-called r-process~\cite{Lattimer} 
and it could reproduce the solar abundance pattern of the r-process heavy elements
~\cite{Wanajo:2014wha,Sekiguchi:2015dma}. The radio-actively powered emission from these elements 
could be a strong electromagnetic transient~\cite{Li:1998bw,Tanvir:2013,Berger:2013}. 
Motivated by these facts, building a physically reliable model of BNS mergers is in rapid progress. 

In this paper, we focus on exploring the role of the magnetic field because it is one of the universal features of NSs. 
The observations of binary pulsars indicate that the surface dipole magnetic-field strength is in the range of 
$10^{9.7-12.2}$ G~\cite{Lorimer:2008se}. Rasio and Shapiro have pointed first that the Kelvin-Helmholtz (KH) 
instability could significantly amplify the magnetic-field strength at the merger~\cite{Rasio}. 
Price and Rosswog suggested for the first time that this could be indeed the case~\cite{Price:2006fi}. 
This instability develops in a shear layer which appears when the two stars come into contact. 

It has been controversial whether this mechanism works in practice and 
several preliminary simulations have been reported~\cite{Price:2006fi,Liu,Anderson,Giacomazzo,Giacomazzo:2014qba,Dionysopoulou:2015tda,Palenzuela:2015dqa}. The issue is that, because the growth rate of the KH instability is proportional to 
the wave number of the mode, 
high-resolution simulations together with a careful convergence study is necessary to explore this instability. 
Recently, the authors of Ref.~\cite{Kiuchi:2014} have performed
general relativistic magnetohydrodynamics (GRMHD) simulation of the BNS merger with
significantly higher resolution (by a factor $\sim 2.5$) than any previous simulation. 
They revealed that, only for a sufficiently high numerical resolution, the KH instability activates as an amplifier 
of the magnetic field at the merger.

However, it is still an open question to what extent the magnetic field is amplified during the merger 
under realistic conditions because an initial magnetic-field strength 
employed in Ref.~\cite{Kiuchi:2014} was assumed to be of magnetar class, i.e., $10^{14.5-16}$ G. 
Because the magnetic-field amplification was saturated at the maximum field strength of $\sim 10^{17}$ G, 
the previous simulations followed this amplification process by a factor of 10. 
Several local box simulations have suggested that the magnetic-field energy may 
be amplified by several orders of magnitude until reaching 
an equipartition level even if we assume moderate initial magnetic-field strength~\cite{Obergaulinger:2010gf,Zrake:2013mra}. 
In this paper, we go a step further to thoroughly explore the amplification of the magnetic field by the KH instability. 
Initially setting moderately strong realistic magnetic fields of maximum strength $10^{13}$ G, we perform GRMHD simulations of BNS mergers 
on the Japanese supercomputer ``K'' increasing the resolution by a factor $4$ with respect to previous 
simulations by Ref.~\cite{Kiuchi:2014}.

This paper is organized as follows. In Sec.~II, we briefly mention the method, the grid setup, and 
the initial models of the BNSs. We also describe how we increase the grid resolution in the shear layer during the merger. 
Sections III and IV are devoted to presenting numerical results. Section V is for the discussion 
and the summary is given in Sec.~VI. 


\section{Method, grid setup, and initial models.}
The code is the same as that in Refs.~\cite{Kiuchi:2012,Kiuchi:2014}; Einstein's equation is formulated 
based on a Baumragte-Shapiro-Shibata-Nakamura-puncture formulation~\cite{SN,BS,Capaneli,Baker} 
and solved by fourth-order finite differencing. GRMHD is 
formulated in a conservative form and solved by a high resolution shock capturing scheme together with a 
third-order cell reconstruction~\cite{Kurganov}. We implement a fixed mesh refinement algorithm to cover a wide dynamical 
range of BNS mergers. Specifically, a refinement domain labeled by $l$ is a cuboid box of 
$x_{(l)} \in [-N \Delta x_{(l)}, N \Delta x_{(l)}]$, $y_{(l)} \in [-N \Delta y_{(l)}, N \Delta y_{(l)}]$, 
and $z_{(l)} \in [0, N \Delta z_{(l)}]$ where we assume an orbital plane symmetry. 
Given a grid resolution, a constant integer $N$ specifies the size of the refinement domain. 
In our algorithm, $N$ is identical for all the refinement levels. 
$\Delta x_{(l)}, \Delta y_{(l)}$, and 
$\Delta z_{(l)}$ are a grid resolution in the refinement domain $l$ and we assume $\Delta x_{(l)}=\Delta y_{(l)}=\Delta z_{(l)}$. 
The grid resolution in a coarser refinement domain $l-1$ is $\Delta x_{(l-1)} = 2 \Delta x_{(l)}$ with $l=2,3,\cdots$, and $l_{\rm max}$. 
$l_{\rm max}$ is the number of the refinement levels. 
The divergence free condition as well as the flux conservation of the magnetic field in a refinement boundary are preserved 
by the Balsara's algorithm~\cite{Balsara:2011}. 

\begin{figure*}[t]
\hspace{-40mm}
\begin{minipage}{0.27\hsize}
\begin{center}
\includegraphics[width=10.0cm,angle=0]{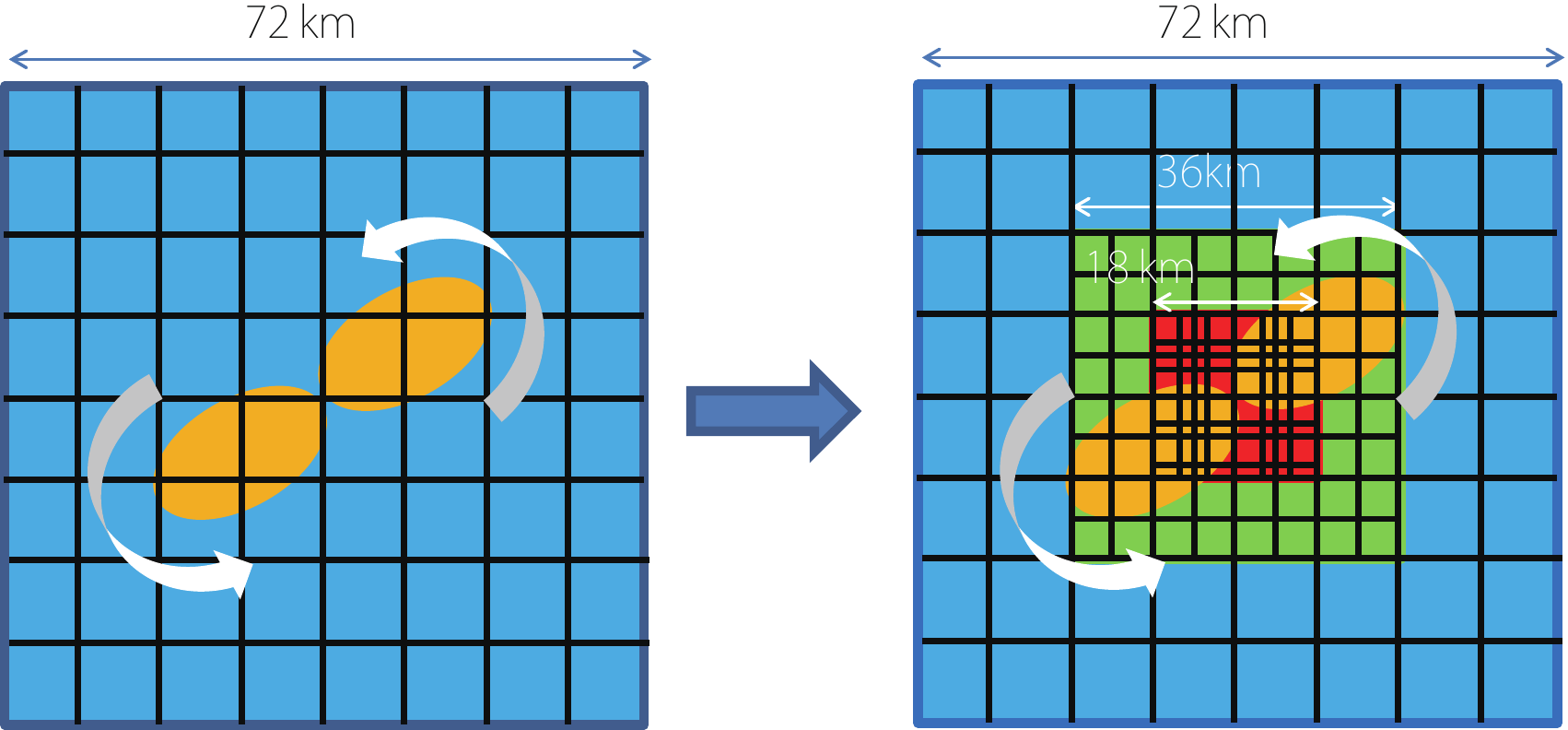}
\end{center}
\end{minipage}
\caption{\label{fig0}
Schematic picture of the refinement domains on the orbital plane for the initial grid configuration with $l=l^0_{\rm max}$ (left) 
and for the final grid configuration with $l=l_{\rm max}-2(=l_{\rm max}^0),l_{\rm max}-1,$ and $l_{\rm max}$ (right). Two ellipses in the vicinity of the domain center represent 
the BNS just before the merger. 
}
\end{figure*}

To improve the grid resolution for a shear layer which appears when the two stars come into contact, we add new refinement domains 
at about $1$ ms before the merger. The size of the new finest refinement domain 
is determined by the initial grid configuration; typically, the {\it initial} finest domain is a cuboid of $\approx (72000~{\rm m})^3/2$. 
If we add one refinement domain, the {\it final} finest domain is a cuboid of 
$\approx (36000~{\rm m})^3/2$. If we add two domains, the final finest domain is that of $\approx (18000~{\rm m})^3/2$ (see Fig.~\ref{fig0} 
for a schematic picture). 
Previous simulations have suggested that the shear layer appears in a central region of 
radius $\sim 10000$ m~\cite{Kiuchi:2014}. 
Therefore, we increase the number of the refinement domains up to two. 
Table~\ref{tab1} shows the grid setup. We achieve $17.5$ m resolution which is much finer than the highest-resolution 
in our previous simulation~\cite{Kiuchi:2014}.  

As a fiducial model, we choose an equal-mass irrotational binary of total mass $2.8M_{\odot}$. 
As an EOS to model the NS, 
we assume the H4 EOS~\cite{H4}. The initial orbital angular frequency is $G m_0 \Omega/c^3=0.0221$ where $G$ is the 
gravitational constant, $m_0$ is the sum of the gravitational mass in isolation, and $c$ is the speed of light. 
At about $1$ ms before the onset of the merger, we add a seed magnetic field whose configuration is given by 

\begin{align}
&A_i = \left( - ( y - y_{\rm NS} ) \delta^x_i + ( x - x_{\rm NS} ) \delta^y_i \right) A_{\rm b} \nonumber\\
&\times ~{\rm max}(P-P(0.04\rho_{\rm max}),0)^2,  
\end{align}
where $i=x,y,$ and $z$, $P$ is the pressure, and $\rho_{\rm max}$ is the maximum rest-mass density. 
$x_{\rm NS}$ and $y_{\rm NS}$ denote the coordinate center of the NS. 
$A_{\rm b}$ determines the maximum magnetic-field strength, which is realized at the stellar center.
The observations of binary pulsars suggest that the {\it surface dipole} magnetic-field 
strength is in the range $\sim~10^{9.7}-10^{12.2}$ G~\cite{Lorimer:2008se}. However, the interior magnetic field
strength is mostly unconstrained since currents at the surface could be shielding a much stronger or weaker field
in the interior.  If shielding is not invoked, theoretical models of magnetized neutron star equilibria
result in a typical interior magnetic field ~2--5 times as large as its surface value \citep{Bocquet:1995,Kiuchi:2008,Colaiuda:2008}.
We thus set the initial maximum magnetic-field strength 
to be $10^{13}$ G, which is approximately compatible with the upper observational limit mentioned above and a 
reasonable choice to mimic the realistic magnetic-field strength 
of the BNSs. 
To explore the saturation of the magnetic-field energy amplified by the KH instability, we artificially increase 
the initial magnetic-field strength up to $10^{15}$ G in some of our simulations. 

Note that, up to 1 ms before the onset of the merger, the simulations are essentially the same as those in Ref.~\cite{Kiuchi:2014} 
because the magnetic field does not affect the inspiral dynamics. 
Table \ref{tab2} summarizes the models with respect to the initial magnetic-field strength.  
During the evolution, we use a piece-wise polytrope prescription to model the H4 EOS~\cite{rlof2009} with 
a thermal part consisting of a gamma-law EOS with the thermal index of $1.8$. 

\begin{table}
\centering
\caption{\label{tab1} 
Grid setup for all the run. $l_{\rm max}^0(l_{\rm max})$: The number of the refinement levels of 
the initial (final) grid configuration. 
$\Delta x_{(l^0_{\rm max})}(\Delta x_{(l_{\rm max})})$: The grid spacing of the finest refinement level of 
the initial (final) grid configuration. 
$N$: The grid number in one positive direction. 
}
\begin{tabular}{ccccccccc}
\hline\hline
$l_{\rm max}^0$               &
$l_{\rm max}$                 &
$\Delta x_{(l^0_{\rm max})}$ [m]  & 
$\Delta x_{(l_{\rm max})}$ [m] & 
$N$                        \\
\hline
7 & 7 & 70  & 70   & 514 \\
7 & 8 & 70  & 35   & 514 \\
7 & 9 & 70  & 17.5 & 514 \\
7 & 7 & 110 & 110  & 322 \\
7 & 8 & 110 & 55   & 322 \\
7 & 9 & 110 & 27.5 & 322 \\
7 & 7 & 150 & 150  & 242 \\
7 & 8 & 150 & 75   & 242 \\
7 & 9 & 150 & 37.5 & 242 \\
\hline\hline
\end{tabular}
\end{table}

\begin{table}
\centering
\caption{\label{tab2}The initial maximum magnetic-field strength of the BNS.}
\begin{tabular}{cc}
\hline\hline
Model & $\log_{10}(B_{\rm max}~[{\rm G}])$ \\
\hline
B13 & 13 \\
B14 & 14 \\
B15 & 15 \\
\hline\hline
\end{tabular}
\end{table}

\section{Dynamics, magnetic-field amplification, and resolution study}

In this section, we describe the dynamics, magnetic-field amplification, and resolution study for the B13 run. 

\subsection{Kelvin-Helmholtz vortex formation}
Figure~\ref{fig1} plots the rest-mass density profiles with the velocity field 
(the 1st row) and the vorticity profiles (the 2nd row) on the orbital plane for the $\Delta x_{(l_{\rm max})}=17.5$~m run.
The vorticity is defined as the spatial components of the following 2-form 
\begin{align}
\omega_{\mu\nu}=\nabla_\mu (hu_\nu) - \nabla_\nu (hu_\mu), 
\end{align}
where $h$, $u_\mu$, and $\nabla_\mu$ are the specific enthalpy, the four velocity, 
and the covariant derivative with respect to the spacetime metric, respectively. 
We define the merger time $t_{\rm mrg}$ to be the time at which the gravitational-wave amplitude becomes maximum.

Just before the merger, a shear layer appears at the contact interface between the two stars 
as shown in Fig.~\ref{fig1}(a1) (see also Fig.~\ref{figvx} and visualization~\cite{viz}). This contact interface 
is subject to the KH instability. 
At the merger, the KH instability develops and subsequently vortices are formed in the shear layer 
[Figs.~\ref{fig1}(a2) and (b2)].
Figures~\ref{fig1}(c1)--(c4) plot the thermal component of the specific internal energy $\epsilon_{\rm th}$. 
A hot region appears in the shear layer. 
This is due to the numerical dissipation of the votrices 
at the finest resolution size (see also Figs.~\ref{fig2} and \ref{fig3}(c1)--(c4) and discussion in Sec.~IIIC).
The shock wave generated by the collision of the two stars also dissipates the vortices.
Note that the KH vortices are not completely dissipated and they cascade into smaller-scale turbulence-like motion. 

Subsequently, a hypermassive neutron star (HMNS) is formed. 
In its early phase, two dense cores are transiently formed and 
they collide several times [Fig.~\ref{fig1}(a3)]. Because of their mutual interaction, 
the shear layer continuously appears between the two cores and additional KH vortices are successively generated [Fig.~\ref{fig1}(b3)]. 
The two cores eventually merge to a single core and the shear layer disappears due to the shock heating driven by stellar oscillations,
leaving behind a highly turbulence-like flow, which decays in a dissipative timescale [Figs.~\ref{fig1}(a4), (b4), 
and (c4)].

We can interpret these results in the light of the KH linear stability analysis~\citep{Chandrasekhar:61}.
The growth rate of the KH instability is proportional to the wavenumber of the unstable mode. For
a shear flow with thickness $D$, there is a cut-off wavenumber, $k_{\rm cutoff}\sim 2\pi/D$ above which
modes do not grow~\citep{Chandrasekhar:61}. 
In practice, this means that the fastest growing KH mode has a wavenumber, $k_{\rm KH}$ close
to the cut-off and its growth rate is $\sigma_{\rm KH} \sim v_{\rm flow} k_{\rm KH}$. 
The right panel of Fig.~\ref{figvx} plots the profile of the $x$ component of the velocity 
along the $y$ axis with $x=0$ km at $t-t_{\rm mrg}=-0.90$ ms. 
This figure clearly shows that $D$ decreases with increasing the grid resolution. Therefore, $k_{\rm cutoff}$ 
increases as a function of $1/\Delta x_{(l_{\rm max})}$. We found that for $\Delta x_{(l_{\rm max})}=17.5$ m run 
$D \approx 130$ m and $v_{\rm flow}\approx 0.1c$. With these values, the growth timescale is 
$\sigma_{\rm KH}^{-1} \approx 6\times 10^{-3}$ ms which is much shorter than the dynamical timescale of $\sim 0.1$ ms. 
In reality, the scale of the shear layer should depend on 
other factors such as the pressure scale height at the surface of the NS and the tidal deformation. 
Therefore, it is not trivial that $D$ goes to zero in the continuum limit. 
Nonetheless, the shear layer should be sufficiently thin and consequently the growth rate of the KH instability 
should be so large that the small-scale turbulence-like motion develops within the dynamical timescale.

\subsection{Magnetic-field amplification}
Due to the KH instability, turbulence-like motion is generated and this indicates that a part of the bulk kinetic energy 
is converted into the energy of turbulence-like motion [see also Figs.~\ref{fig1}(b1)--(b4)]. 
In the presence of magnetic fields, a part of this energy is transferred to the magnetic-field energy, which results 
in an efficient amplification of the magnetic field. Figures~\ref{fig1}(d1)--(d4) show the magnetic-field strength 
profiles on the orbital plane. A strong magnetic field of $\sim 10^{14}$ G is rapidly generated at the contact interface 
[Fig.~\ref{fig1}(d2)] and it is amplified in the shear layer between the two cores [Fig.~\ref{fig1}(d3)]. 
At the end of the simulation, a strongly magnetized core is formed [Fig.~\ref{fig1}(d4)], with rms field values
  of $\sim 10^{15.5}$~G and peak values of $\sim 10^{17}$~G.

During this {\it kinematic phase}, the magnetic field is still weak and does not influence
  the dynamics; the magnetic field is being stretched by the overturning
  turbulence-like motion, which results in an exponential growth. This kinematic phase should end up with 
  any of the following mechanisms. 
  First, once the amplitude of the magnetic field reaches an equipartition level, the magnetic field starts playing a dynamical effect. 
  Second, the turbulence-like motion decays within the diffusion timescale which is limited by the numerical resolution used. 
  Third, it is the ability of the global flow to regenerate the shear flow and produce additional large-scale eddies, 
  as described in the previous section. These limitations are discussed in subsequent sections.

\subsection{Resolution study}
\begin{figure*}[t]
\hspace{-50mm}
\begin{minipage}{0.27\hsize}
\begin{center}
\includegraphics[width=9.0cm,angle=0]{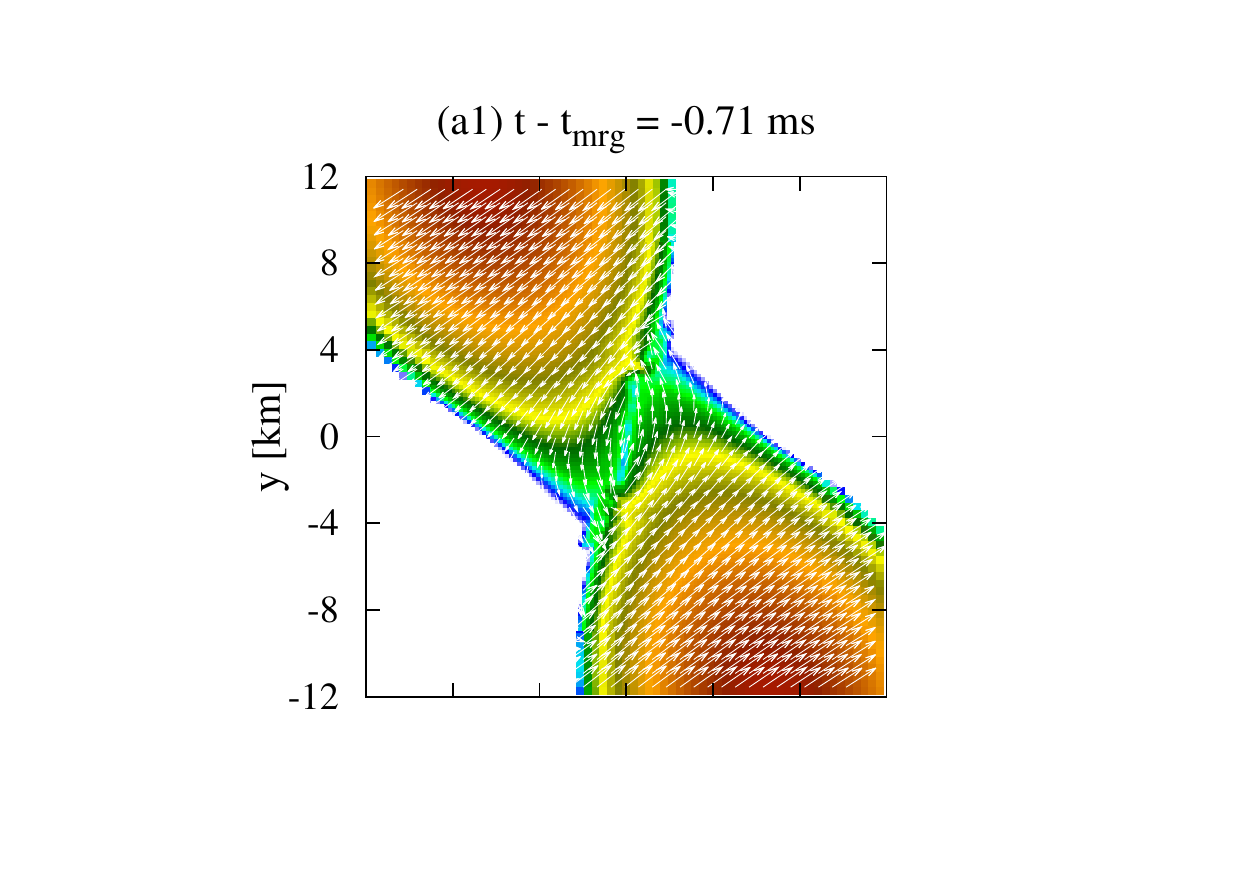}
\end{center}
\end{minipage}
\hspace{-12mm}
\begin{minipage}{0.27\hsize}
\begin{center}
\includegraphics[width=9.0cm,angle=0]{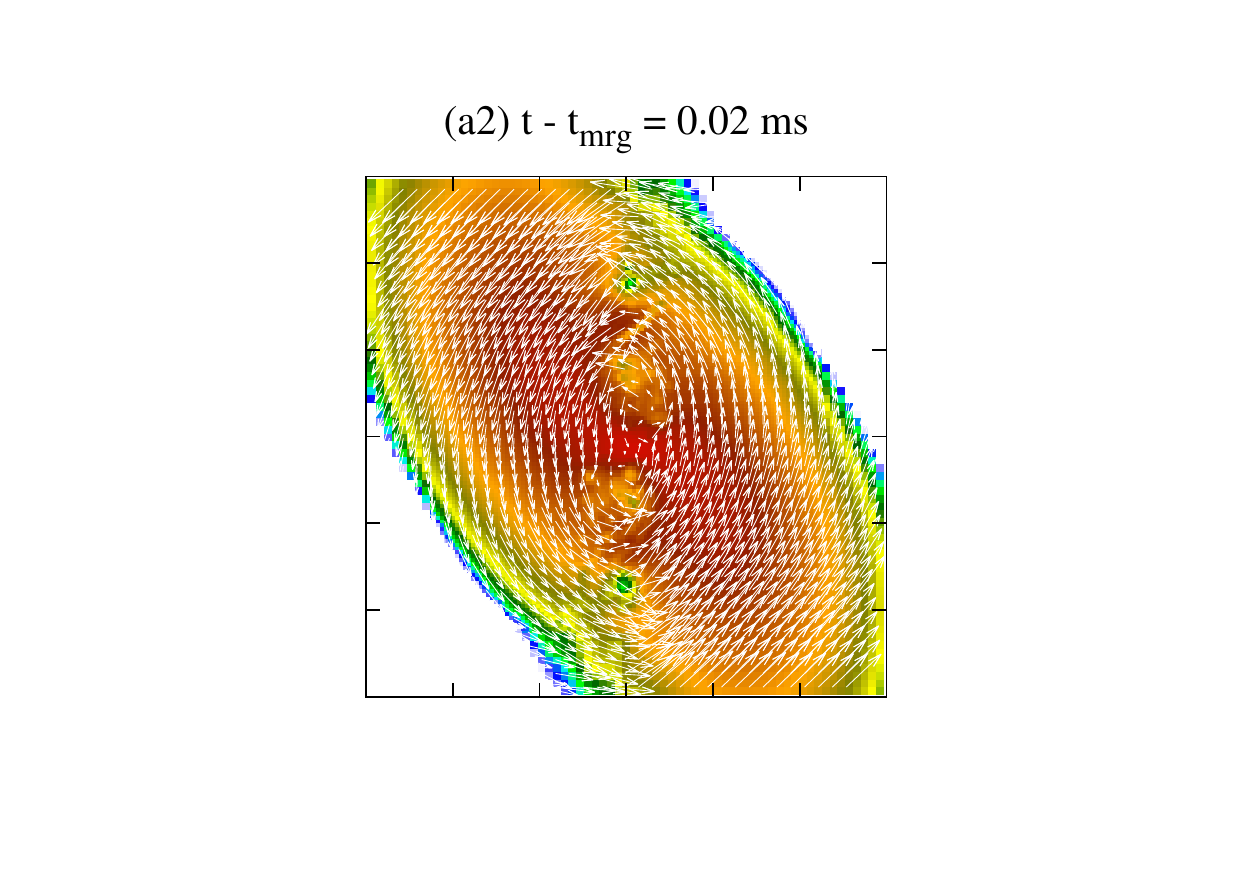}
\end{center}
\end{minipage}
\hspace{-12mm}
\begin{minipage}{0.27\hsize}
\begin{center}
\includegraphics[width=9.0cm,angle=0]{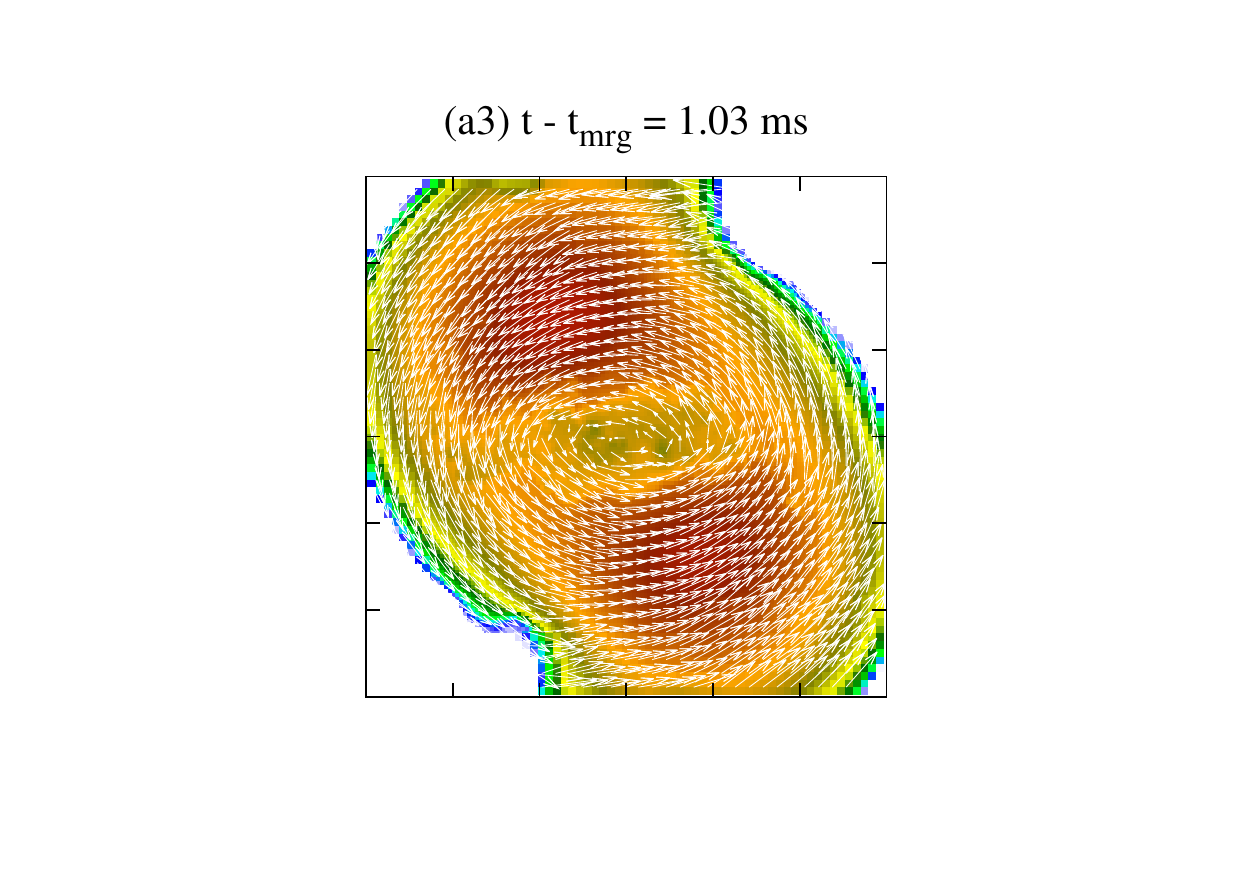}
\end{center}
\end{minipage}
\hspace{-12mm}
\begin{minipage}{0.27\hsize}
\begin{center}
\includegraphics[width=9.0cm,angle=0]{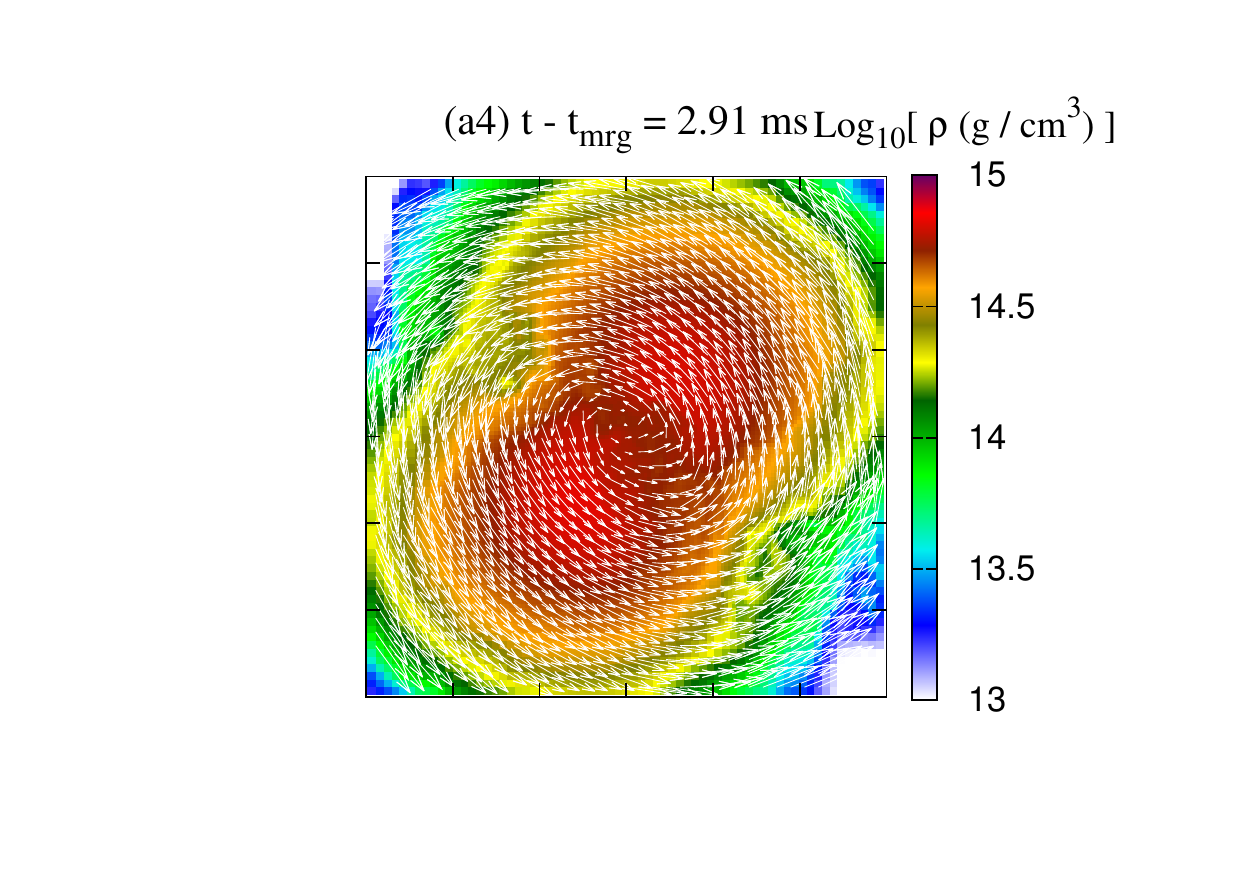}
\end{center}
\end{minipage}\\
\vspace{-19mm}
\hspace{-50mm}
\begin{minipage}{0.27\hsize}
\begin{center}
\includegraphics[width=9.0cm,angle=0]{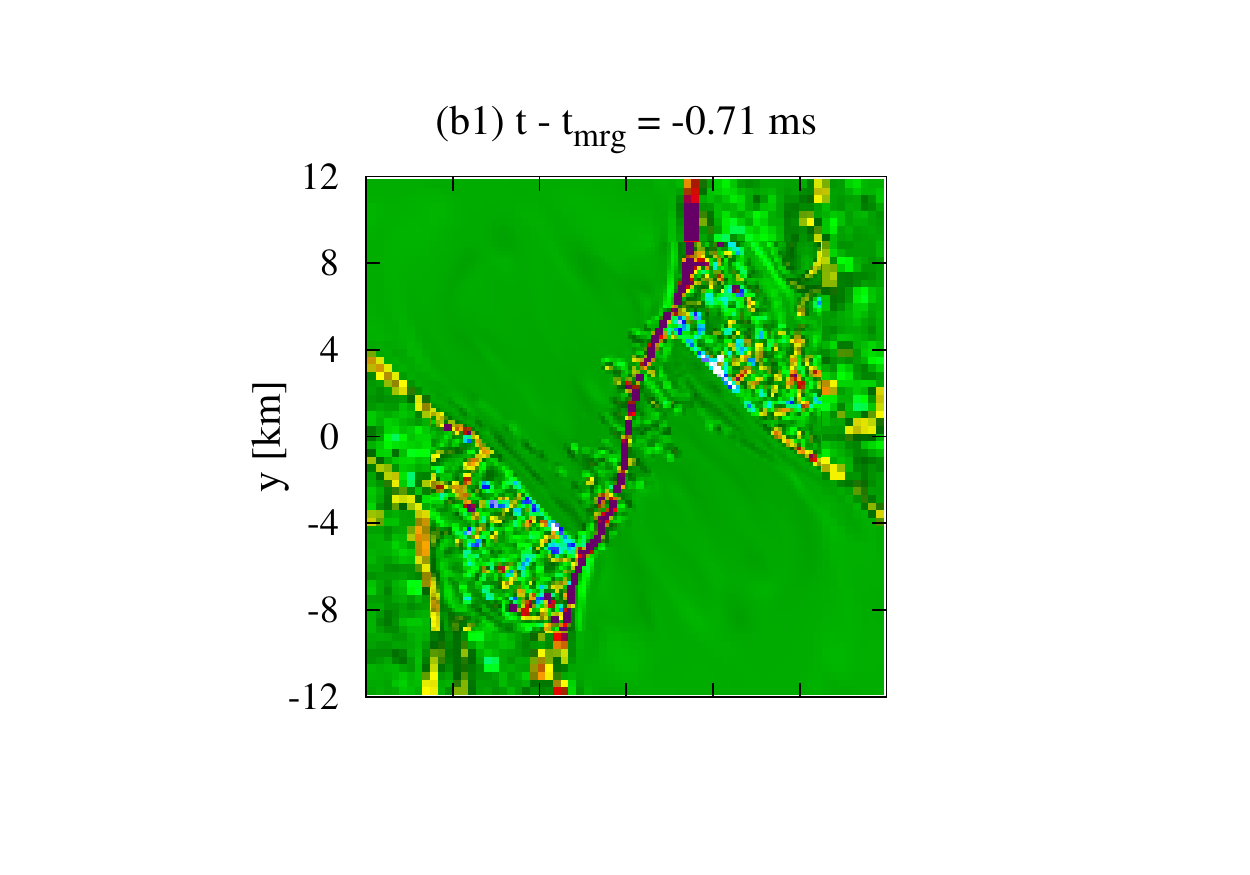}
\end{center}
\end{minipage}
\hspace{-12mm}
\begin{minipage}{0.27\hsize}
\begin{center}
\includegraphics[width=9.0cm,angle=0]{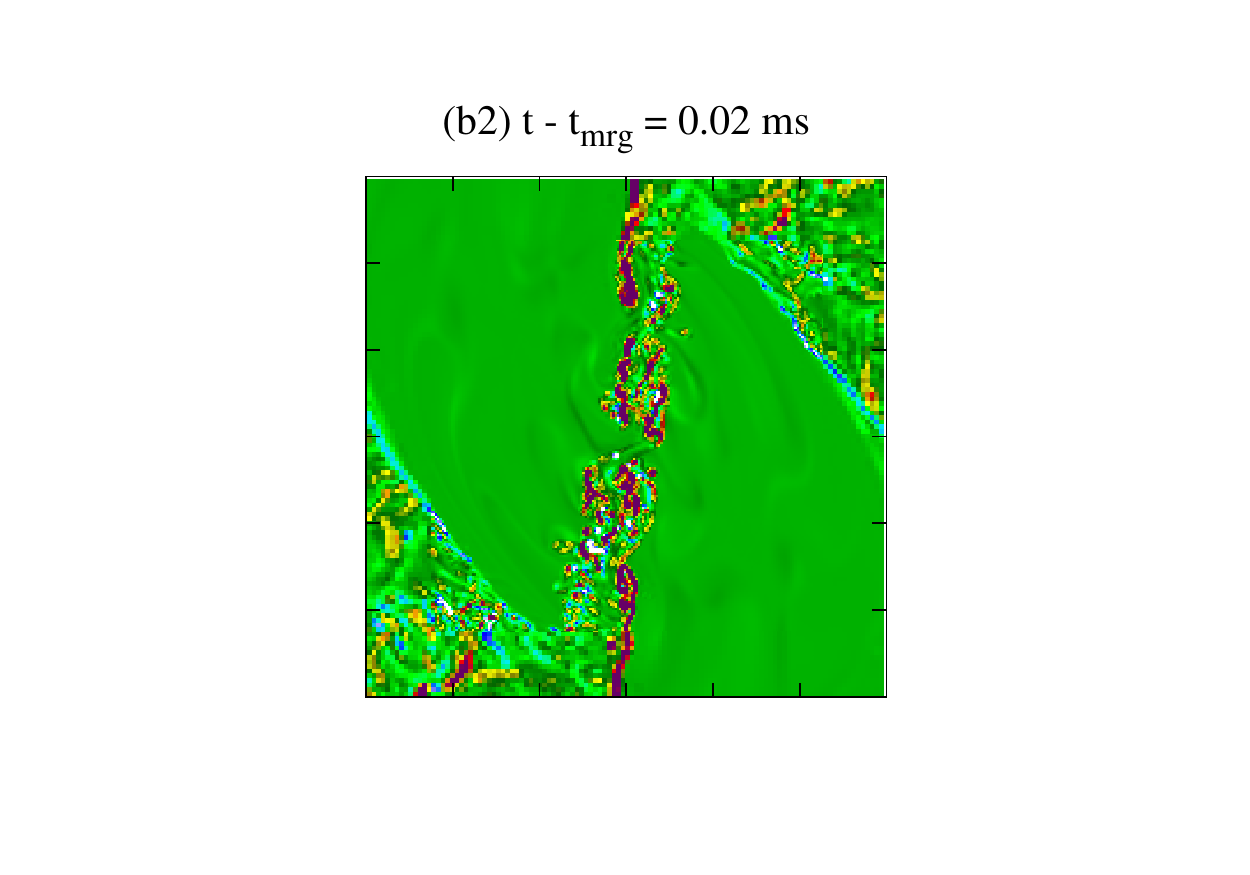}
\end{center}
\end{minipage}
\hspace{-12mm}
\begin{minipage}{0.27\hsize}
\begin{center}
\includegraphics[width=9.0cm,angle=0]{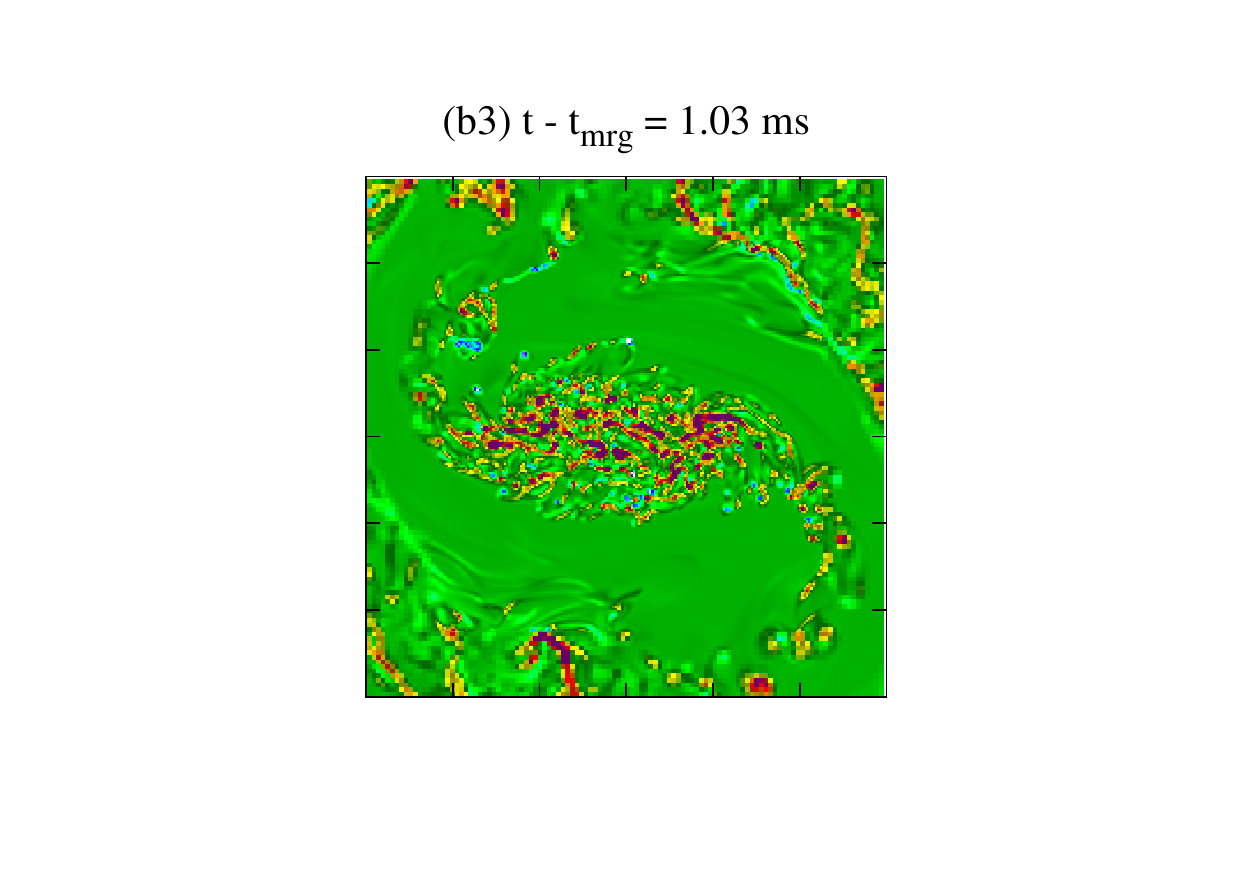}
\end{center}
\end{minipage}
\hspace{-12mm}
\begin{minipage}{0.27\hsize}
\begin{center}
\includegraphics[width=9.0cm,angle=0]{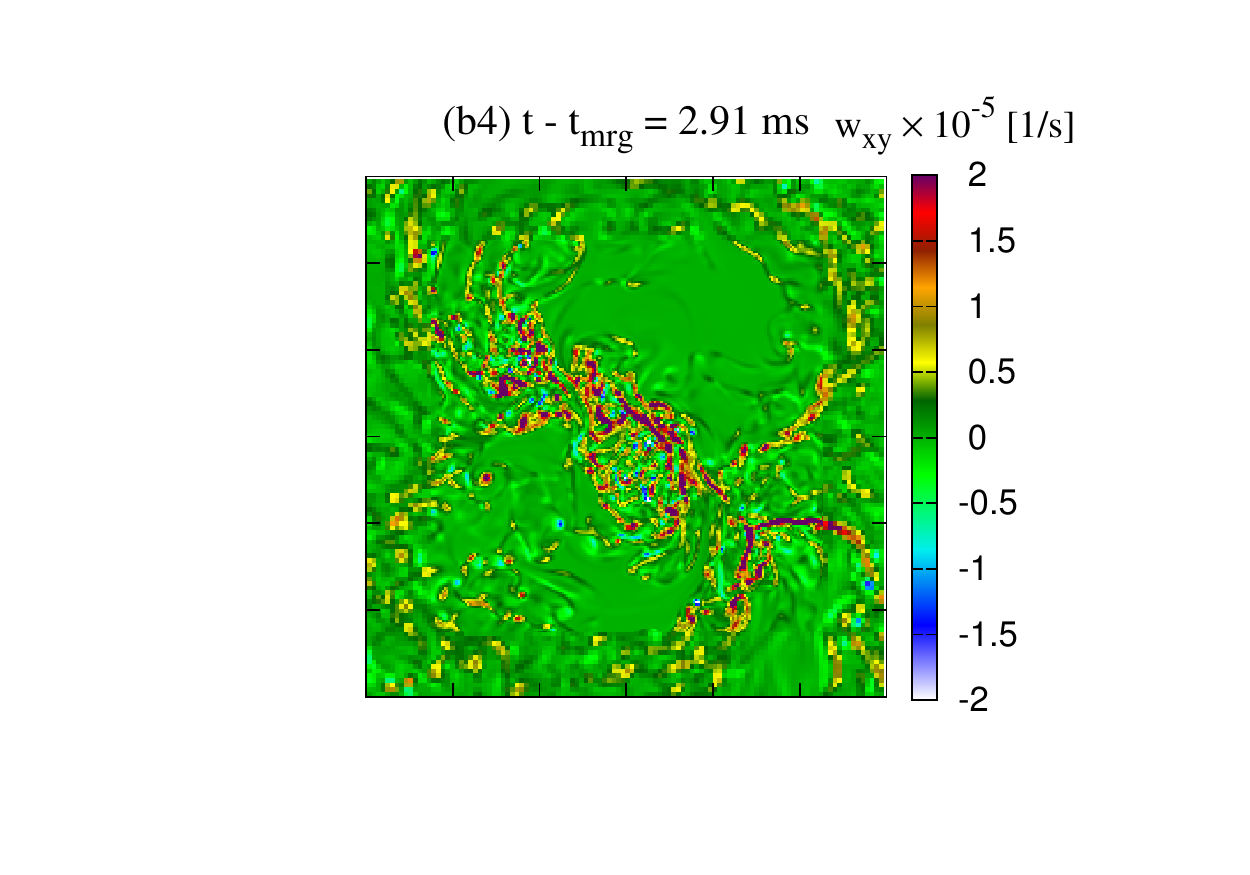}
\end{center}
\end{minipage}\\
\vspace{-19mm}
\hspace{-50mm}
\begin{minipage}{0.27\hsize}
\begin{center}
\includegraphics[width=9.0cm,angle=0]{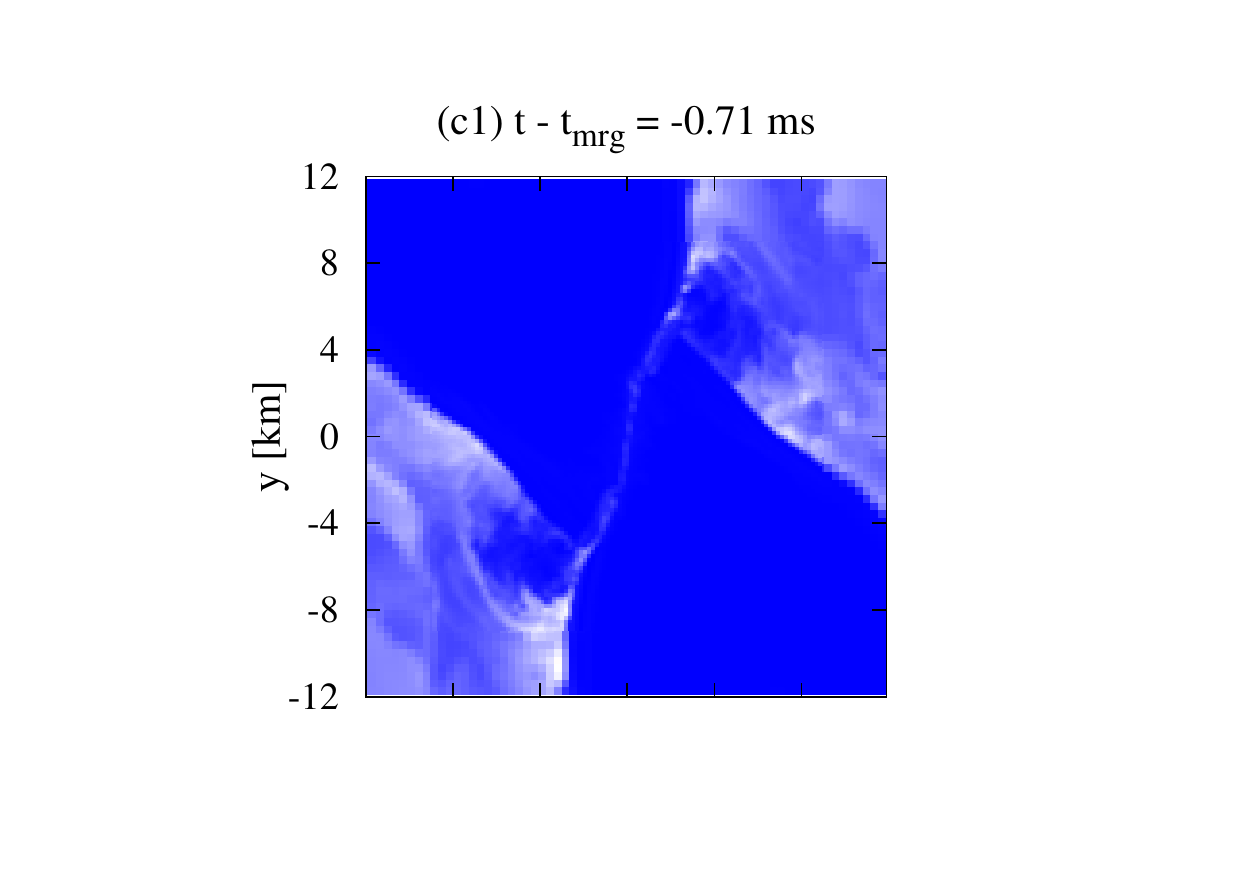}
\end{center}
\end{minipage}
\hspace{-12mm}
\begin{minipage}{0.27\hsize}
\begin{center}
\includegraphics[width=9.0cm,angle=0]{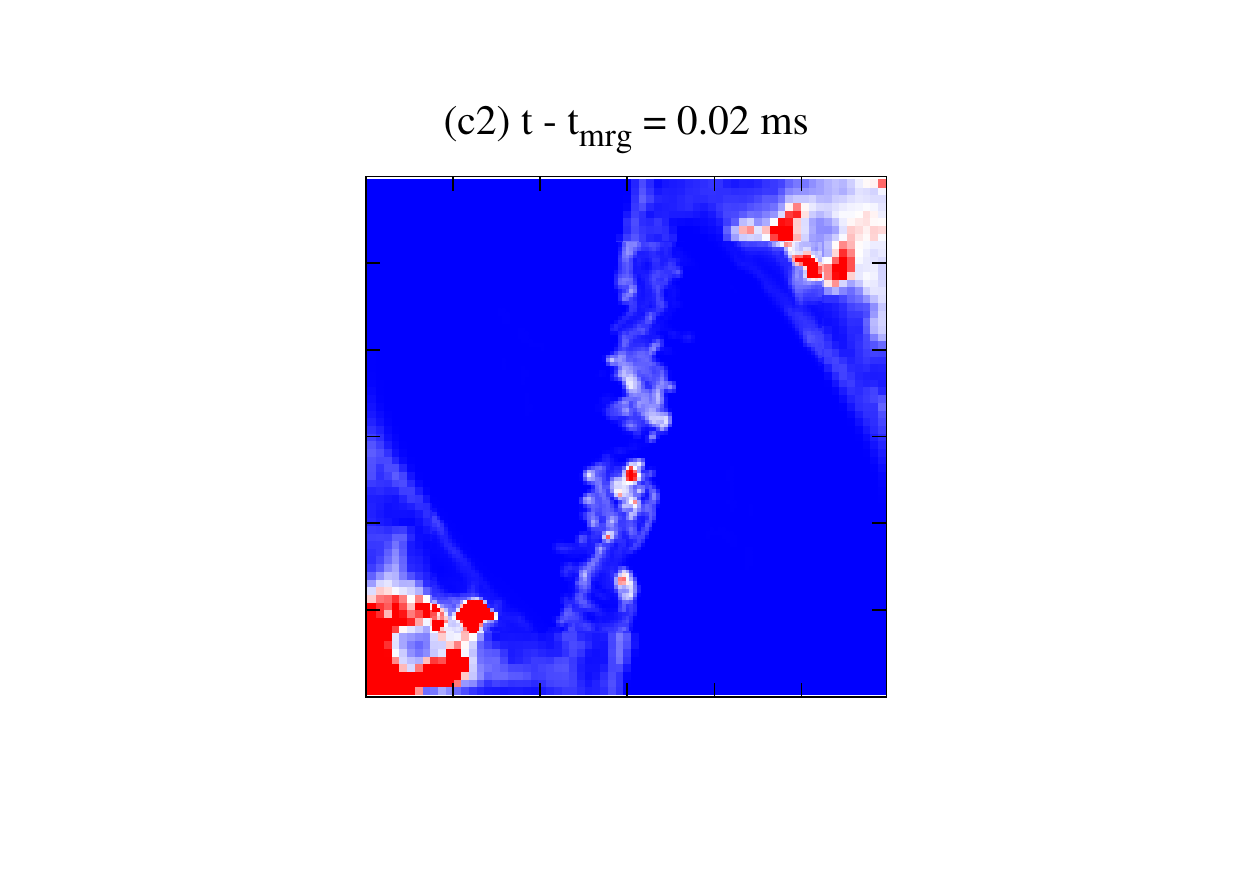}
\end{center}
\end{minipage}
\hspace{-12mm}
\begin{minipage}{0.27\hsize}
\begin{center}
\includegraphics[width=9.0cm,angle=0]{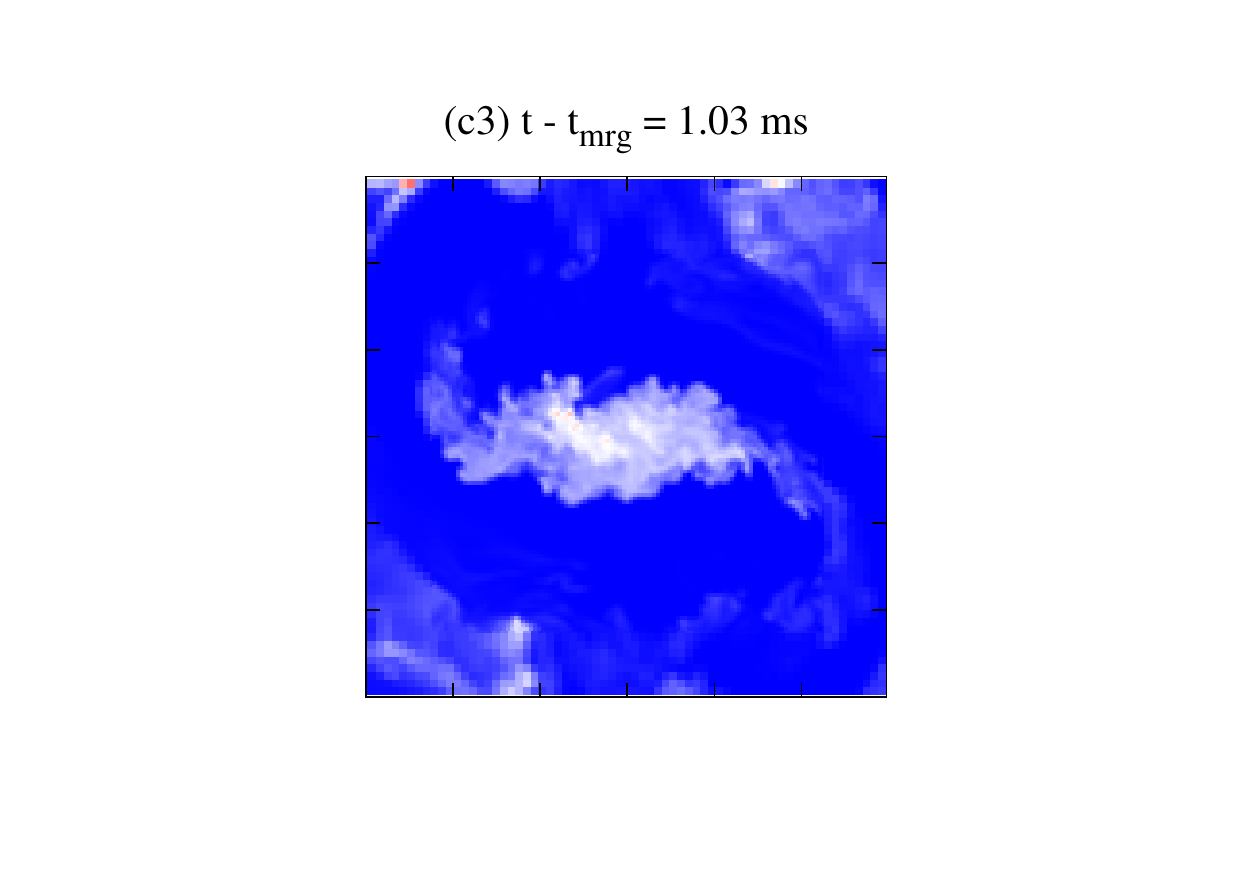}
\end{center}
\end{minipage}
\hspace{-12mm}
\begin{minipage}{0.27\hsize}
\begin{center}
\includegraphics[width=9.0cm,angle=0]{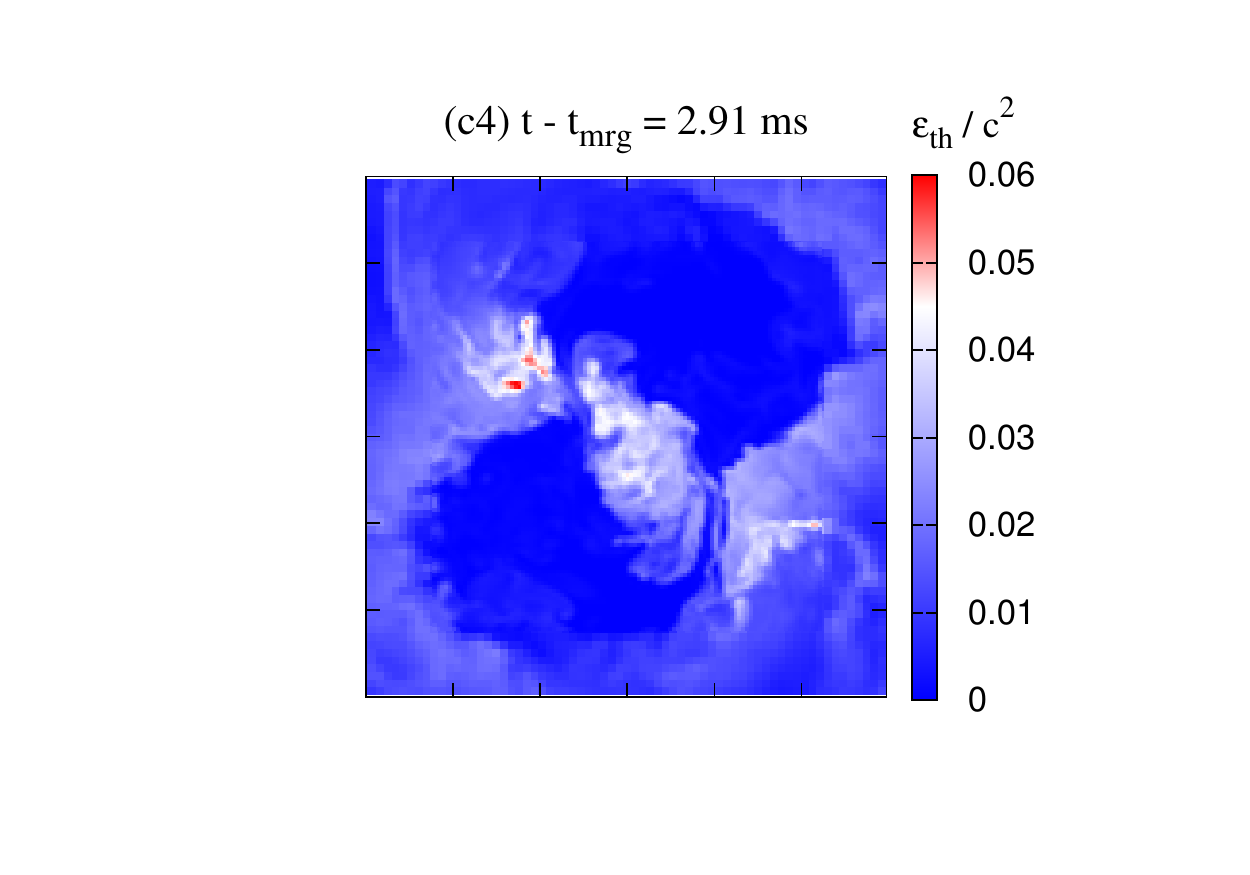}
\end{center}
\end{minipage}\\
\vspace{-19mm}
\hspace{-50mm}
\begin{minipage}{0.27\hsize}
\begin{center}
\includegraphics[width=9.0cm,angle=0]{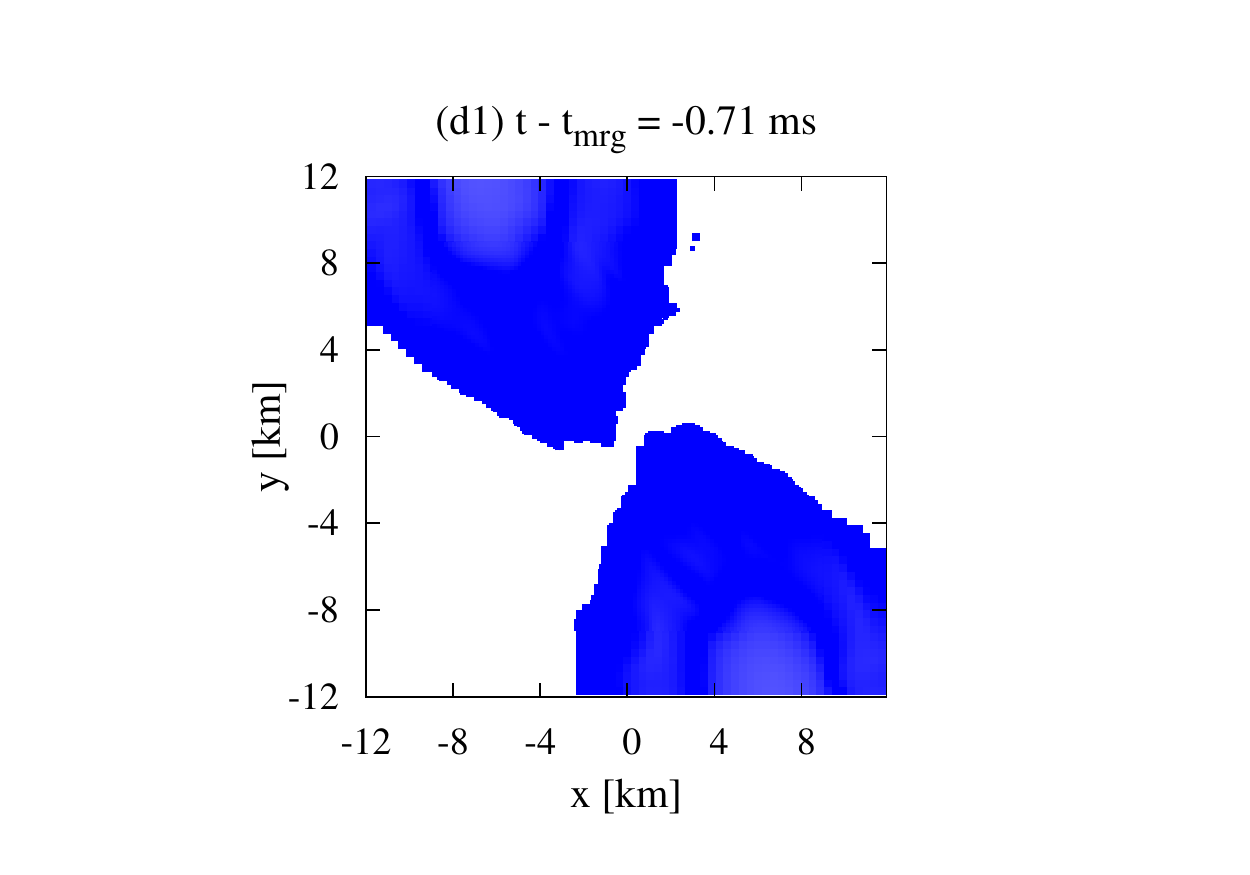}
\end{center}
\end{minipage}
\hspace{-12mm}
\begin{minipage}{0.27\hsize}
\begin{center}
\includegraphics[width=9.0cm,angle=0]{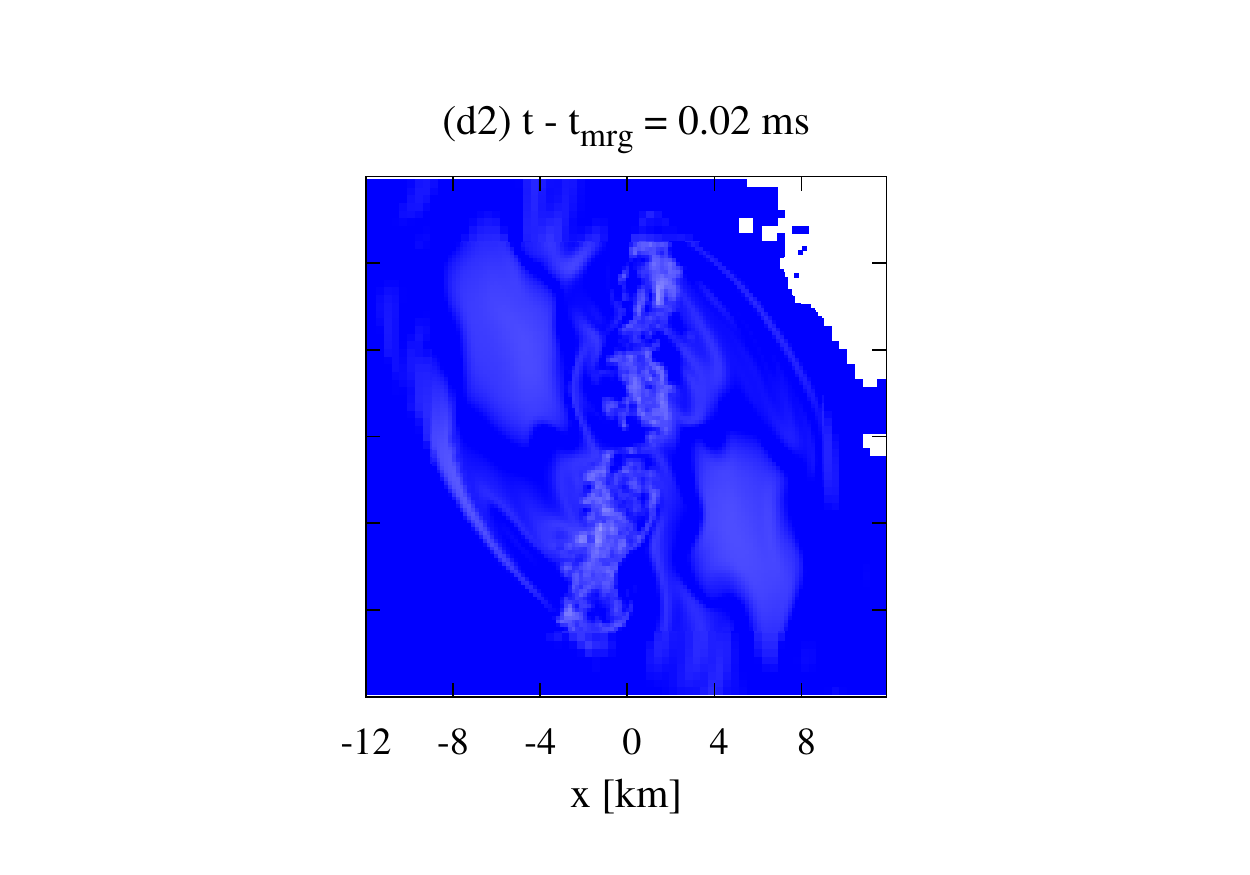}
\end{center}
\end{minipage}
\hspace{-12mm}
\begin{minipage}{0.27\hsize}
\begin{center}
\includegraphics[width=9.0cm,angle=0]{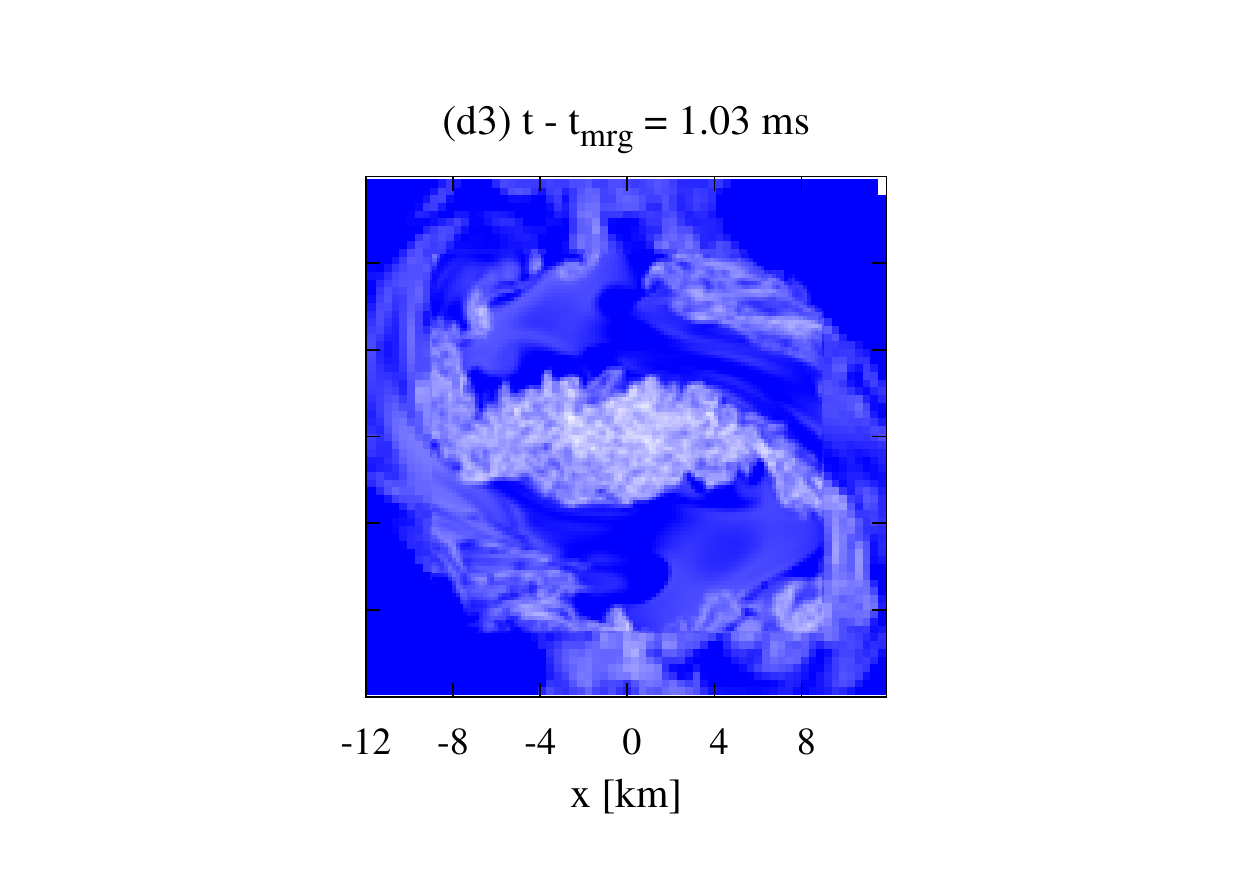}
\end{center}
\end{minipage}
\hspace{-12mm}
\begin{minipage}{0.27\hsize}
\begin{center}
\includegraphics[width=9.0cm,angle=0]{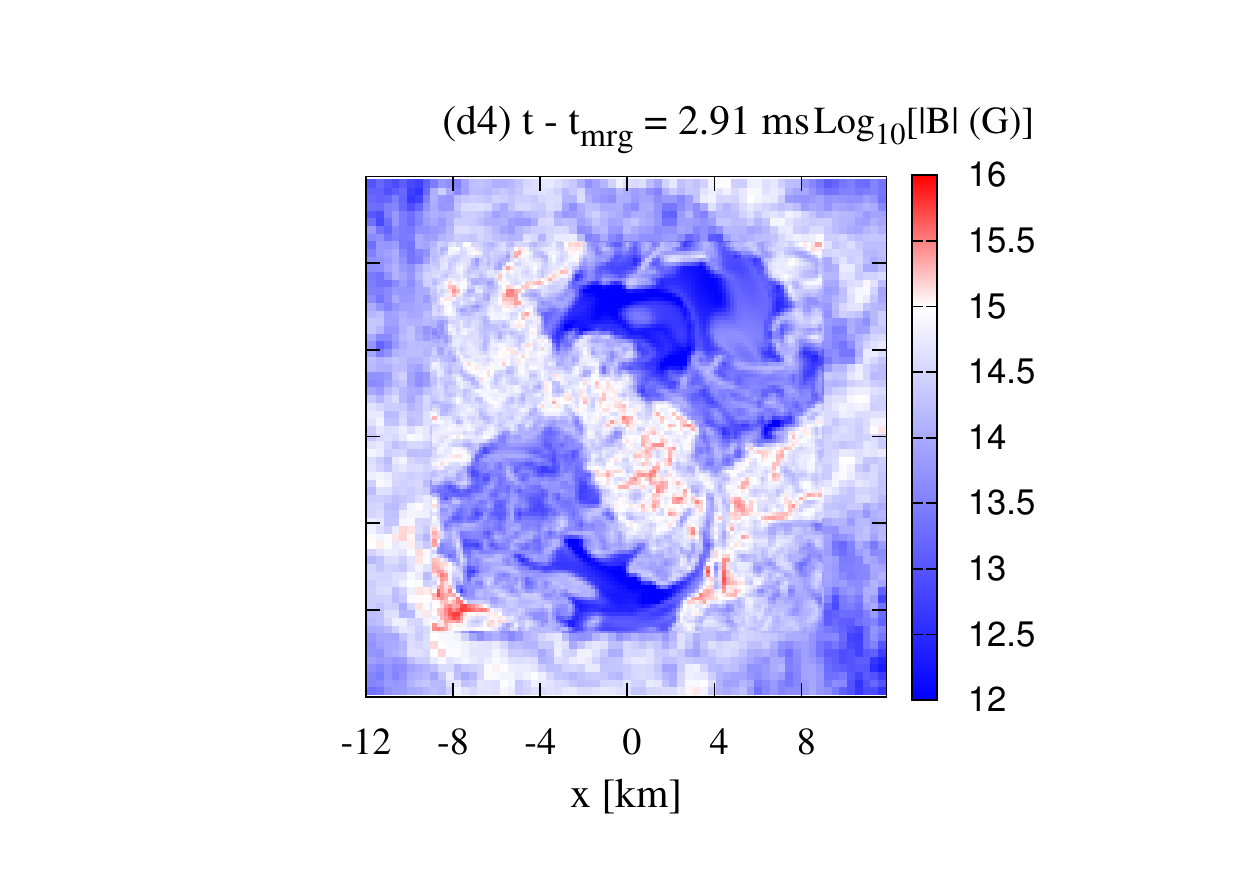}
\end{center}
\end{minipage}
\caption{\label{fig1}
Profiles of the rest-mass density with velocity field (the 1st row), of the vorticity (the 2nd row), 
of the thermal component of the specific internal energy (the 3rd row), and 
of the magnetic-field strength (the 4th row) on the orbital plane. 
The initial magnetic-field strength is $10^{13}$ G and the resolution is $\Delta x_{(l_{\rm max})}=17.5$ m. 
$t_{\rm mrg}$ is the merger time (see the text for details). 
}
\end{figure*}

\begin{figure*}[t]
\hspace{-50mm}
\begin{minipage}{0.27\hsize}
\begin{center}
\includegraphics[width=9.0cm,angle=0]{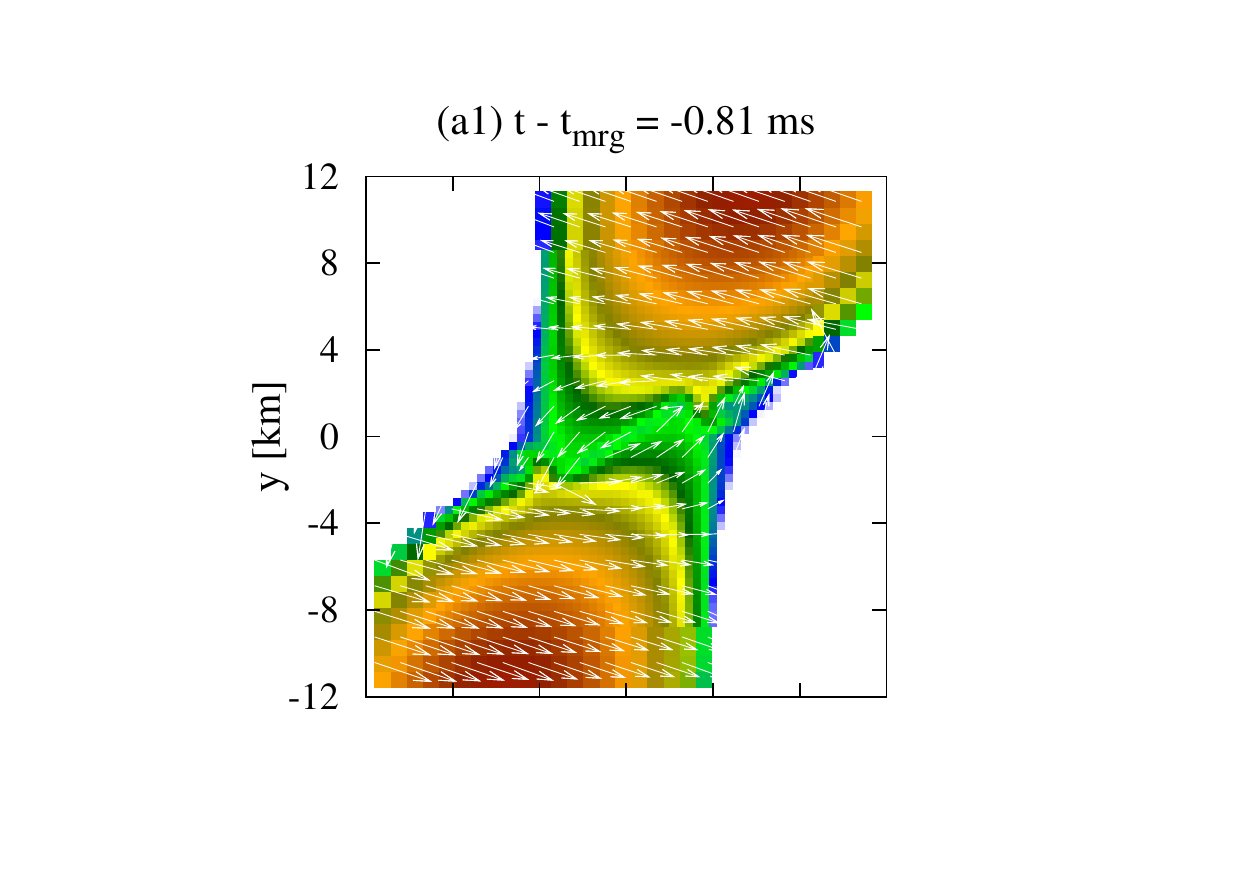}
\end{center}
\end{minipage}
\hspace{-12mm}
\begin{minipage}{0.27\hsize}
\begin{center}
\includegraphics[width=9.0cm,angle=0]{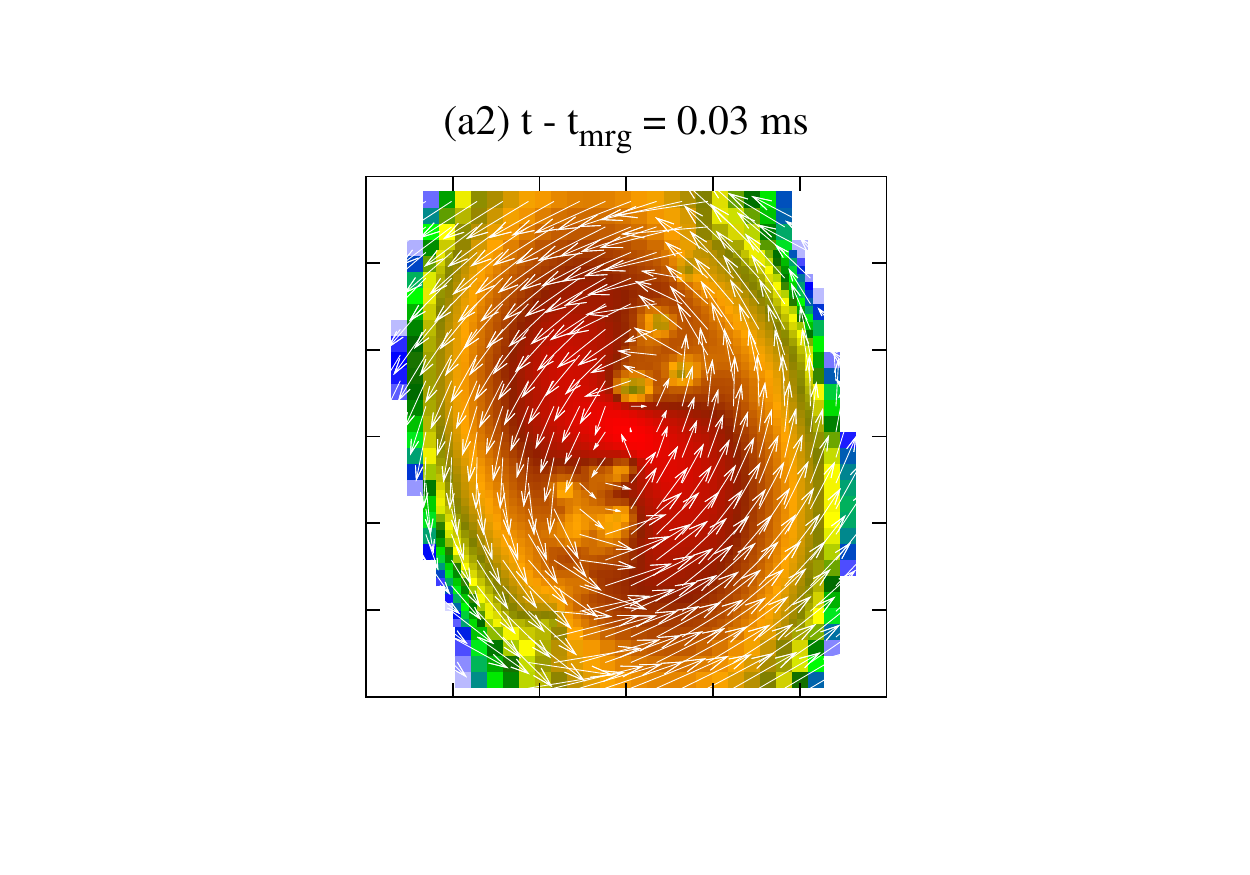}
\end{center}
\end{minipage}
\hspace{-12mm}
\begin{minipage}{0.27\hsize}
\begin{center}
\includegraphics[width=9.0cm,angle=0]{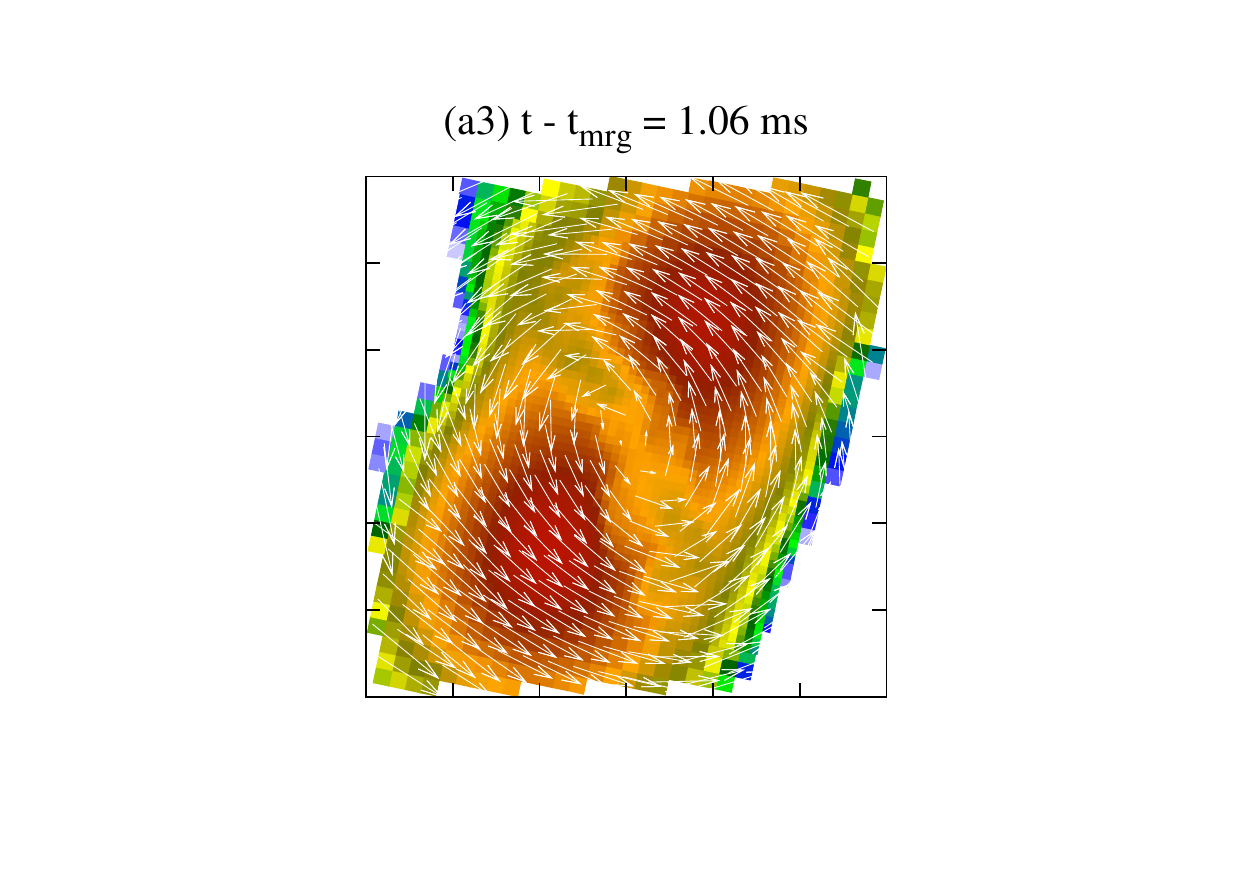}
\end{center}
\end{minipage}
\hspace{-12mm}
\begin{minipage}{0.27\hsize}
\begin{center}
\includegraphics[width=9.0cm,angle=0]{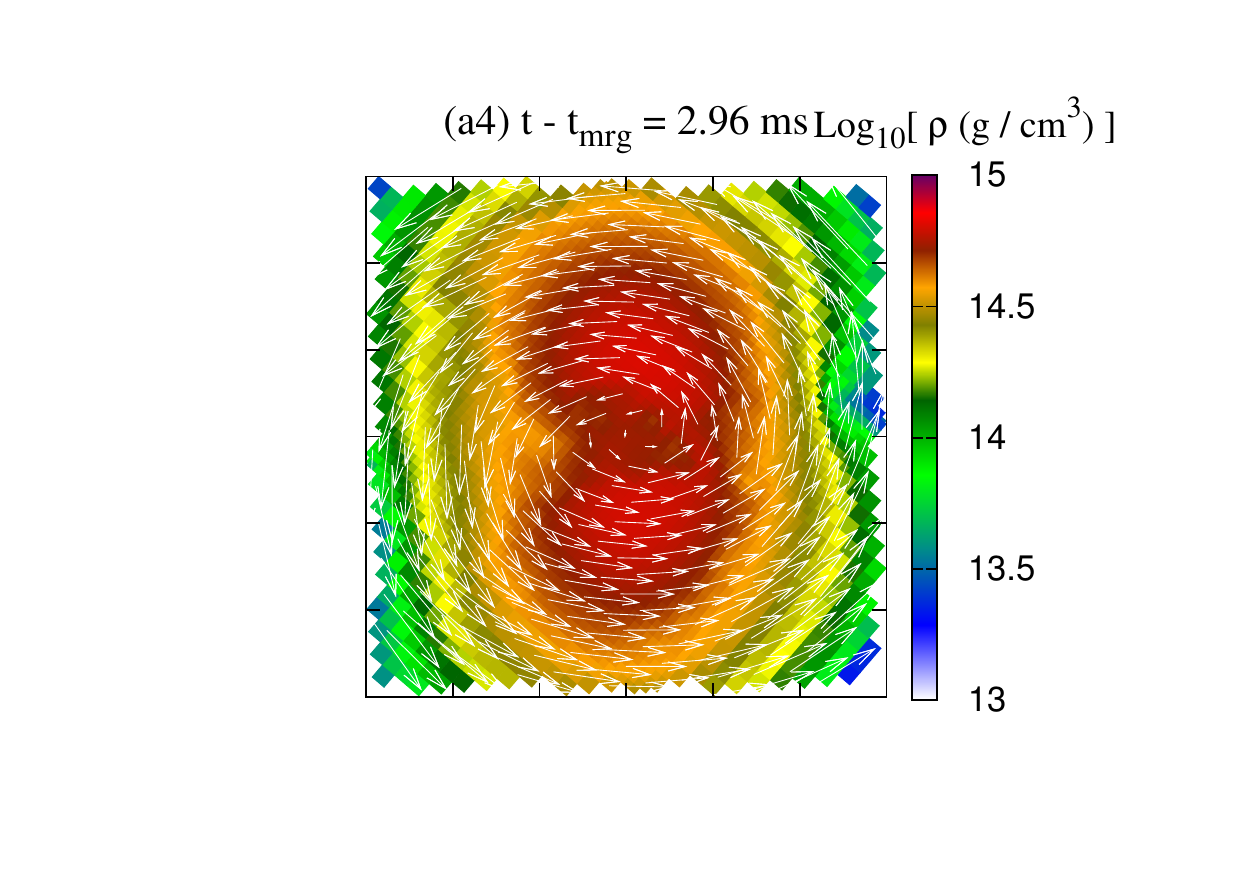}
\end{center}
\end{minipage}\\
\vspace{-19mm}
\hspace{-50mm}
\begin{minipage}{0.27\hsize}
\begin{center}
\includegraphics[width=9.0cm,angle=0]{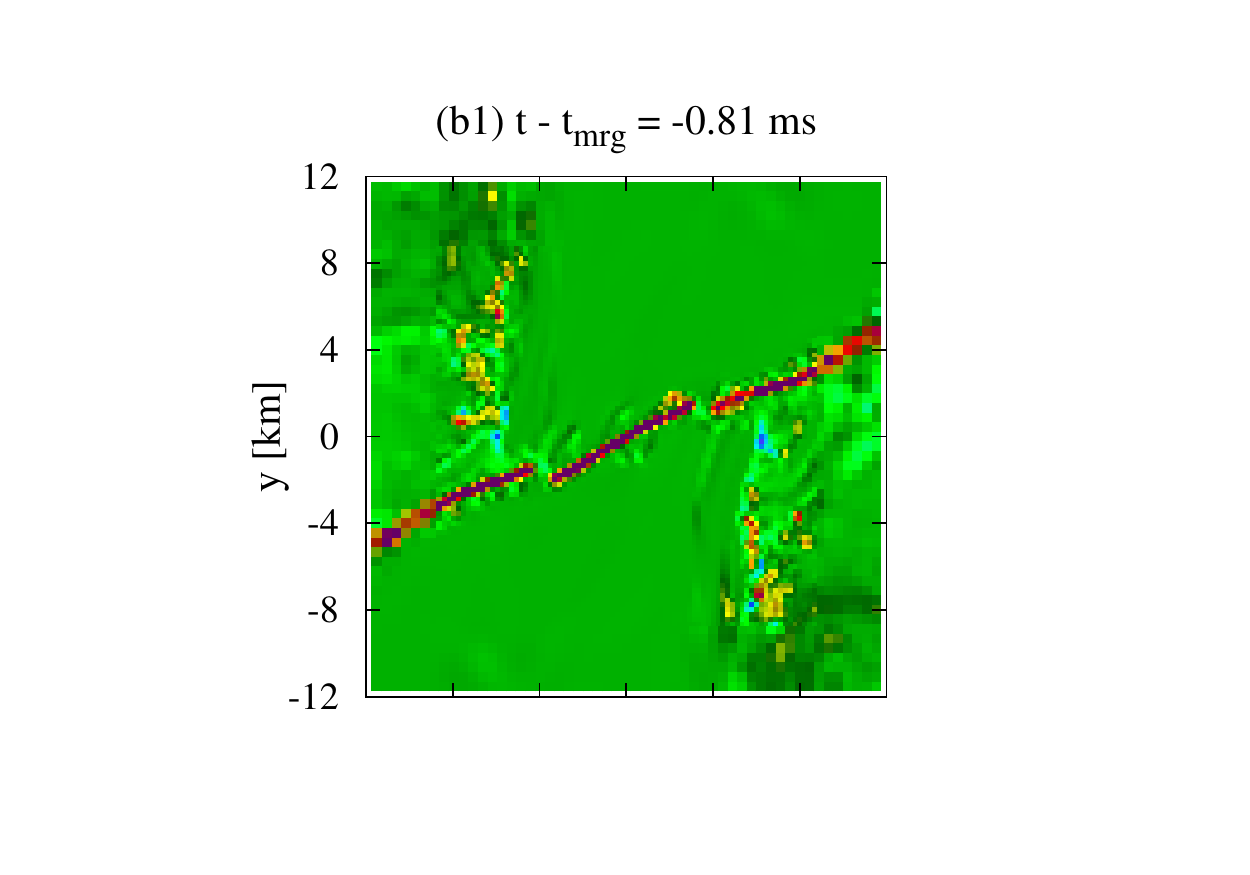}
\end{center}
\end{minipage}
\hspace{-12mm}
\begin{minipage}{0.27\hsize}
\begin{center}
\includegraphics[width=9.0cm,angle=0]{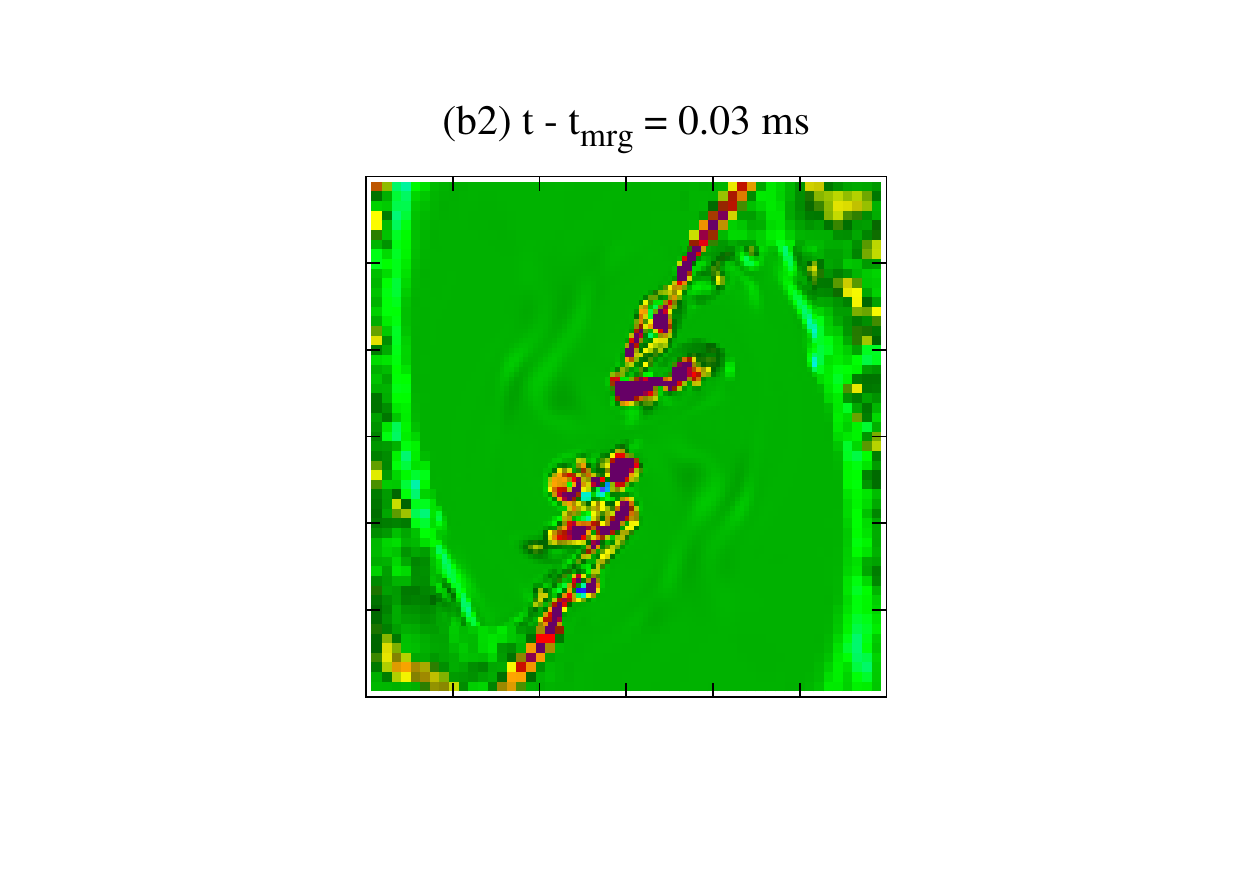}
\end{center}
\end{minipage}
\hspace{-12mm}
\begin{minipage}{0.27\hsize}
\begin{center}
\includegraphics[width=9.0cm,angle=0]{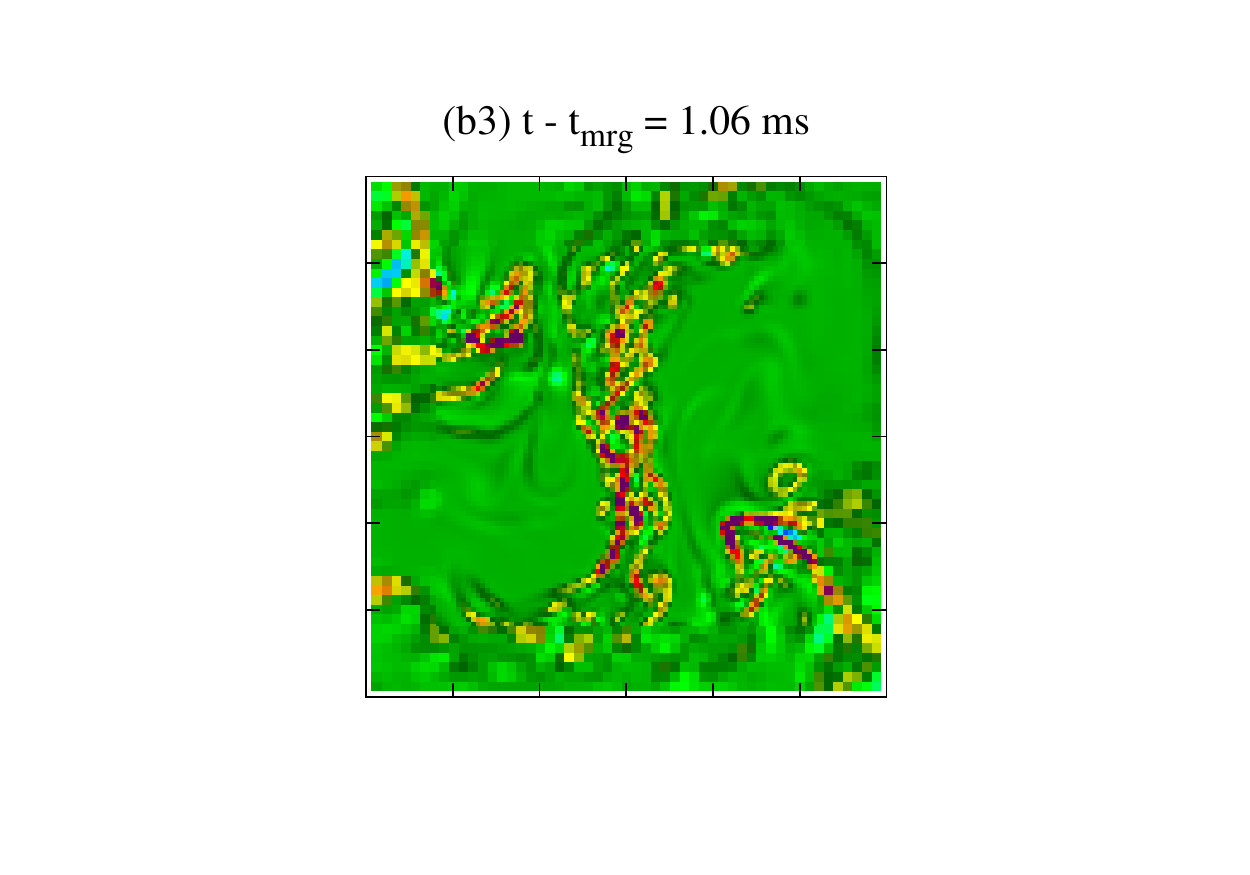}
\end{center}
\end{minipage}
\hspace{-12mm}
\begin{minipage}{0.27\hsize}
\begin{center}
\includegraphics[width=9.0cm,angle=0]{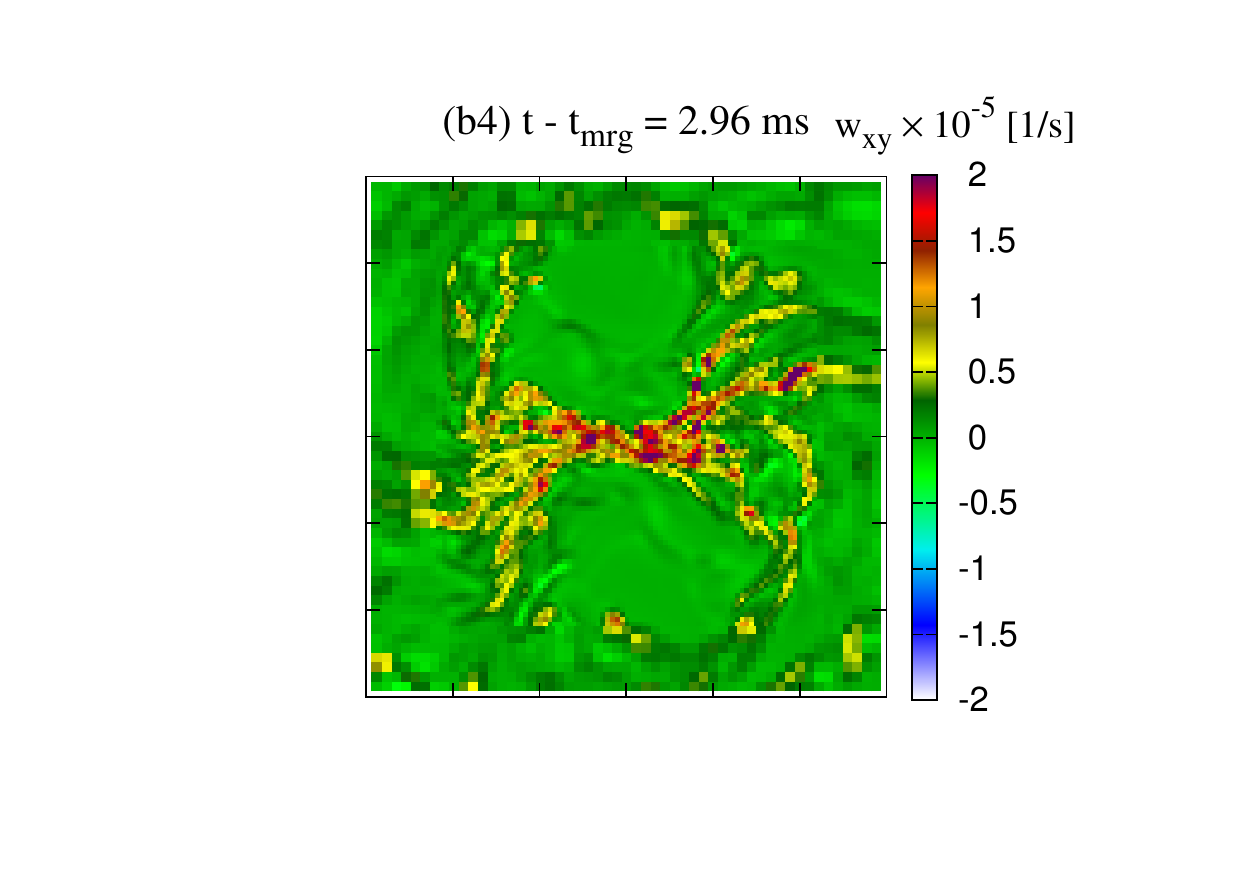}
\end{center}
\end{minipage}\\
\vspace{-19mm}
\hspace{-50mm}
\begin{minipage}{0.27\hsize}
\begin{center}
\includegraphics[width=9.0cm,angle=0]{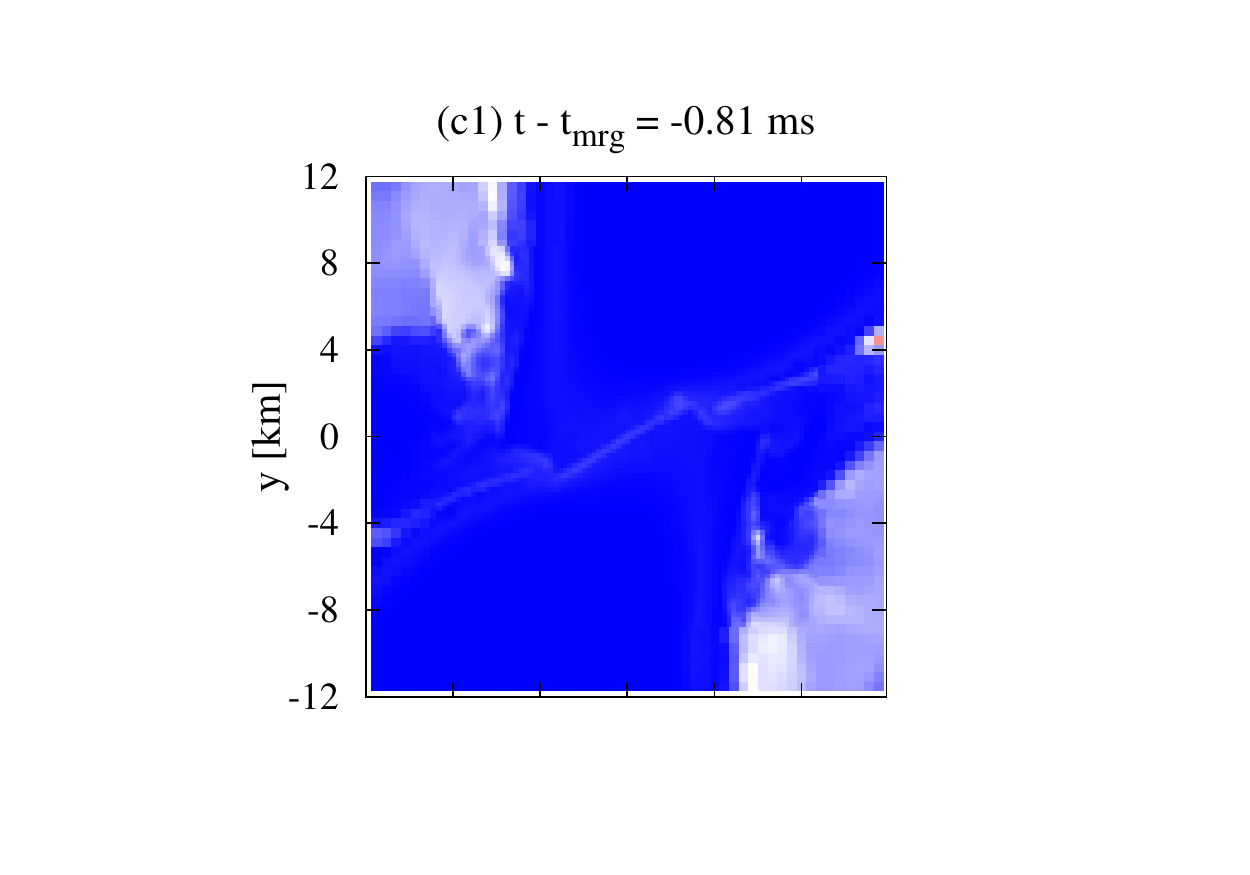}
\end{center}
\end{minipage}
\hspace{-12mm}
\begin{minipage}{0.27\hsize}
\begin{center}
\includegraphics[width=9.0cm,angle=0]{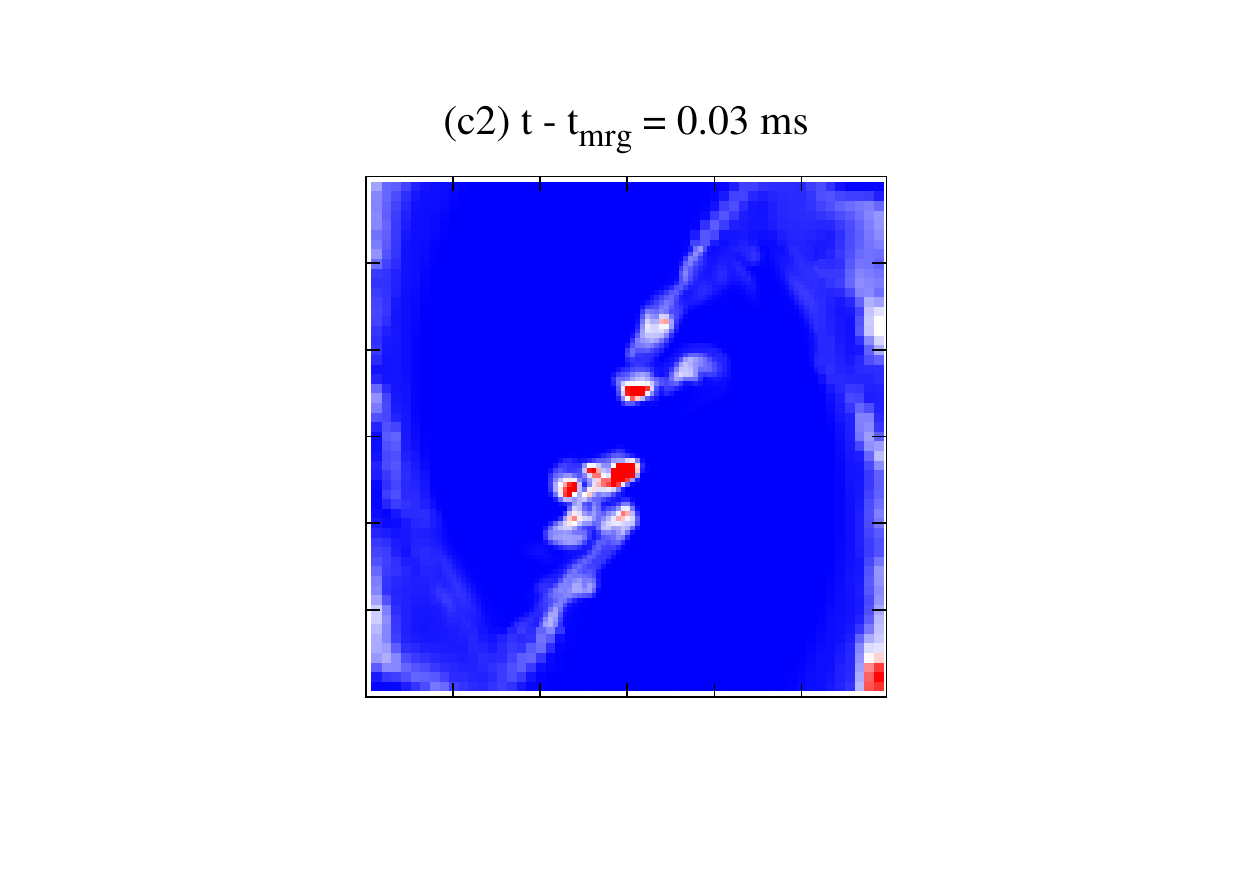}
\end{center}
\end{minipage}
\hspace{-12mm}
\begin{minipage}{0.27\hsize}
\begin{center}
\includegraphics[width=9.0cm,angle=0]{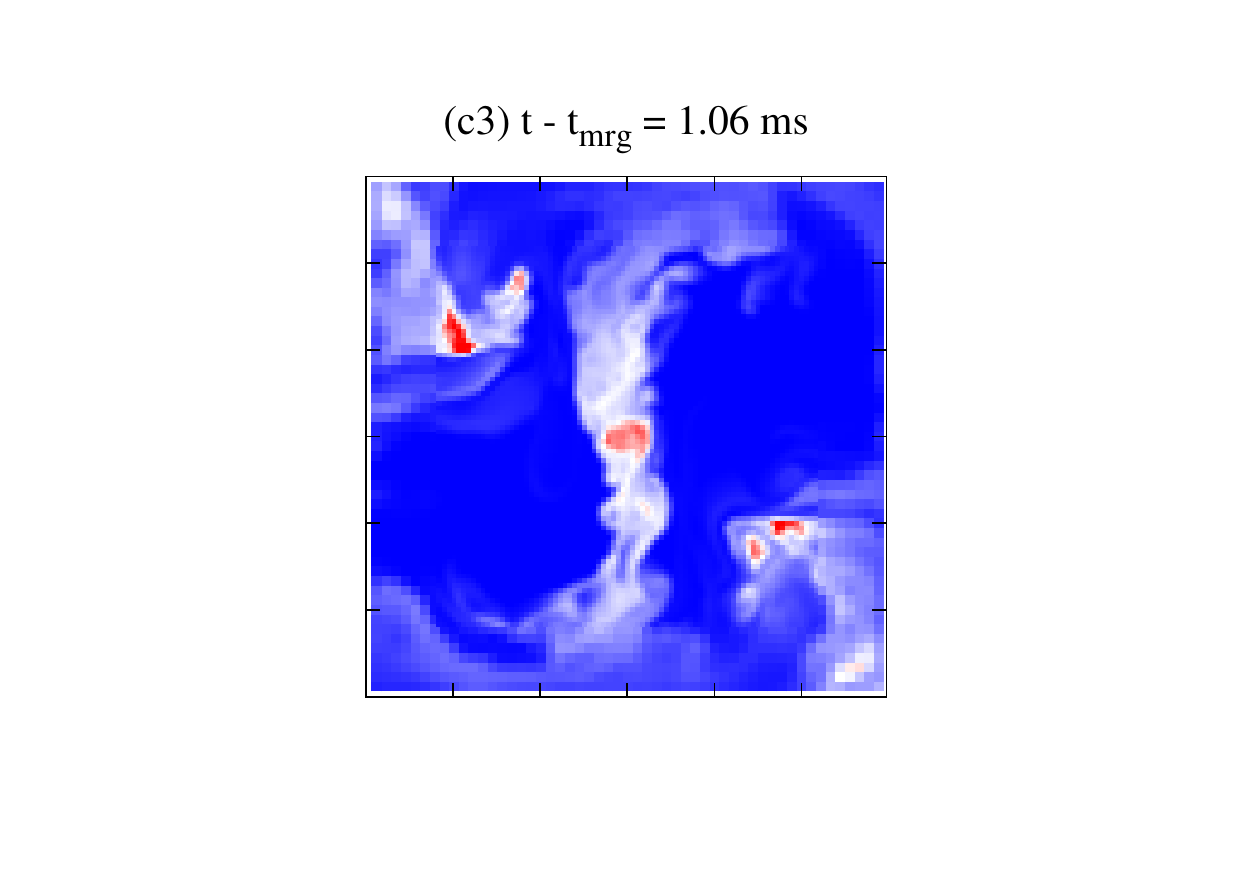}
\end{center}
\end{minipage}
\hspace{-12mm}
\begin{minipage}{0.27\hsize}
\begin{center}
\includegraphics[width=9.0cm,angle=0]{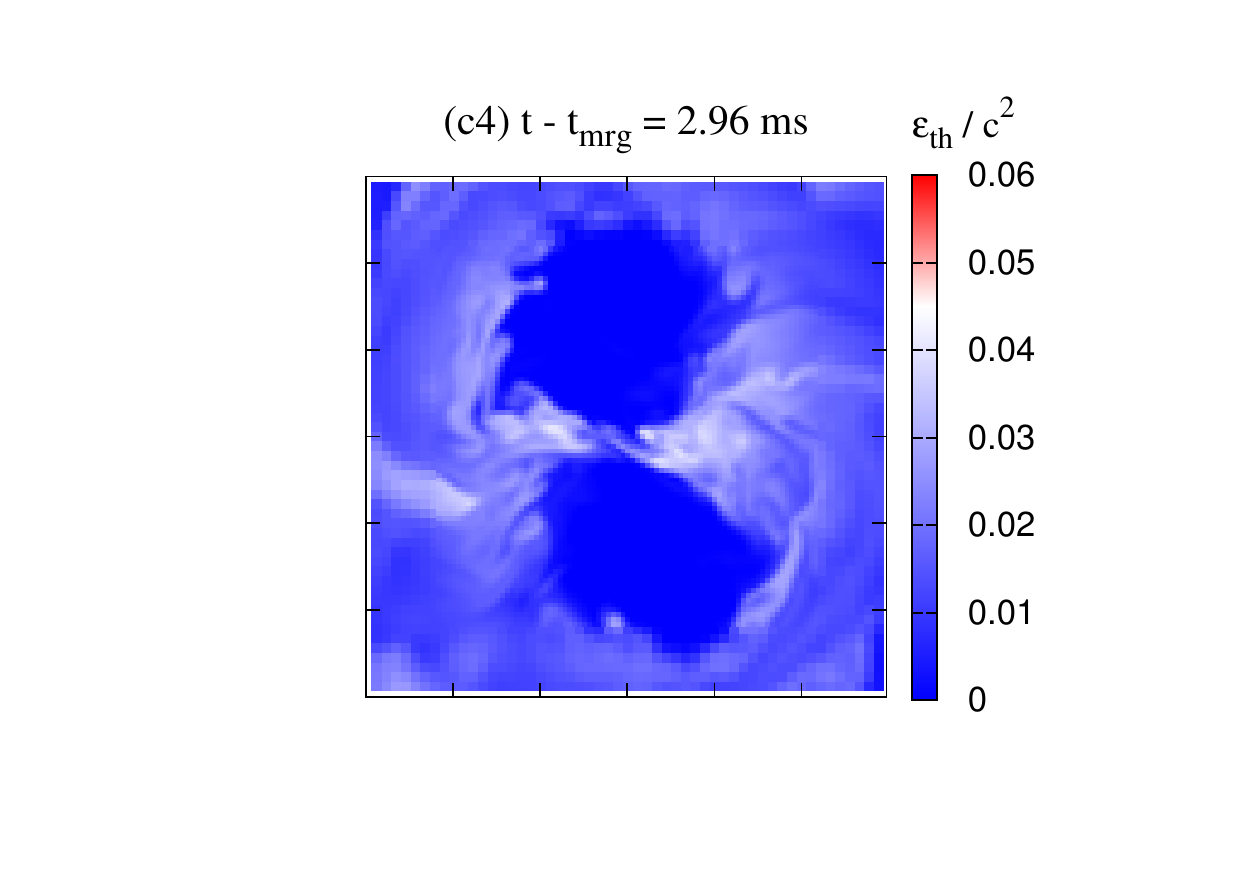}
\end{center}
\end{minipage}\\
\vspace{-19mm}
\hspace{-50mm}
\begin{minipage}{0.27\hsize}
\begin{center}
\includegraphics[width=9.0cm,angle=0]{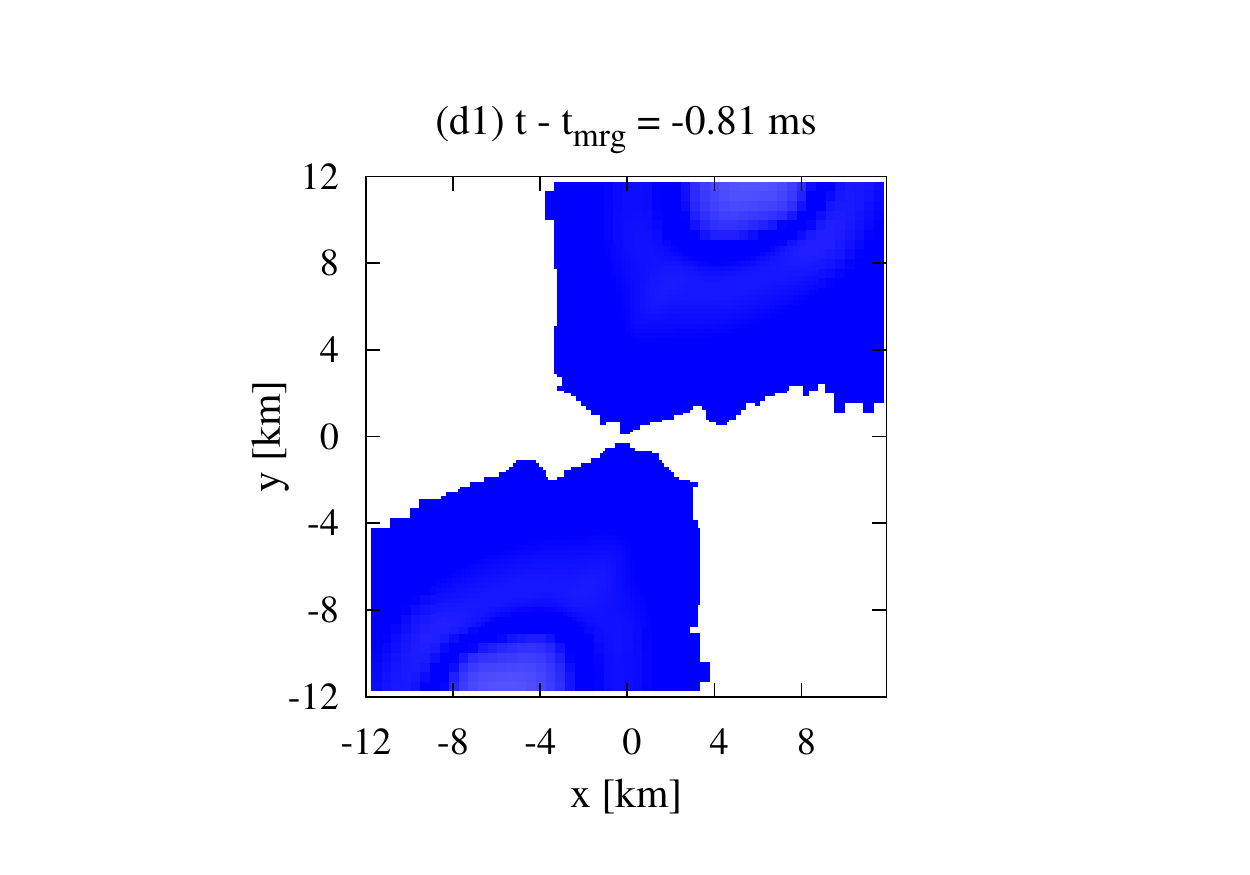}
\end{center}
\end{minipage}
\hspace{-12mm}
\begin{minipage}{0.27\hsize}
\begin{center}
\includegraphics[width=9.0cm,angle=0]{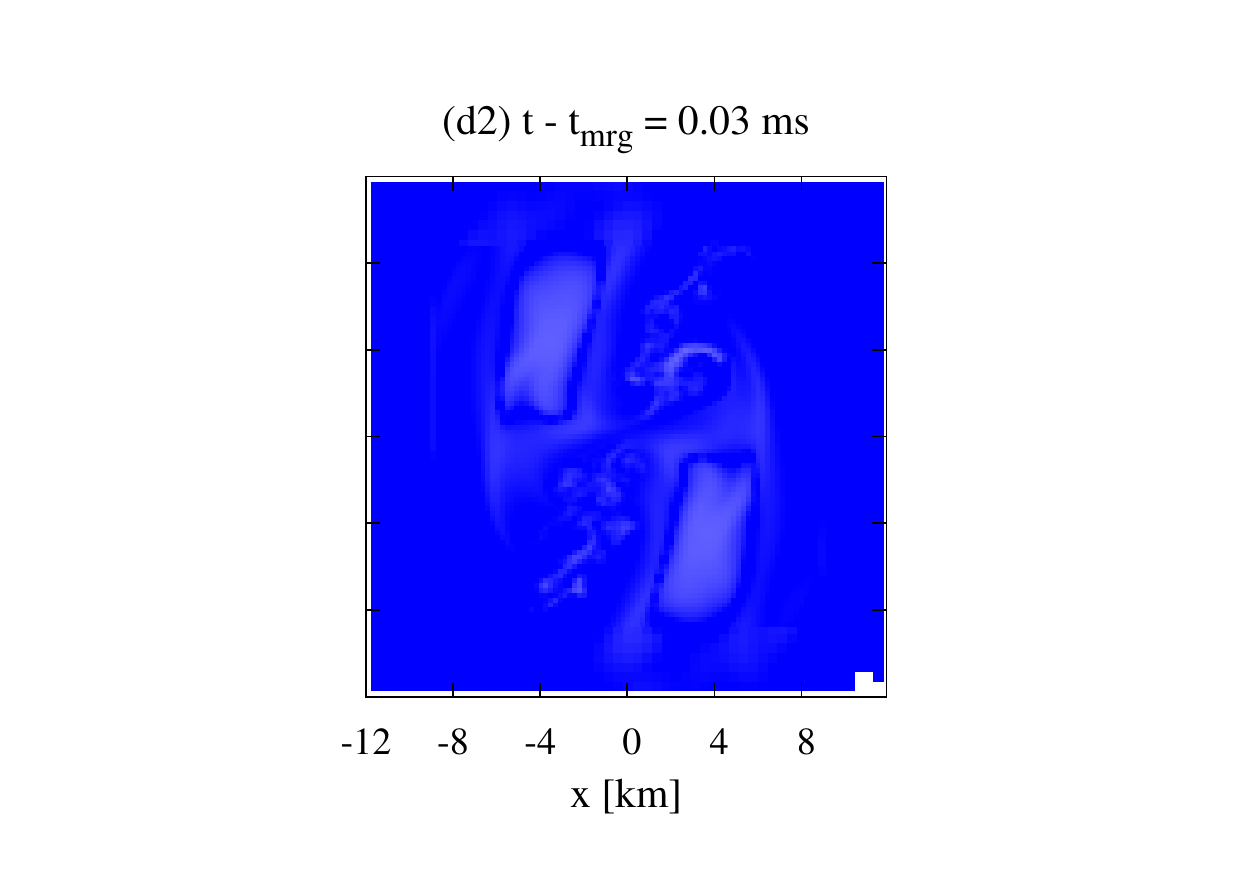}
\end{center}
\end{minipage}
\hspace{-12mm}
\begin{minipage}{0.27\hsize}
\begin{center}
\includegraphics[width=9.0cm,angle=0]{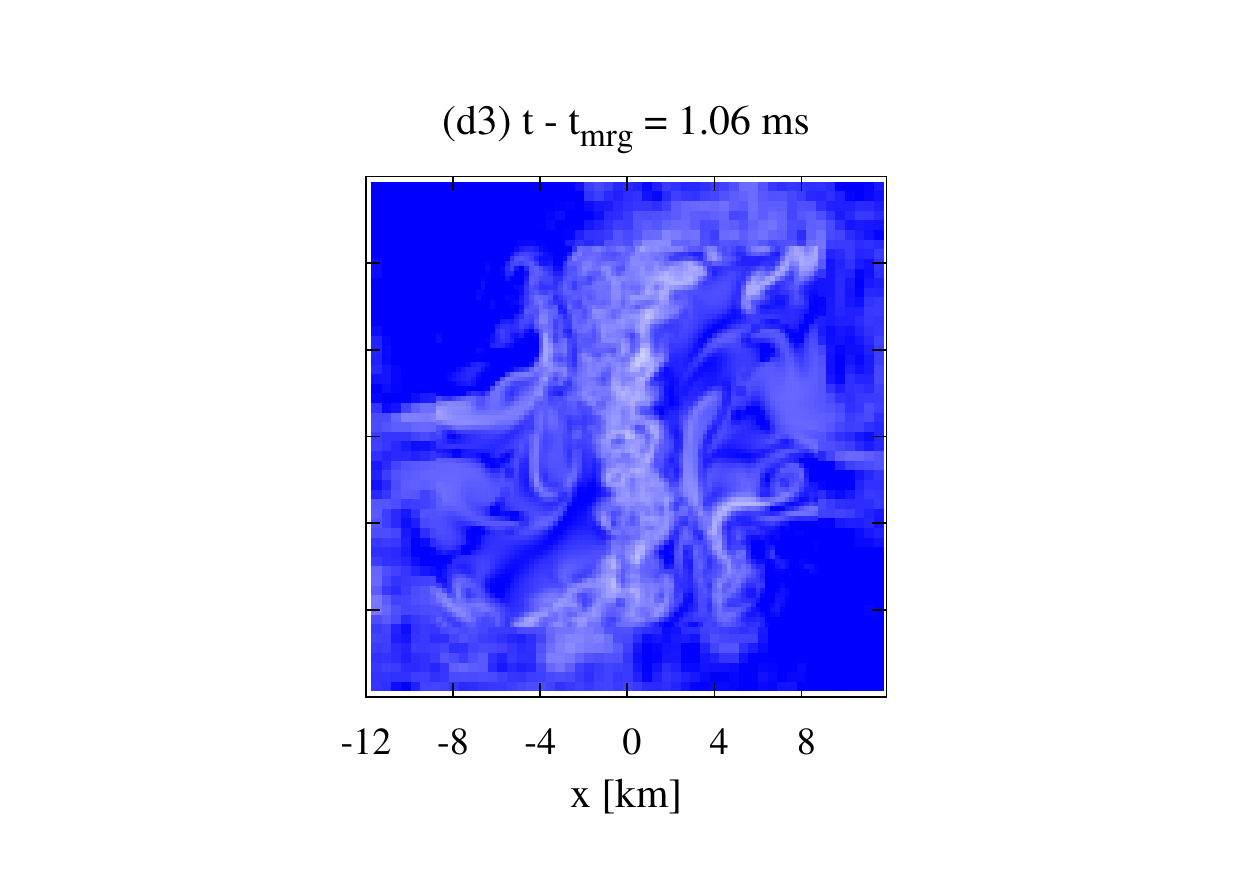}
\end{center}
\end{minipage}
\hspace{-12mm}
\begin{minipage}{0.27\hsize}
\begin{center}
\includegraphics[width=9.0cm,angle=0]{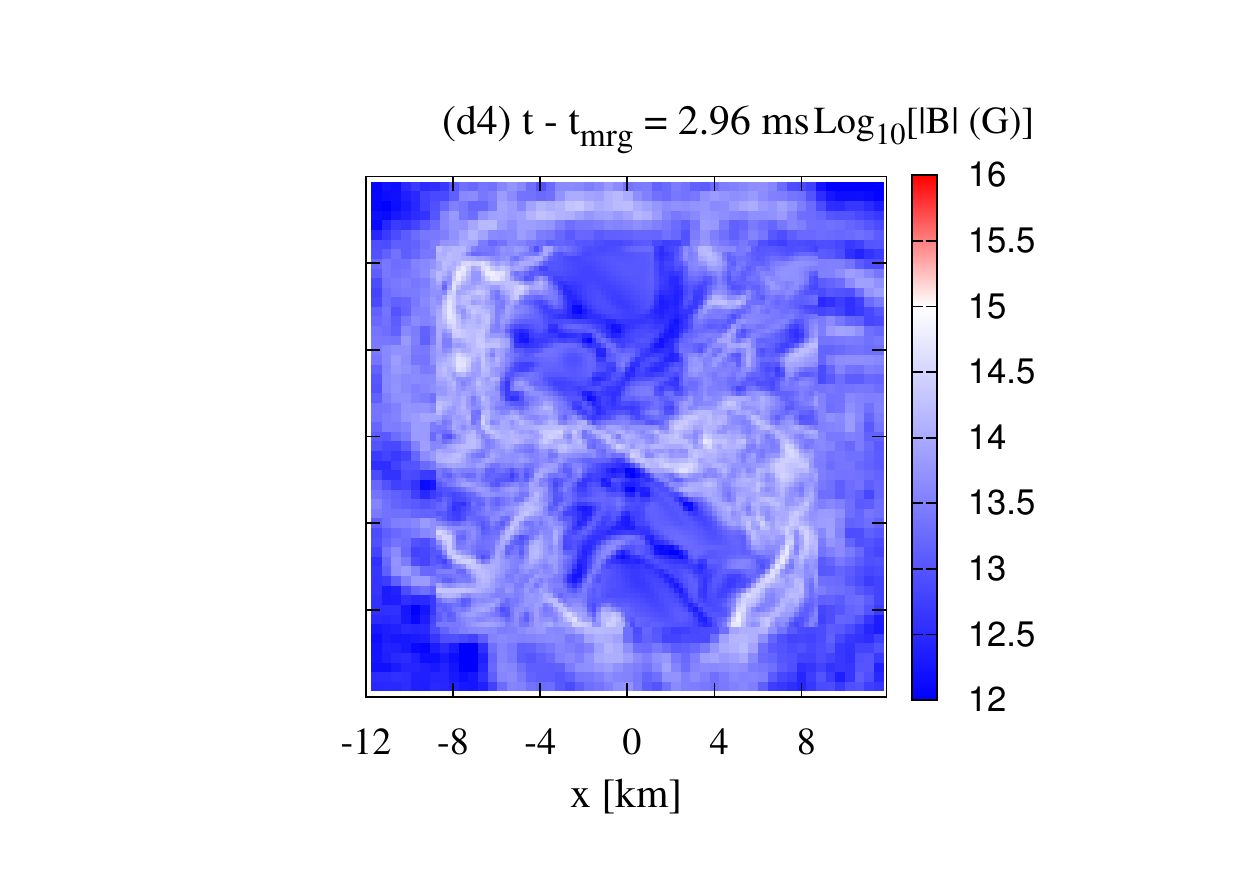}
\end{center}
\end{minipage}
\caption{\label{fig2}
Same as Fig.~\ref{fig1}, but for $\Delta x_{(l_{\rm max})}=37.5$ m.
}
\end{figure*}

\begin{figure*}[t]
\hspace{-50mm}
\begin{minipage}{0.27\hsize}
\begin{center}
\includegraphics[width=9.0cm,angle=0]{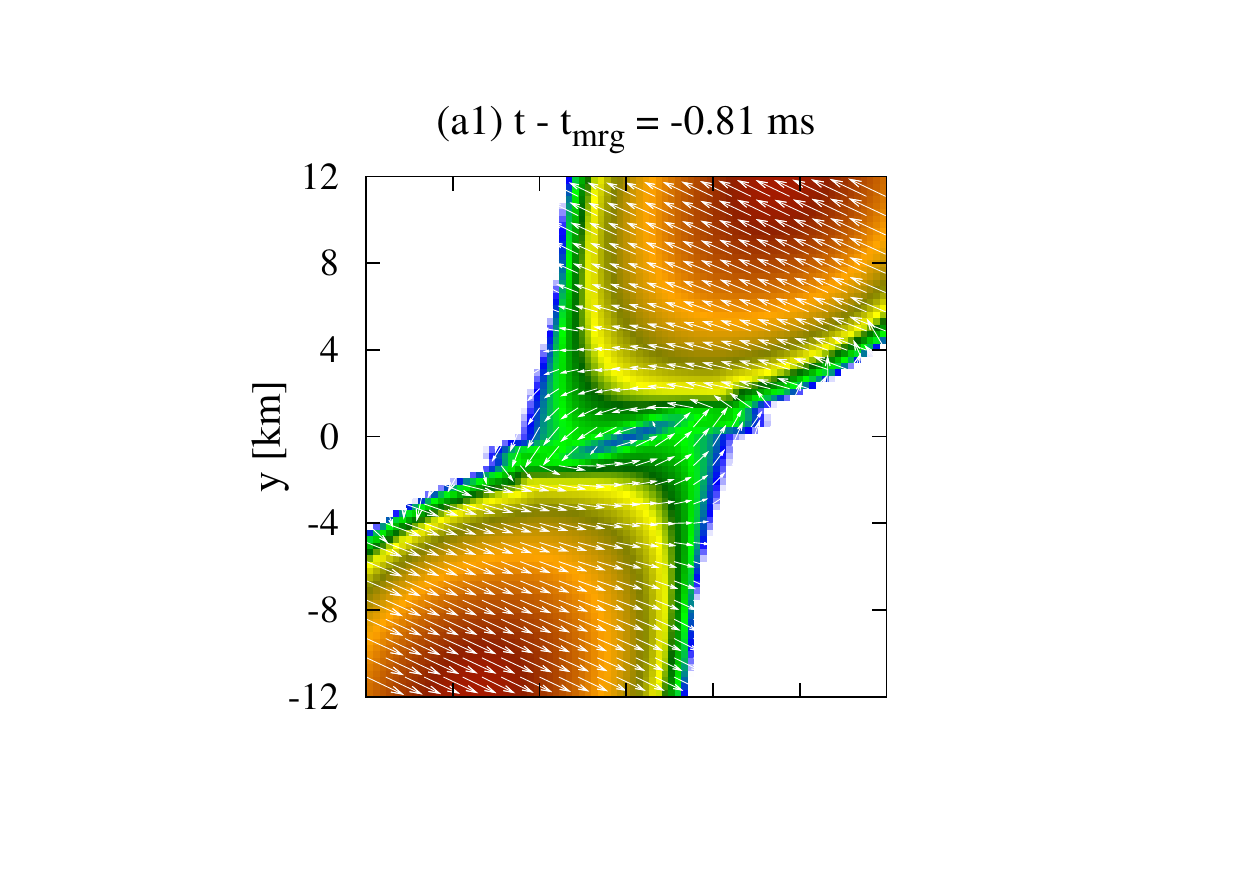}
\end{center}
\end{minipage}
\hspace{-12mm}
\begin{minipage}{0.27\hsize}
\begin{center}
\includegraphics[width=9.0cm,angle=0]{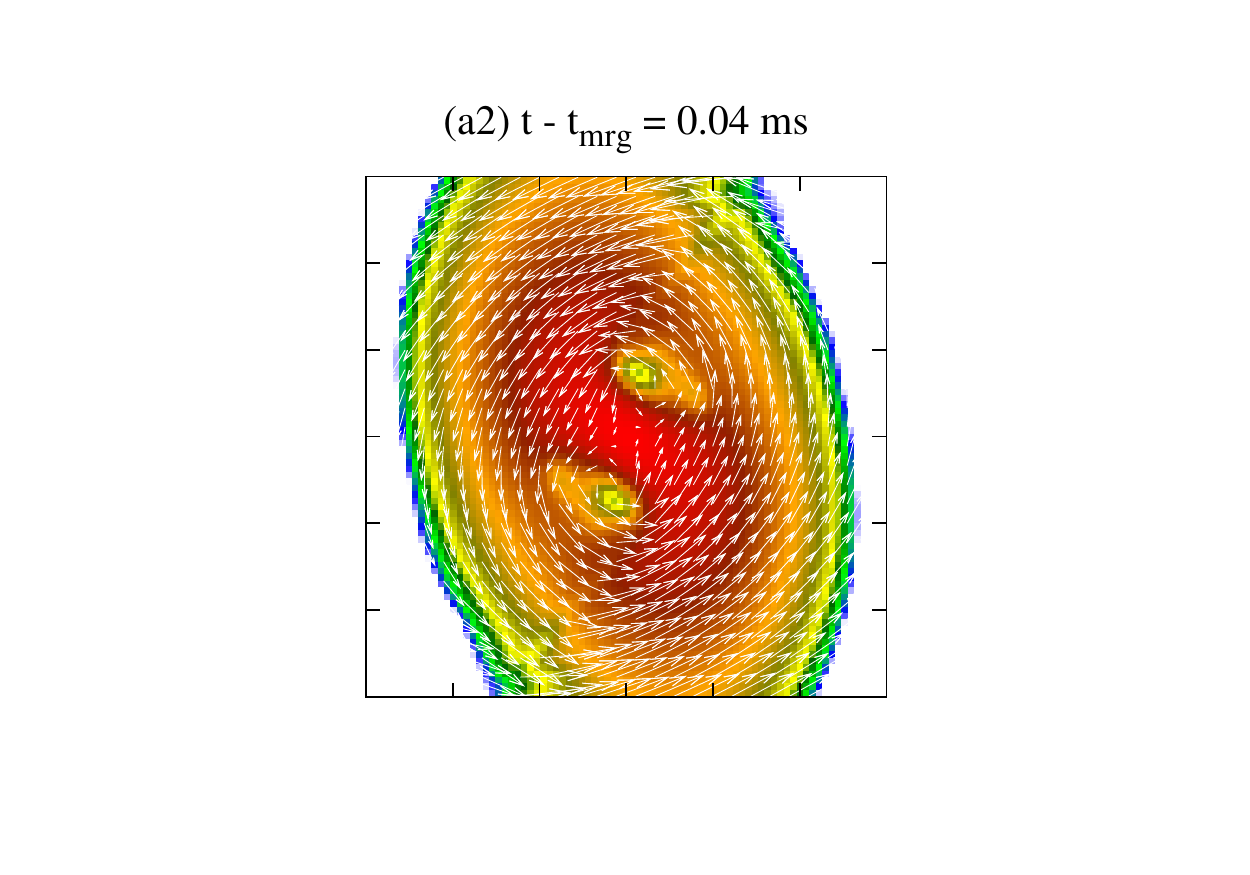}
\end{center}
\end{minipage}
\hspace{-12mm}
\begin{minipage}{0.27\hsize}
\begin{center}
\includegraphics[width=9.0cm,angle=0]{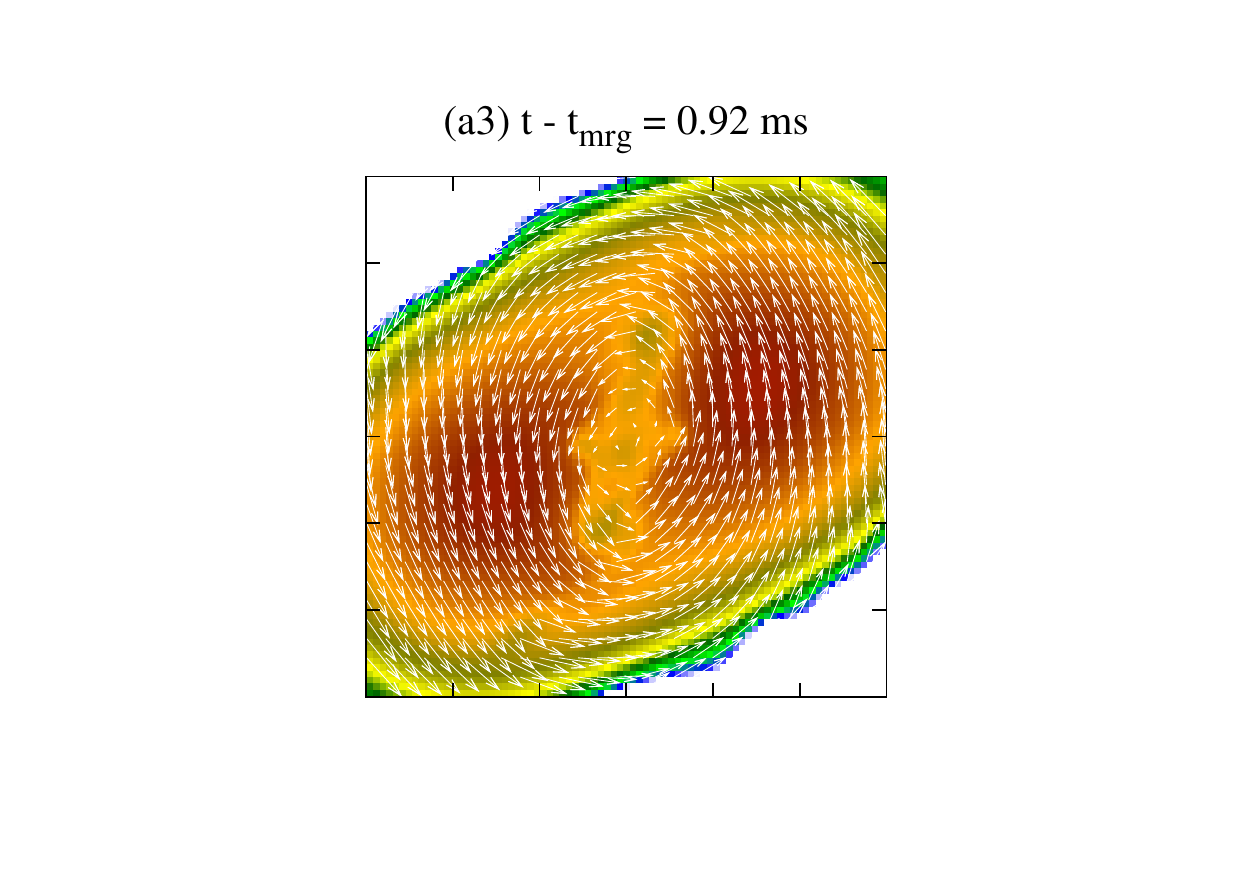}
\end{center}
\end{minipage}
\hspace{-12mm}
\begin{minipage}{0.27\hsize}
\begin{center}
\includegraphics[width=9.0cm,angle=0]{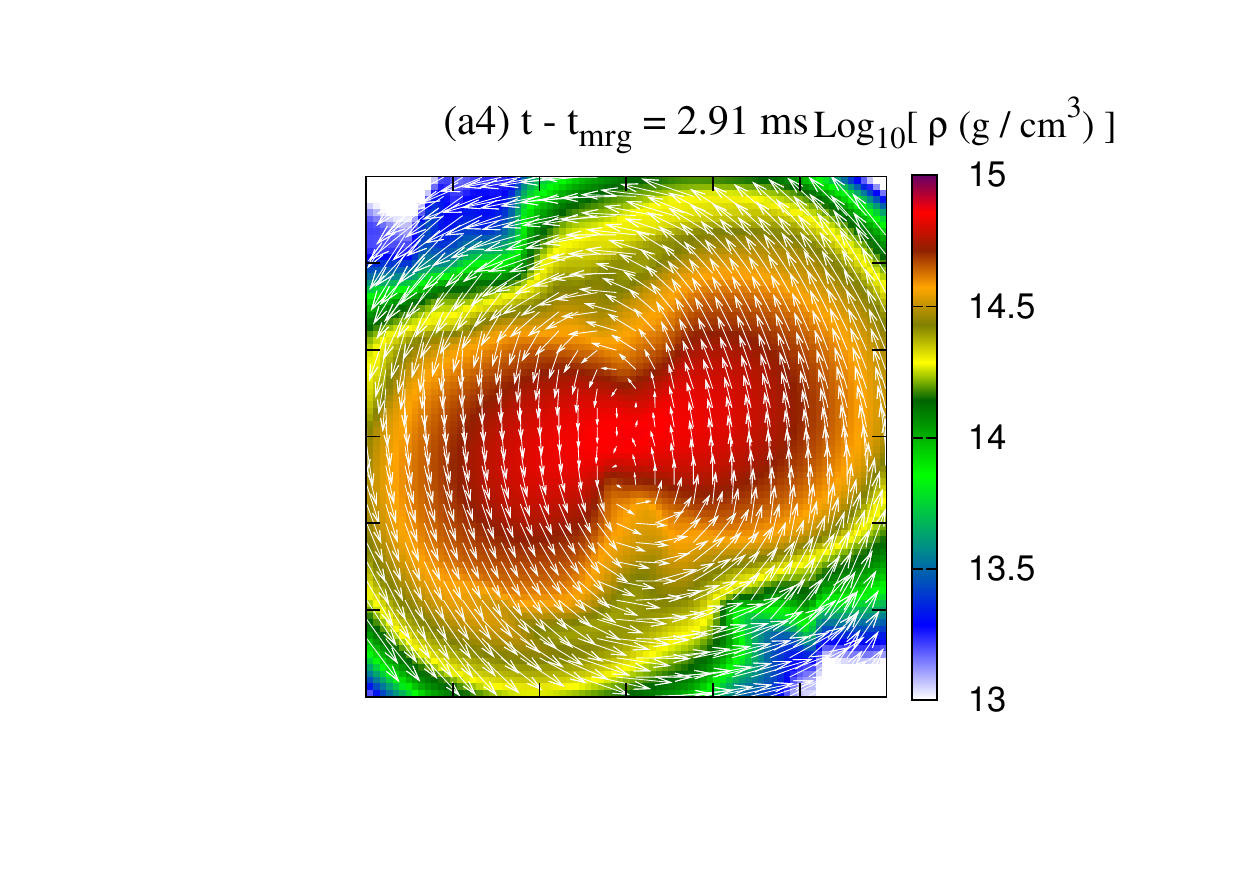}
\end{center}
\end{minipage}\\
\vspace{-19mm}
\hspace{-50mm}
\begin{minipage}{0.27\hsize}
\begin{center}
\includegraphics[width=9.0cm,angle=0]{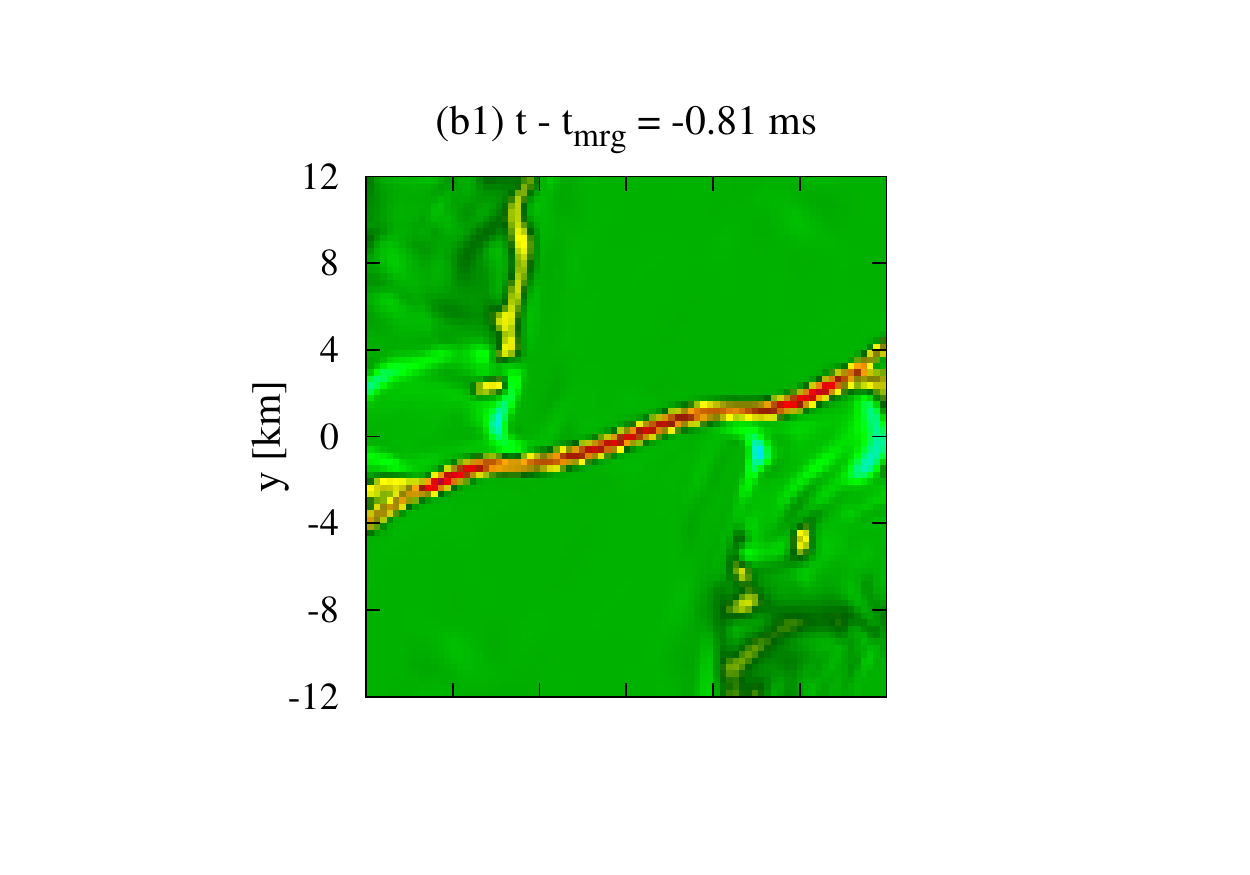}
\end{center}
\end{minipage}
\hspace{-12mm}
\begin{minipage}{0.27\hsize}
\begin{center}
\includegraphics[width=9.0cm,angle=0]{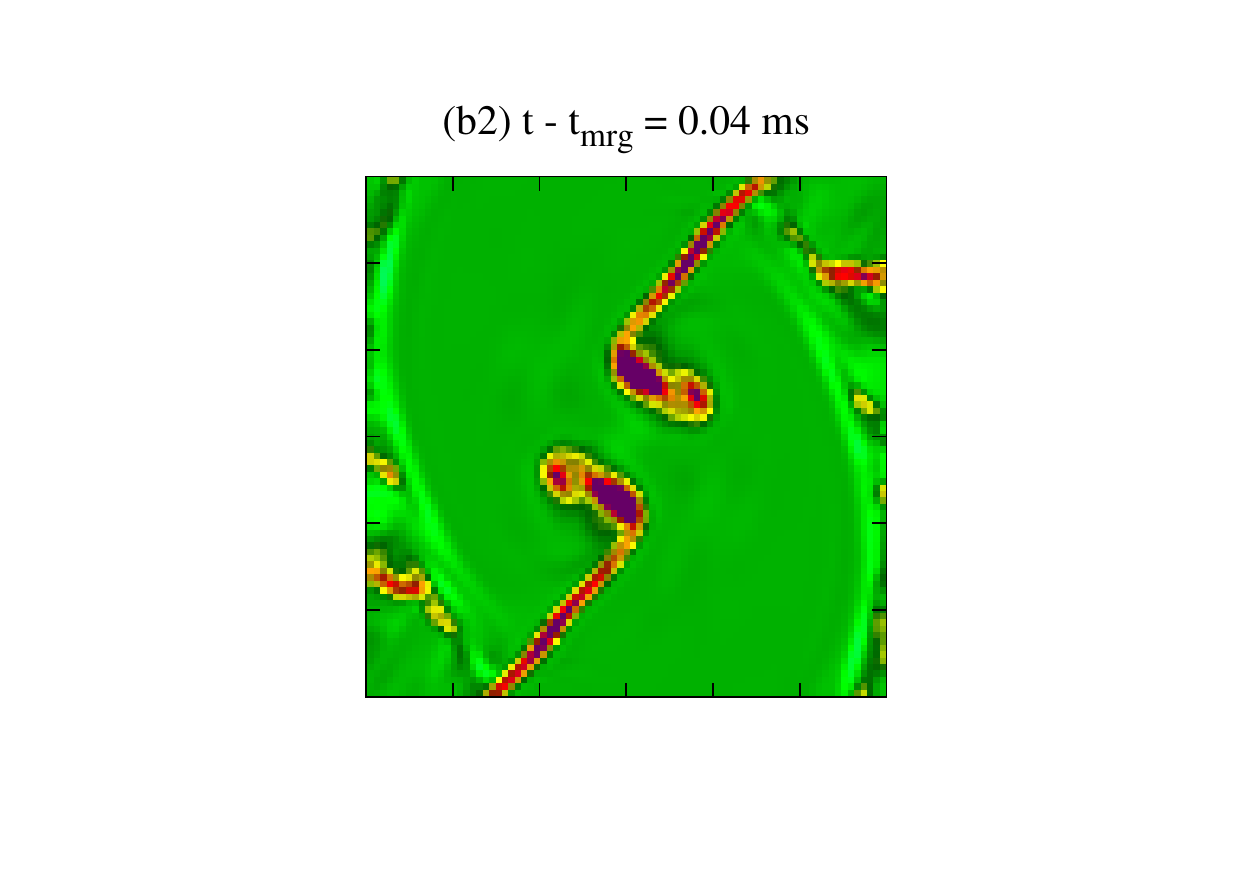}
\end{center}
\end{minipage}
\hspace{-12mm}
\begin{minipage}{0.27\hsize}
\begin{center}
\includegraphics[width=9.0cm,angle=0]{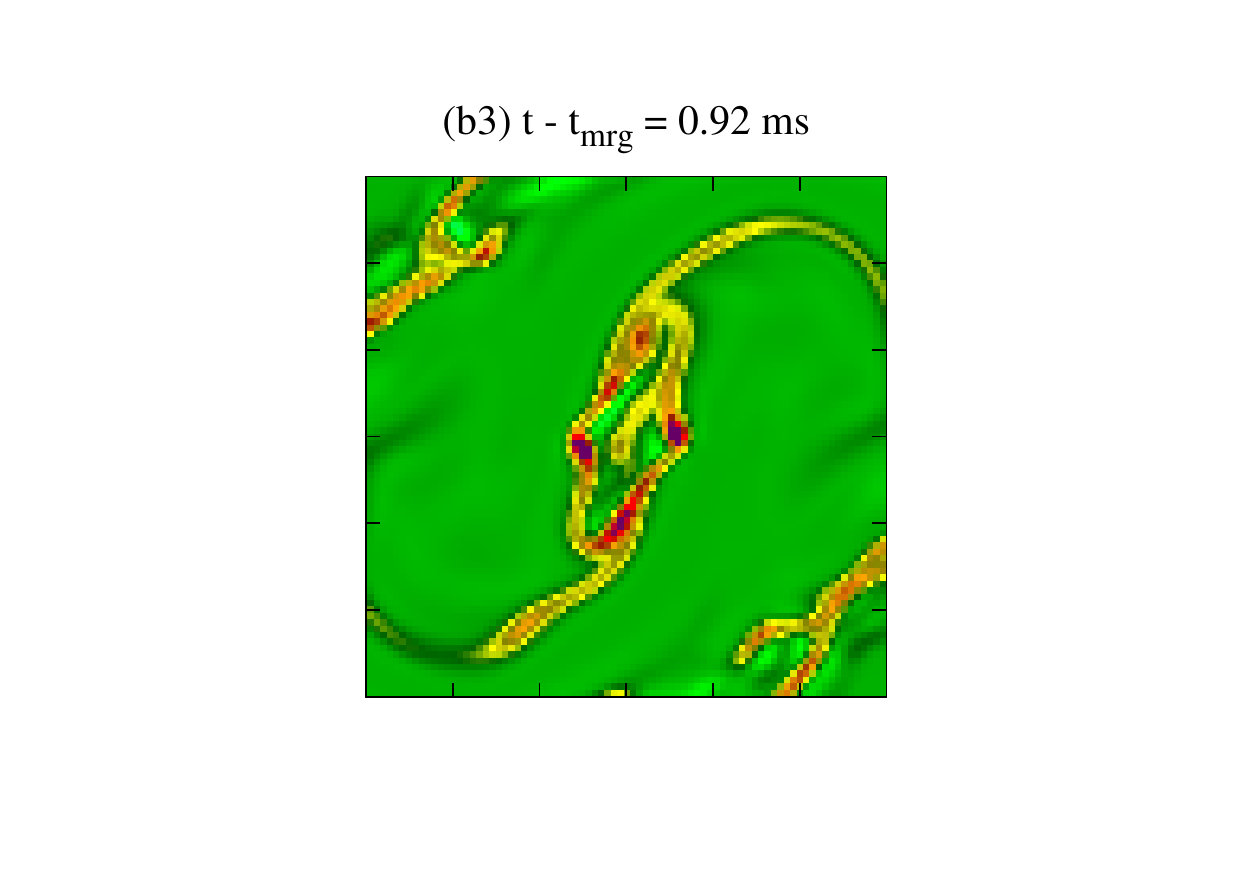}
\end{center}
\end{minipage}
\hspace{-12mm}
\begin{minipage}{0.27\hsize}
\begin{center}
\includegraphics[width=9.0cm,angle=0]{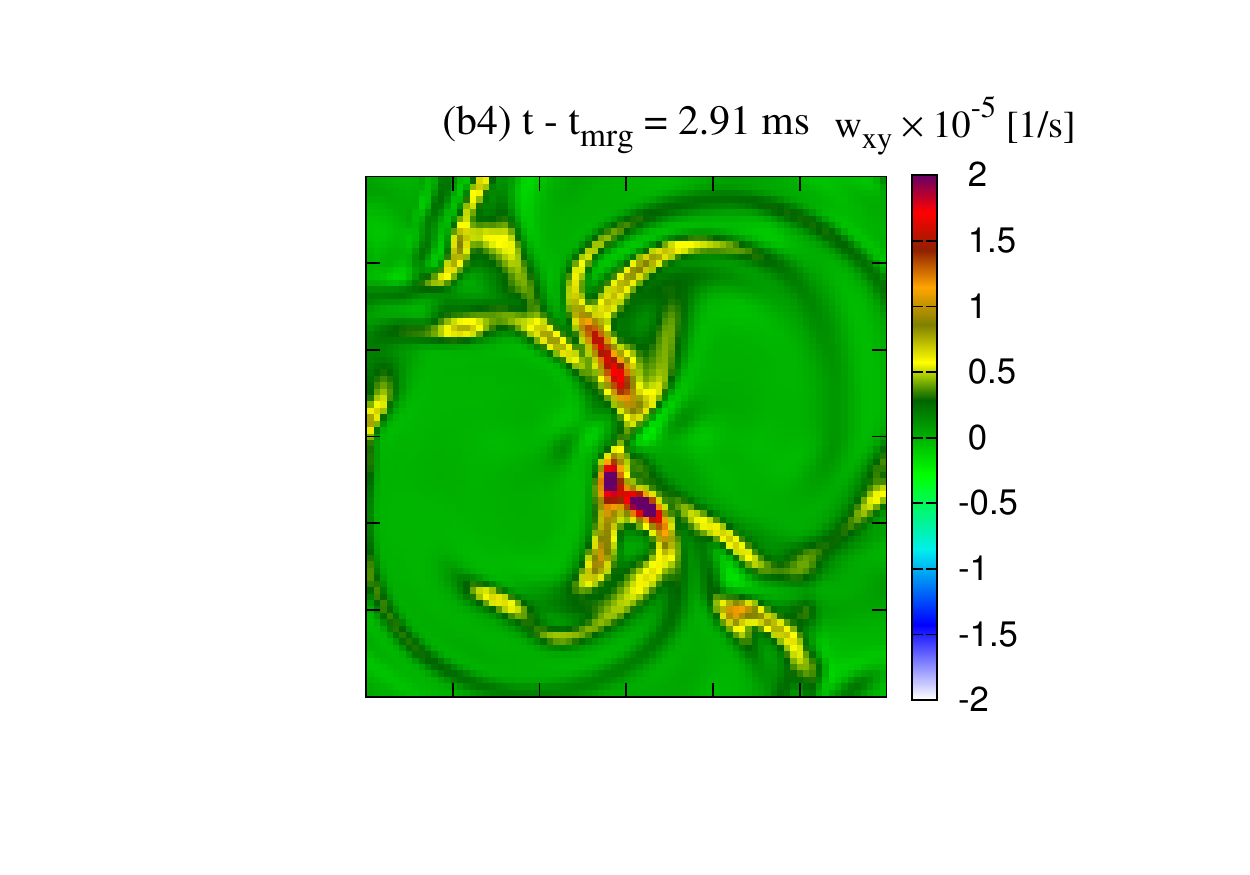}
\end{center}
\end{minipage}\\
\vspace{-19mm}
\hspace{-50mm}
\begin{minipage}{0.27\hsize}
\begin{center}
\includegraphics[width=9.0cm,angle=0]{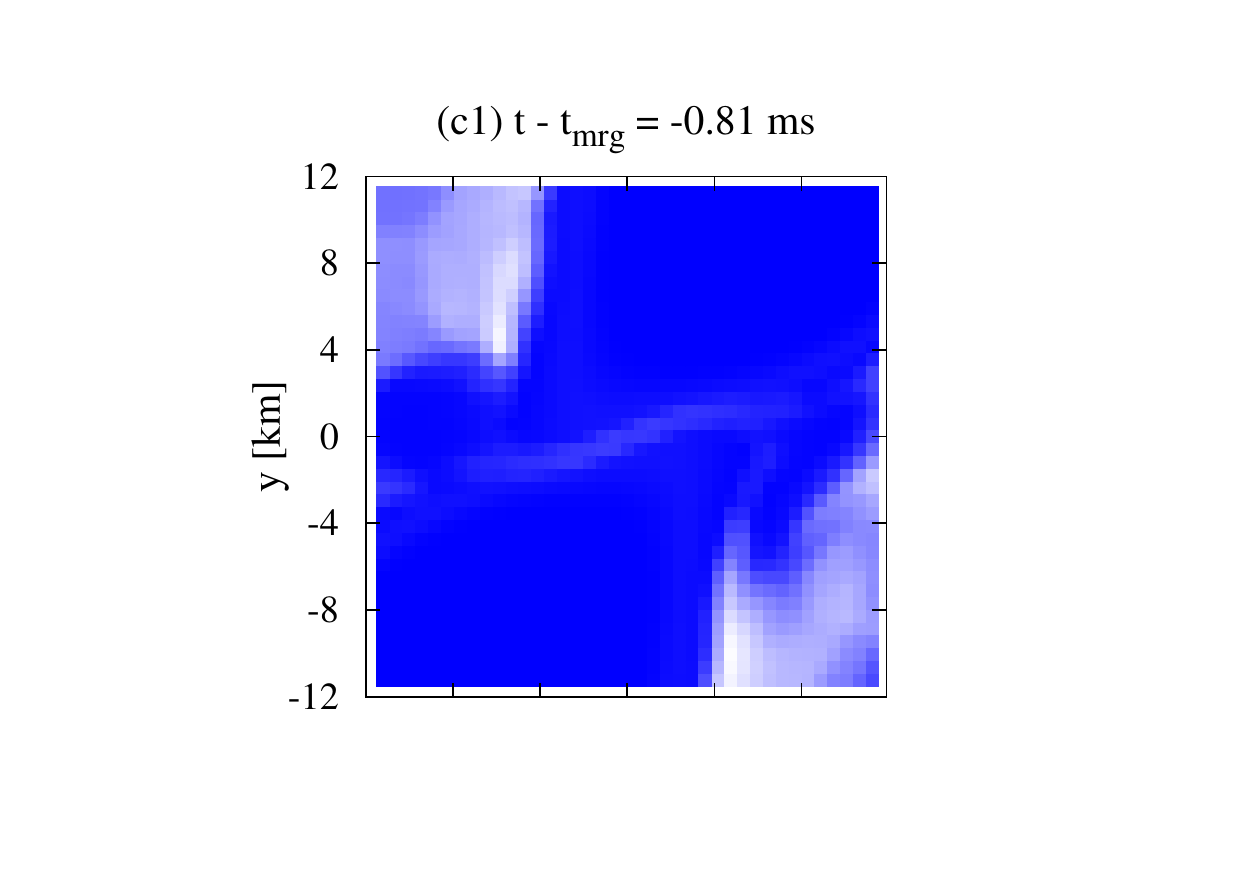}
\end{center}
\end{minipage}
\hspace{-12mm}
\begin{minipage}{0.27\hsize}
\begin{center}
\includegraphics[width=9.0cm,angle=0]{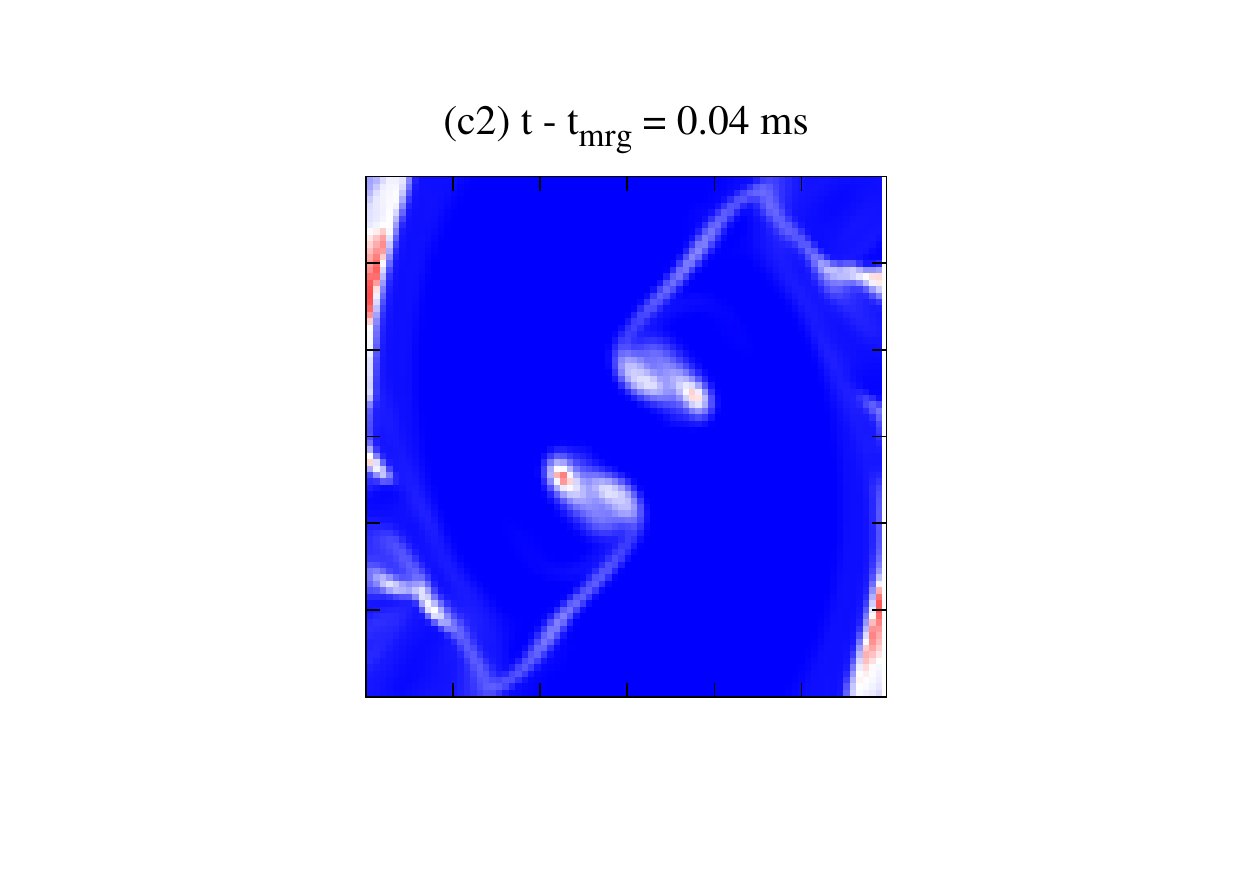}
\end{center}
\end{minipage}
\hspace{-12mm}
\begin{minipage}{0.27\hsize}
\begin{center}
\includegraphics[width=9.0cm,angle=0]{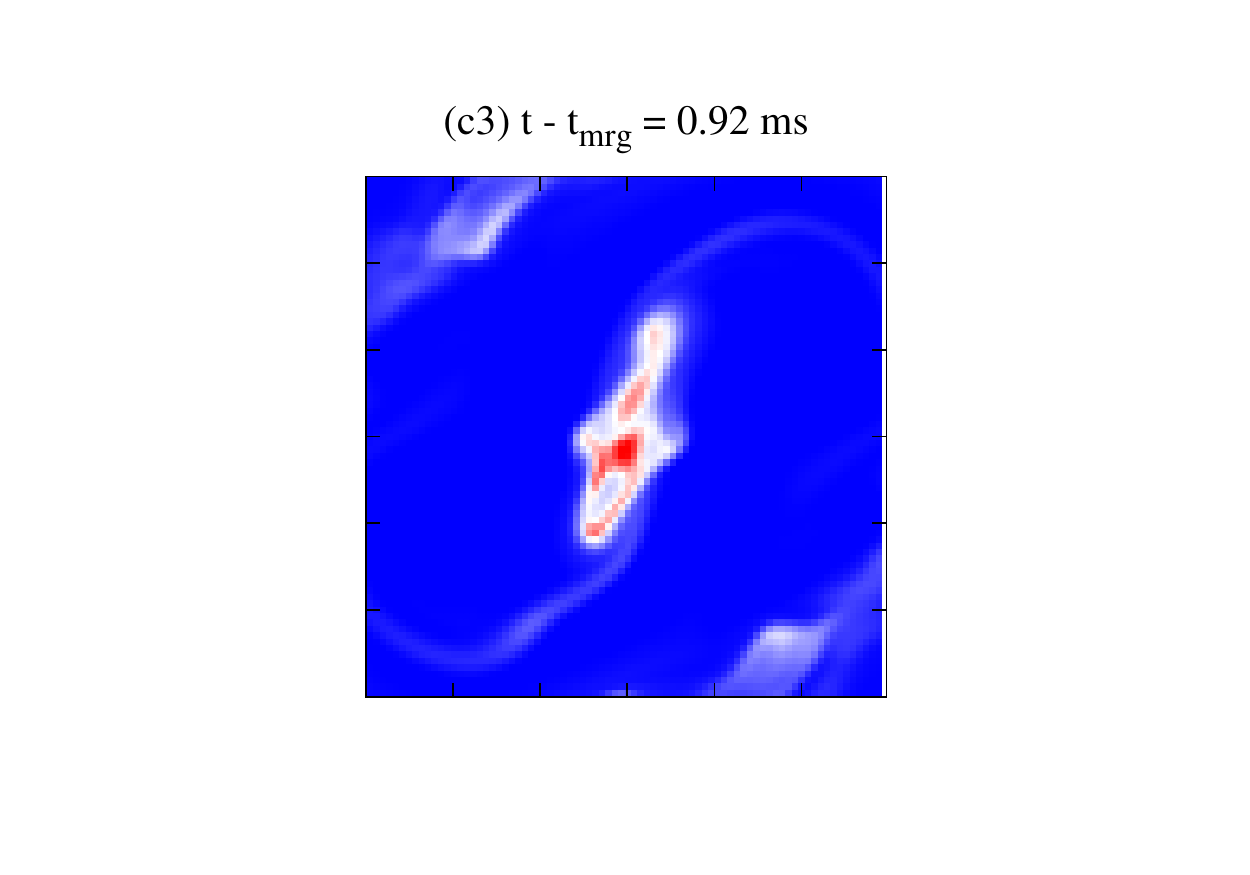}
\end{center}
\end{minipage}
\hspace{-12mm}
\begin{minipage}{0.27\hsize}
\begin{center}
\includegraphics[width=9.0cm,angle=0]{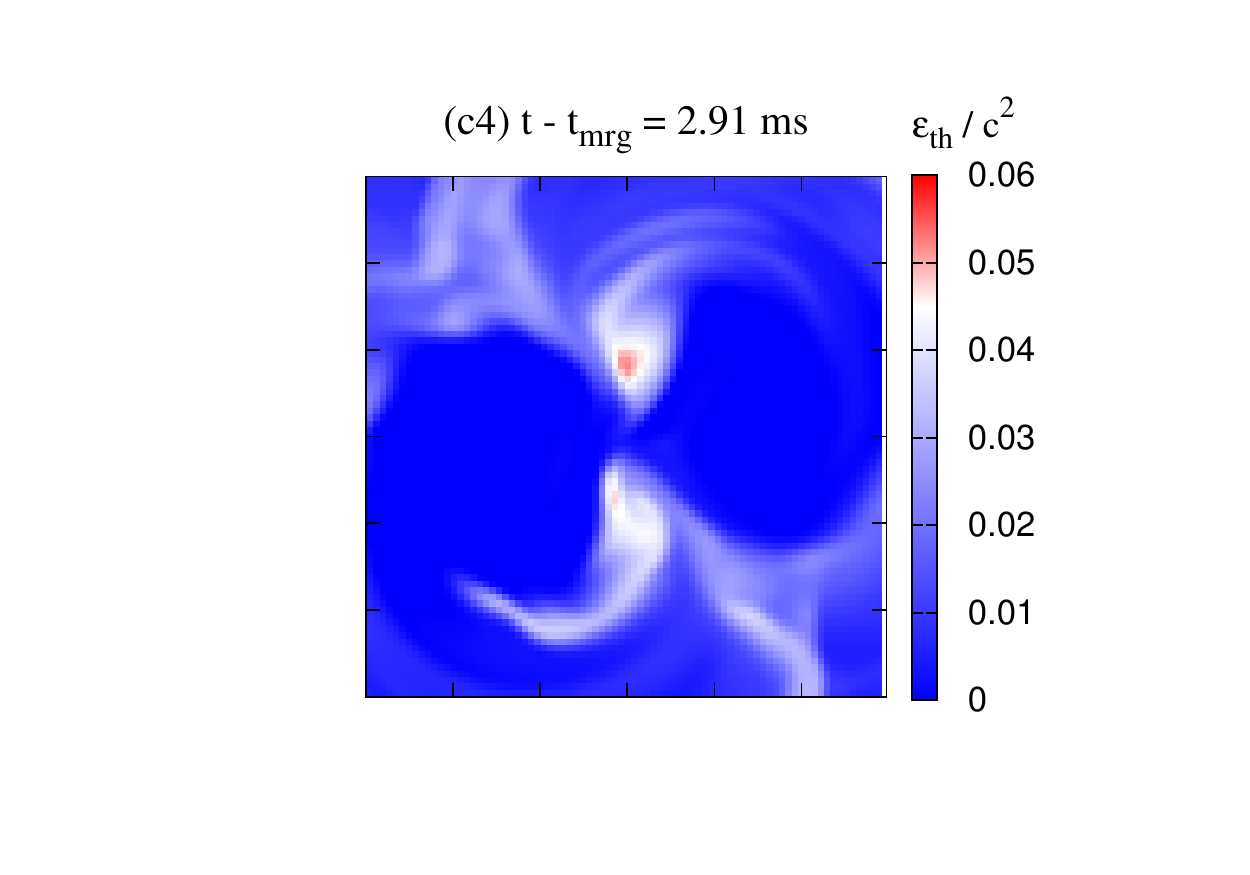}
\end{center}
\end{minipage}\\
\vspace{-19mm}
\hspace{-50mm}
\begin{minipage}{0.27\hsize}
\begin{center}
\includegraphics[width=9.0cm,angle=0]{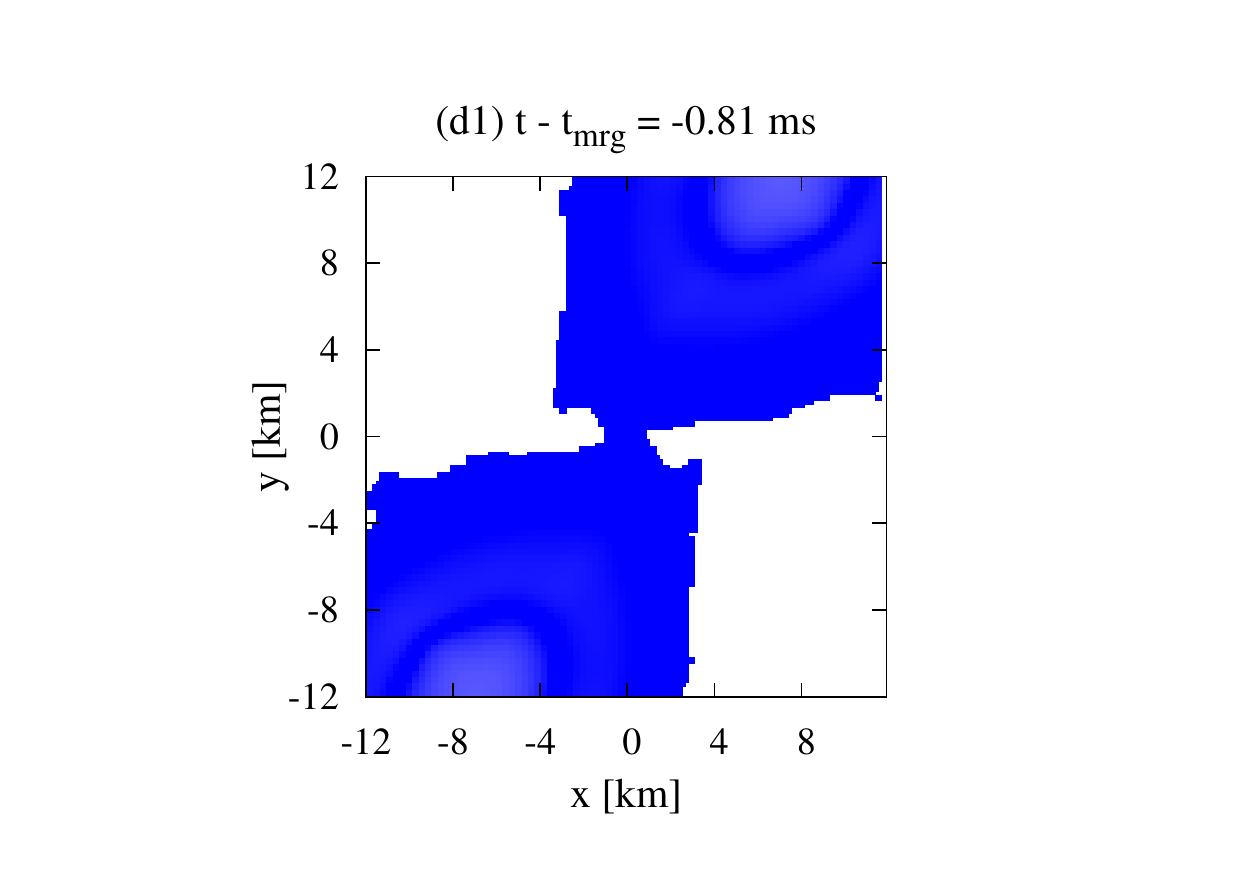}
\end{center}
\end{minipage}
\hspace{-12mm}
\begin{minipage}{0.27\hsize}
\begin{center}
\includegraphics[width=9.0cm,angle=0]{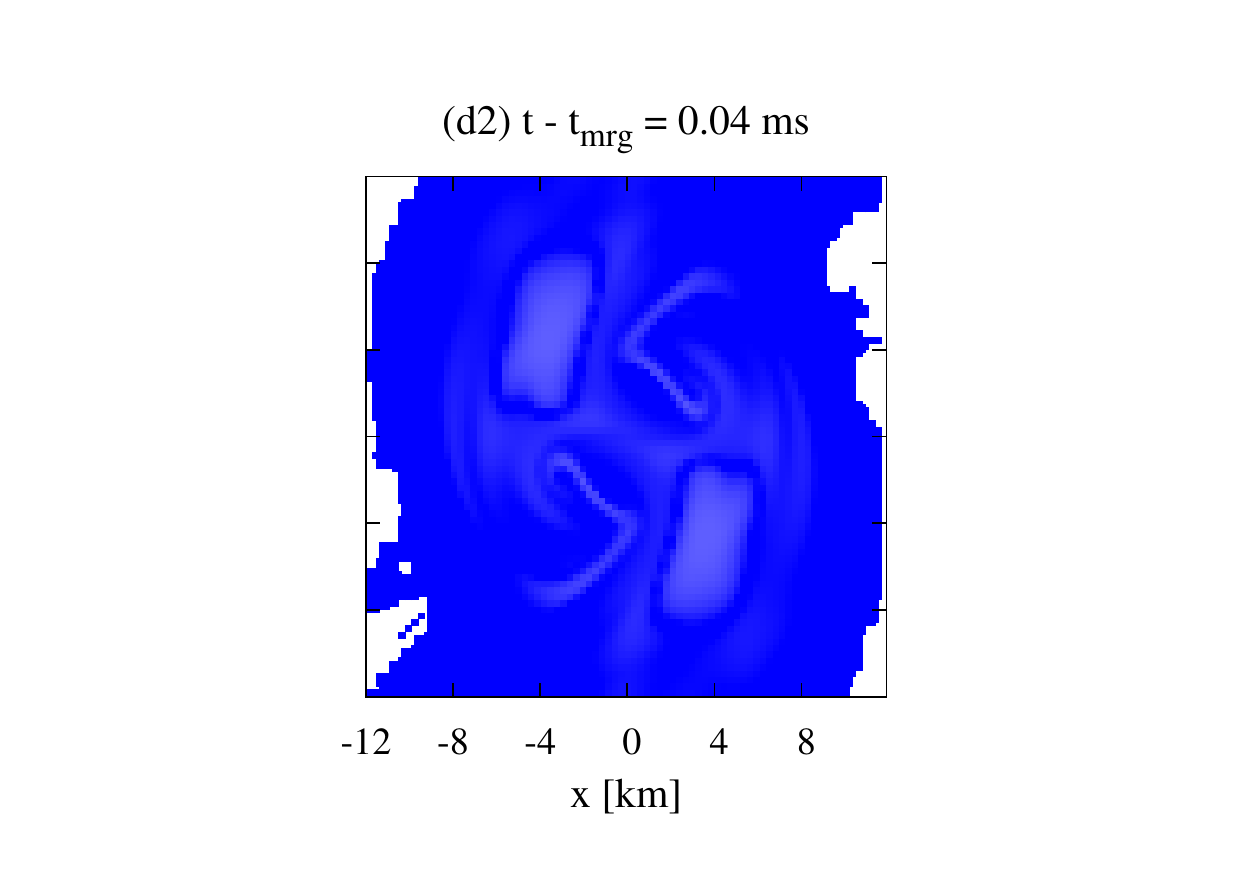}
\end{center}
\end{minipage}
\hspace{-12mm}
\begin{minipage}{0.27\hsize}
\begin{center}
\includegraphics[width=9.0cm,angle=0]{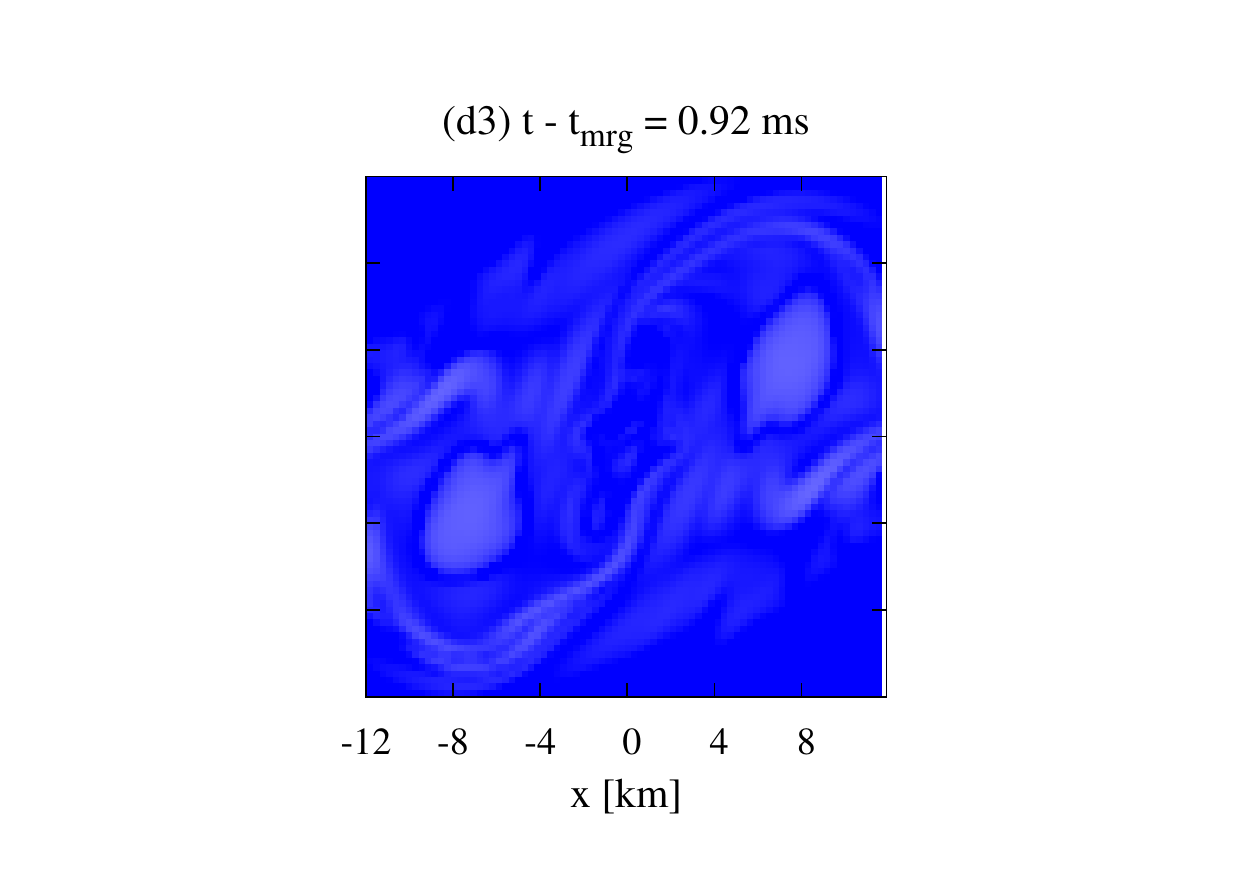}
\end{center}
\end{minipage}
\hspace{-12mm}
\begin{minipage}{0.27\hsize}
\begin{center}
\includegraphics[width=9.0cm,angle=0]{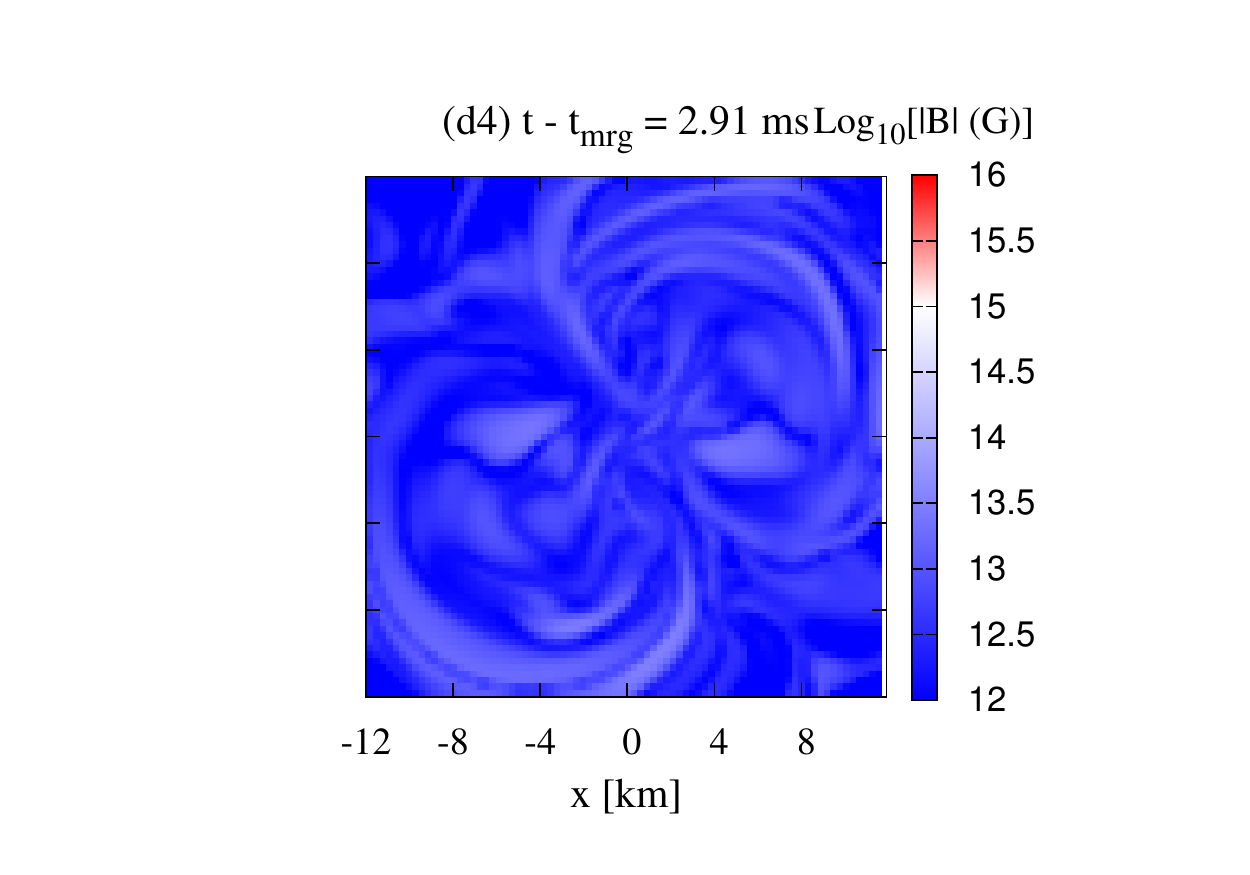}
\end{center}
\end{minipage}
\caption{\label{fig3}
Same as Fig.~\ref{fig1}, but for $\Delta x_{(l_{\rm max})}=150$ m.}
\end{figure*}

\begin{figure*}[t]
\hspace{-40mm}
\begin{minipage}{0.31\hsize}
\begin{center}
\includegraphics[width=11.0cm,angle=0]{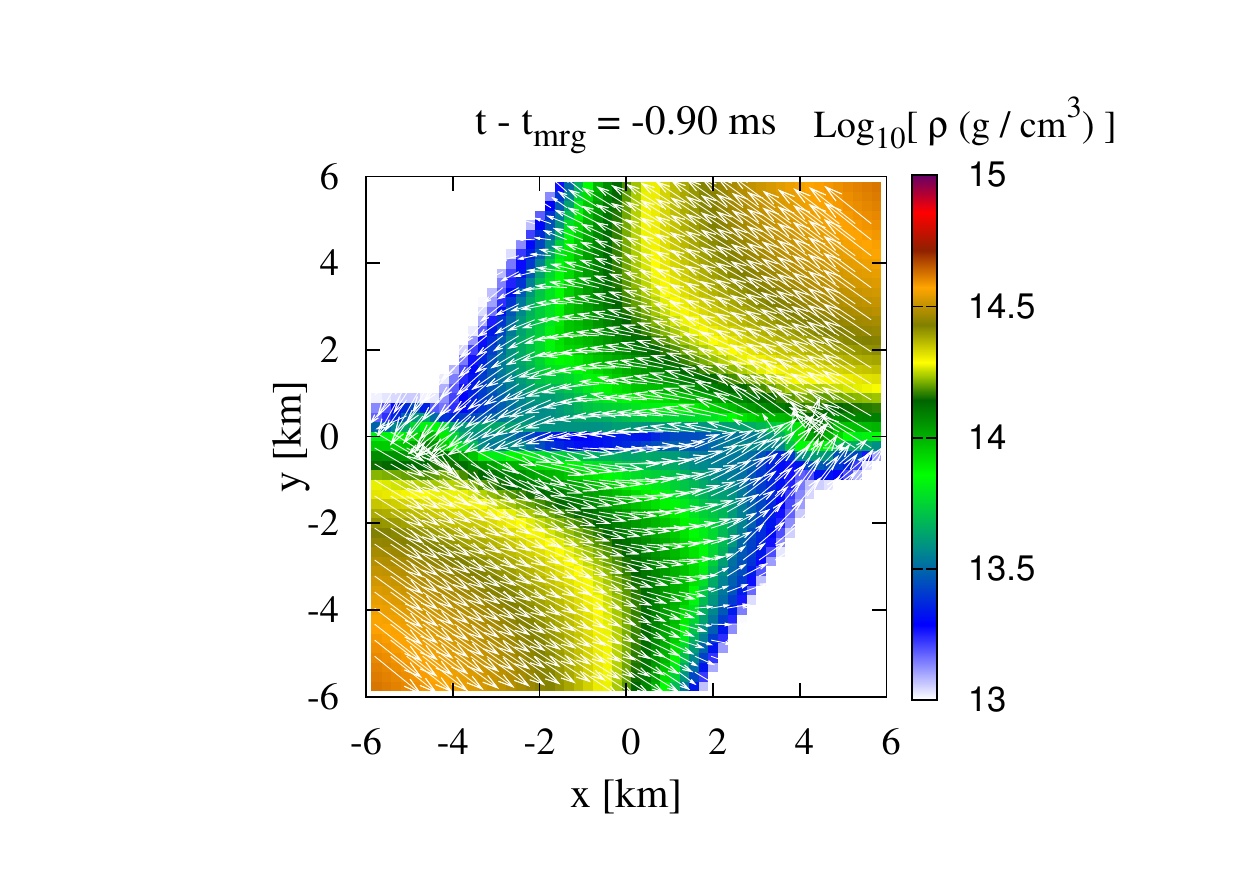}
\end{center}
\end{minipage}
\hspace{40mm}
\begin{minipage}{0.25\hsize}
\begin{center}
\includegraphics[width=8.0cm,angle=0]{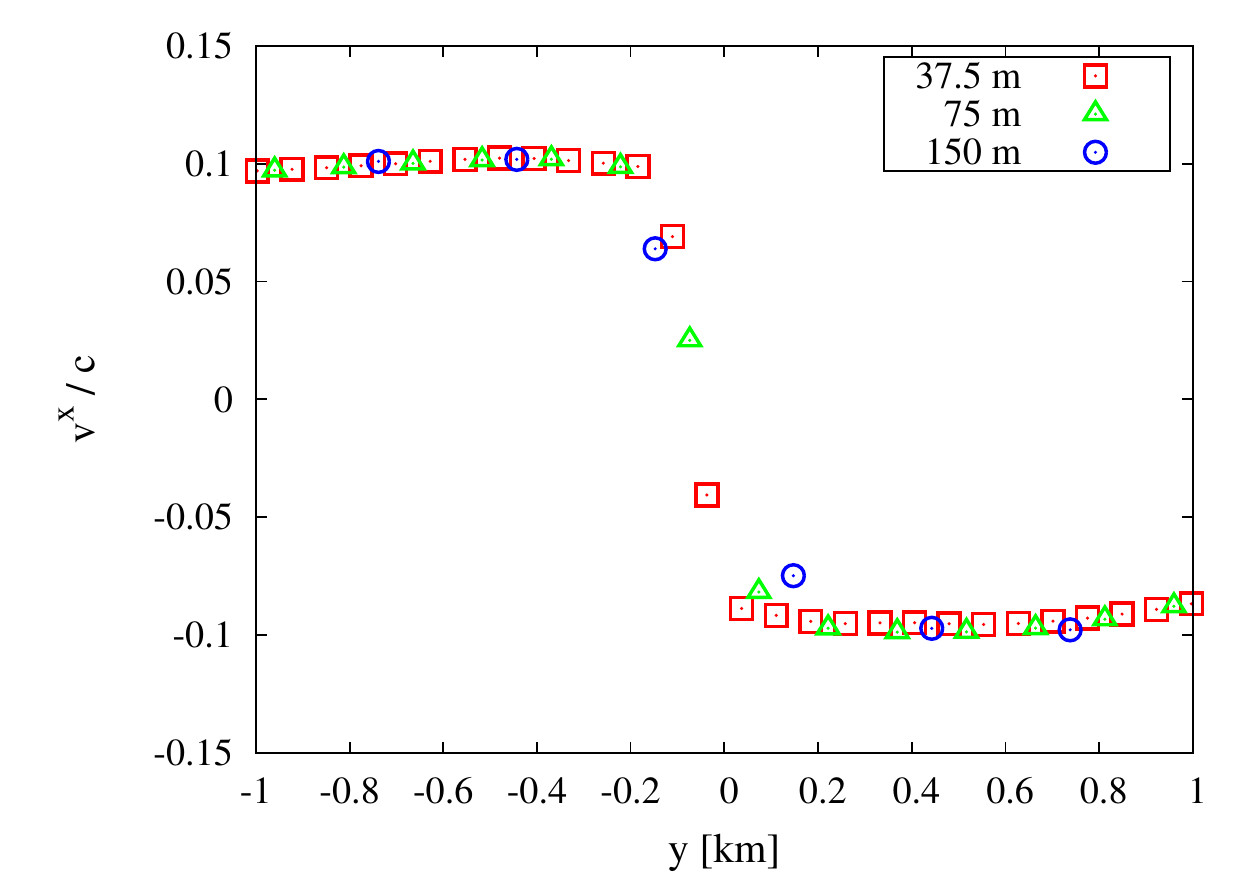}
\end{center}
\end{minipage}
\caption{\label{figvx}
(Left) Profile of the rest-mass density with velocity field on the orbital plane just before the onset of the merger 
and (right) profile of $v^x$ in the left panel along the $y$ axis with $x=0$ km for 
$\Delta x_{(l_{\rm max})} = 37.5$, $75$, and $150$ m runs.}
\end{figure*}

Figures~\ref{fig2} and \ref{fig3} plot the profiles of the rest-mass density, the vorticity, 
the thermal component of the specific internal energy, and the magnetic-field 
strength on the orbital plane for $\Delta x_{(l_{\rm max})}=37.5$ and $150$ m runs, respectively. 
For $\Delta x_{(l_{\rm max})}=37.5$ m run, the qualitative features agree with those discussed 
in the previous subsections, but 
the formation of vortices and the magnetic-field amplification are likely to be less prominent than those for 
$\Delta x_{(l_{\rm max})}=17.5$ m run. 
[compare Figs.~\ref{fig1}(b2)--(b4) and (d2)--(d4) with Figs.~\ref{fig2}(b2)--(b4) and (d2)--(d4)]. 
The difference is clearer in Fig.~\ref{fig3} 
with $\Delta x_{(l_{\rm max})}=150$ m. 
The scale of the vortices is smaller and the magnetic-field is amplified more efficiently in the higher-resolution run. 
This feature is consistent with the linear perturbation analysis of the KH instability; 
the growth rate of the fastest growing mode is inversely proportional to the minimum grid spacing 
and hence high-resolutions exhibit faster development of vortices. 

The peak value of $\epsilon_{\rm th}$ in the shear layer decreases 
with increasing the resolution [see Figs.~\ref{fig1}(c2)--(c3), \ref{fig2}(c2)--(c3), and \ref{fig3}(c2)--(c3)]. 
As discussed below, the cascade of the turbulence-like motion converts the 
kinetic energy to the thermal energy at the dissipation scale determined by the grid resolution. 
This finding is consistent with such a dissipation mechanism. 

\begin{figure*}[t]
\begin{minipage}{0.27\hsize}
\begin{center}
\includegraphics[width=8.0cm,angle=0]{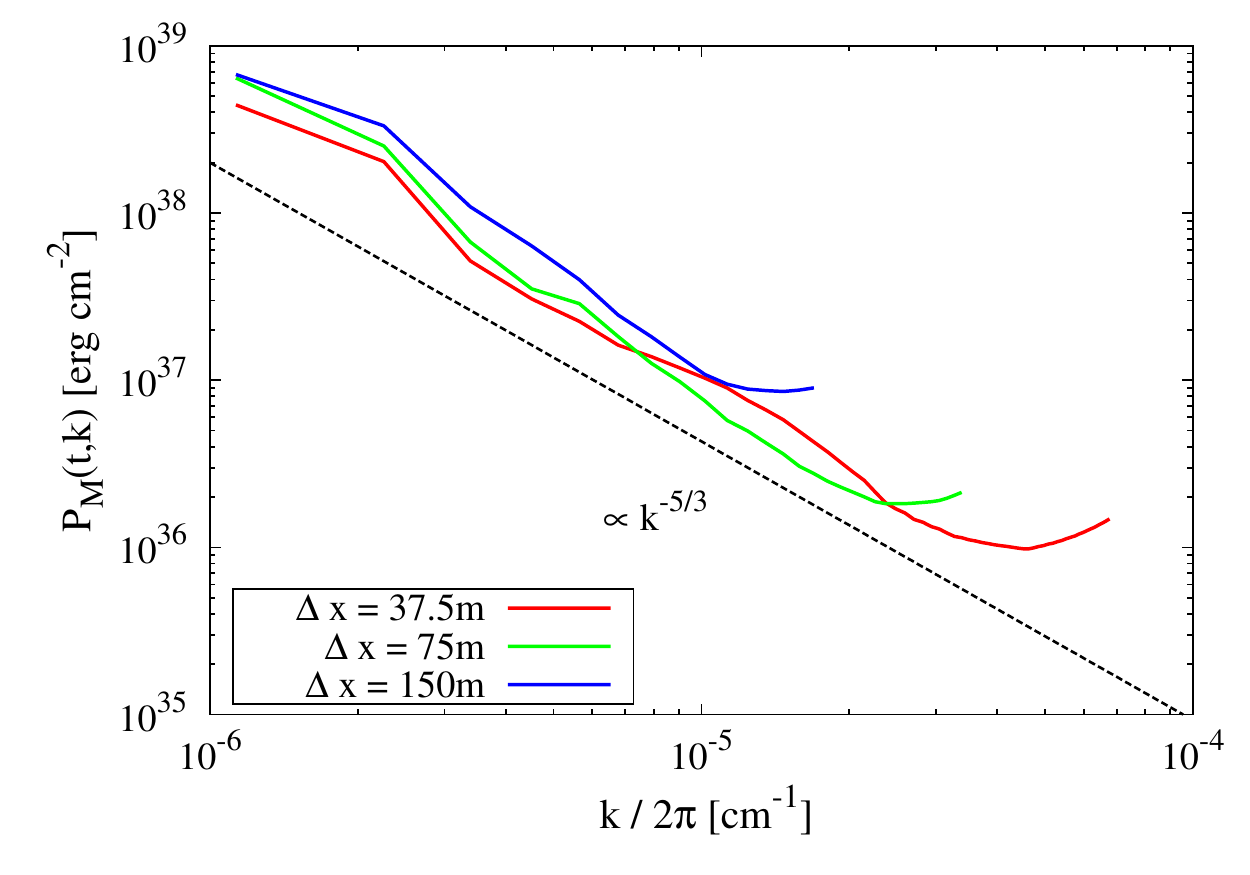}
\end{center}
\end{minipage}
\caption{\label{fig41}
Kinetic energy spectrum of the matter flow for $\Delta x_{(l_{\rm max})}=37.5$, $75$, and $150$ m. 
The spectrum is evaluated at $t-t_{\rm mrg}=2.0$ ms. 
}
\end{figure*}

We calculate the power spectrum of the turbulence-like motion as 
\begin{align}
P_M(t,k) = \frac{1}{2V}\int_V \delta\tilde{\bold{v}}_\rho(t,\bold{k}) \cdot \delta\tilde{\bold{v}}^*_\rho(t,\bold{k})d\Omega_k,
\end{align}
where 
\begin{align}
\delta \tilde{\bold{v}}_\rho(t,\bold{k}) = \int_V e^{-i\bold{k}\cdot\bold{r}} \sqrt{\rho(t,\bold{x})} \delta\bold{v}(t,\bold{x})d^3r.
\end{align}
The bold symbol denotes a spatial three vector. 
$\bold{k}$ is a wavenumber vector and $k=|\bold{k}|$. $d\Omega_k$ is a phase-space volume element in a spherical shell between $k$ and 
$k+dk$, and $V$ is a cuboid region of $x,y[{\rm km}]\in[-4.5,4,5]$, $z[{\rm km}]\in[0.0,4.5]$. We choose 
$\bold{r}$ to be a position vector from the coordinate center. The velocity fluctuation $\delta \bold{v}(t,\bold{x})$ 
is $\bold{v}(t,\bold{x})-\langle\bold{v}\rangle({\bold x})$ 
where $\langle\cdot\rangle$ denotes the time average for the duration $0.0~{\rm ms} \le t-t_{\rm mrg} \le 2.0~{\rm ms}$. 
We evaluate $\delta \bold{v}$ at $t-t_{\rm mrg}=2.0$ ms.
With this, $\int P_M(t,k) dk$ corresponds to the kinetic energy density of the turbulence-like flow. 
Figure~\ref{fig41} plots the power spectrum of $\Delta x_{(l_{\rm max})}=37.5$, $75$, and $150$ m runs, respectively. 
The amplitude at large scale in the lower-resolution run is higher than that in the higher-resolution run 
and the spectrum extends to a higher wavenumber in the higher-resolution run. This indicates that 
the energy of large-scale turbulence-like motion cascades to that of small-scale motion. 
The figure also indicates that there is a kinetic energy sink at the end of the turbulence cascade due to the 
numerical dissipation. 
The power excess at the highest $k$ is the so-called {\it bottleneck} effect \citep{Falkovich:93}
and is observed always in 3D simulations of turbulence (see, e.g., \citep{Porter:98,Gotoh:02,Kaneda:03}). 
Because energy budget of the KH instability should be the bulk kinetic energy, all the findings show that 
a part of the bulk kinetic energy is converted into the kinetic energy of the turbulence-like motion and subsequently 
thermalized at the dissipation scale.

\begin{figure*}[t]
\hspace{-40mm}
\begin{minipage}{0.27\hsize}
\begin{center}
\includegraphics[width=9.0cm,angle=0]{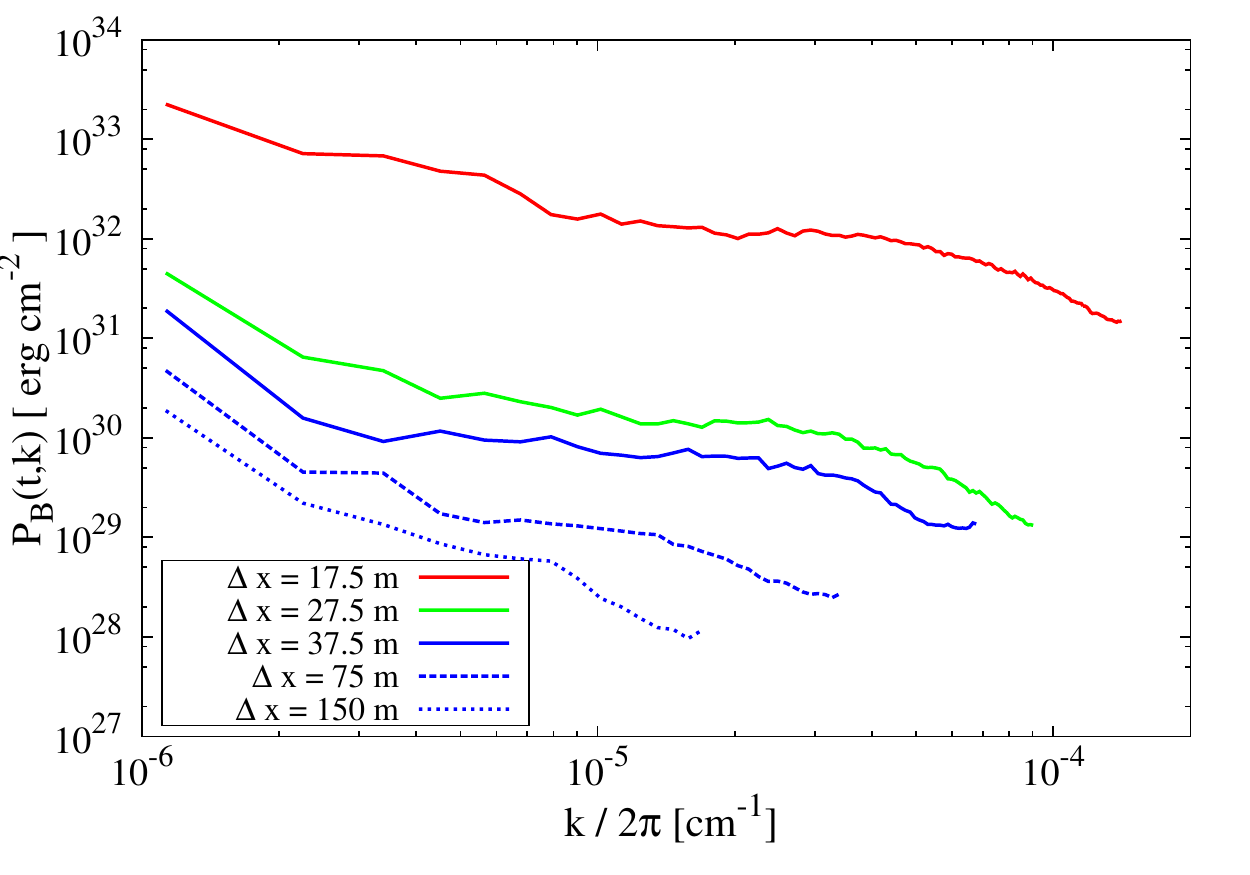}
\end{center}
\end{minipage}
\hspace{40mm}
\begin{minipage}{0.27\hsize}
\begin{center}
\includegraphics[width=9.0cm,angle=0]{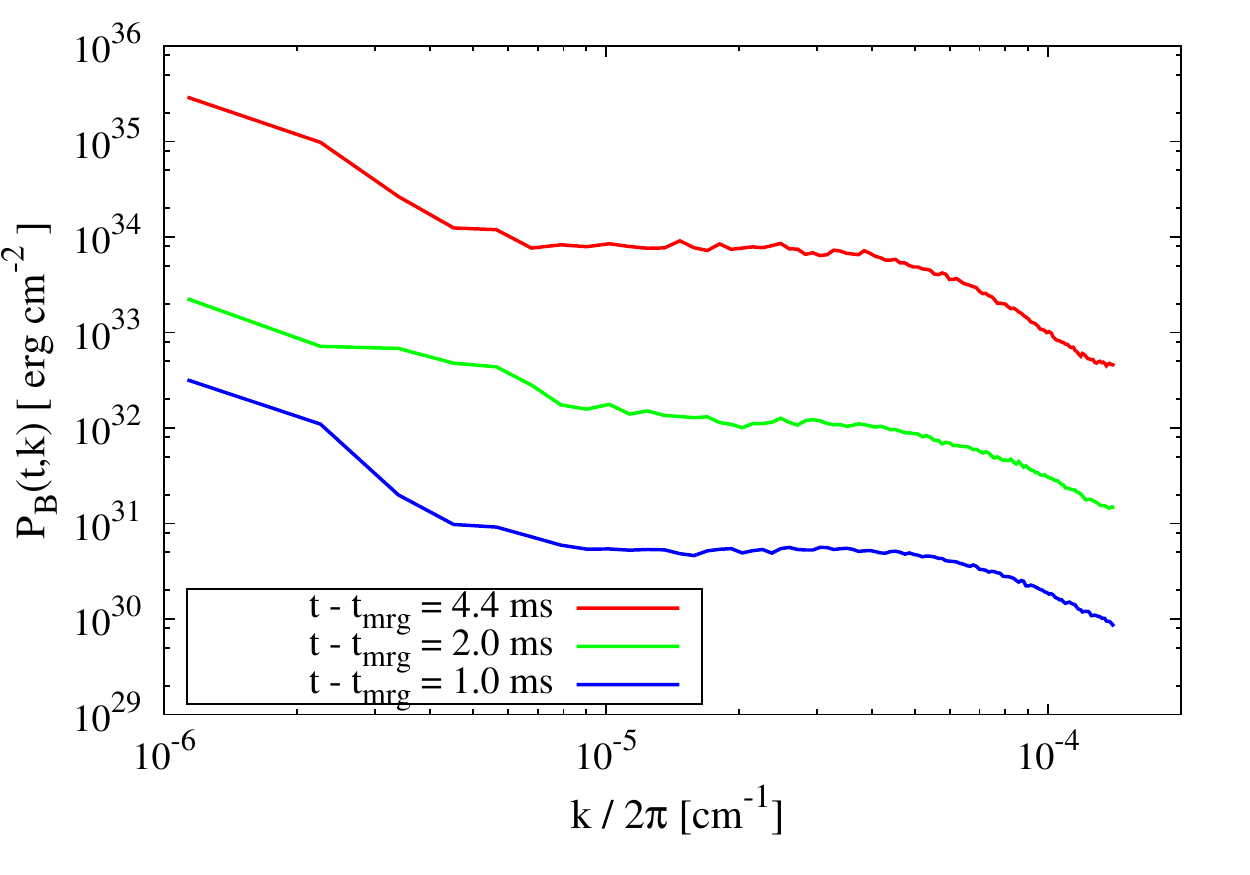}
\end{center}
\end{minipage}
\caption{\label{fig42}
Energy spectrum of the magnetic field for 
$\Delta x_{(l_{\rm max})}=17.5$, $27.5$, $37.5$, $75$, and $150$ m runs at $t-t_{\rm mrg}=2.0$ ms (left) 
and for $\Delta x_{(l_{\rm max})}=17.5$ m run at $t-t_{\rm mrg} = 1.0,2.0,$ and $4.4$ ms.
}
\end{figure*}

Figure~\ref{fig42} plots the power spectrum of the magnetic-field energy:
\begin{align}
P_B(t,k) = \frac{1}{8\pi V}\int_V \tilde{b}(t,\bold{k})\tilde{b}^*(t,\bold{k})d\Omega_k,
\end{align}
where
\begin{align}
\tilde{b}(t,\bold{k}) = \int_V e^{-i\bold{k}\cdot\bold{r}} b(t,\bold{x})d^3r.
\end{align}
$b=\sqrt{b^\mu b_\mu}$ and $b^\mu$ is a magnetic field measured in a fluid rest frame. 
We evaluate $b$ for $\Delta x_{(l_{\rm max})}=17.5$, $27.5$, $37.5$, $75$, and $150$ m at $t-t_{\rm mrg}=2.0$ ms as shown in 
the left panel of Fig.~\ref{fig42}, respectively. As expected, the amplitude is higher in the higher-resolution run. 
This indicates that the kinetic energy of the turbulence-like motion is converted to the magnetic-field energy more efficiently 
in the higher-resolution runs. 

The right panel is for $\Delta x_{(l_{\rm max})}=17.5$ m run at $t-t_{\rm mrg}=1.0,2.0$ and $4.4$ ms, respectively. 
The spectrum is relatively flat in the inertial range of the turbulent cascade and its amplitude increases with time. 
These features resemble the behavior of the spectrum in local simulations of 
a large-scale dynamos during the kinematic amplification phase (see, e.g., Fig.~2 in Ref.~\citep{Brandenburg:01}).

In local simulations, the kinematic phase ends when the magnetic-field strength reaches an equipartition level and a slow
  saturation phase occurs in which large-scale field is generated (see discussion in Ref.~\citep{Brandenburg:05}).
  Checking whether a large-scale dynamo is active
  in our simulations would require longer simulations of the saturated state.
  However, it is unclear if this will be indeed the case; the shear layer
  is pumping kinetic energy at sub km-size scales in the first few ms after the merger, feeding the turbulent
  cascade and making it viable dynamo action; however this initial phase eventually ends as the two cores merge
  and turbulence-like motion will decay in diffusive timescales,
  in which the dynamo may not be active anymore.
  We do not see any sign of a $3/2$ Kazantzev spectrum during the kinematic phase, typical of small-scale dynamos,
  as predicted by Ref.~\citep{Zrake:2013mra}.

The fact that the magnetic field power is growing at all scales indicates that 
the kinetic energy of the turbulence-like motion is converted into the magnetic-field energy. 
We estimate the kinetic energy density and the magnetic-field energy density 
for the highest-resolution run at $t-t_{\rm mrg}=4.4$ ms
\begin{align}
\epsilon_K &= \int P_M(t,k)dk \approx 2.3 \times 10^{33}~{\rm erg~cm^{-3}},\\
\epsilon_B &= \int P_B(t,k) dk \approx 6.0 \times 10^{30}~{\rm erg~cm^{-3}}.
\end{align}
These values correspond approximately to an averaged fluctuation velocity of $\delta v \approx 0.05c$ with 
typical density field of $10^{15}~{\rm g/cm^3}$, and 
averaged magnetic-field strength of $\bar{B}\approx 10^{16}~{\rm G}$. 
Even with the highest-resolution run, the equipartition is not achieved and this indicates 
that there is a room for further magnetic-field amplification.

\section{Saturation of the magnetic field}

\begin{figure*}[t]
\hspace{-40mm}
\begin{minipage}{0.27\hsize}
\begin{center}
\includegraphics[width=9.0cm,angle=0]{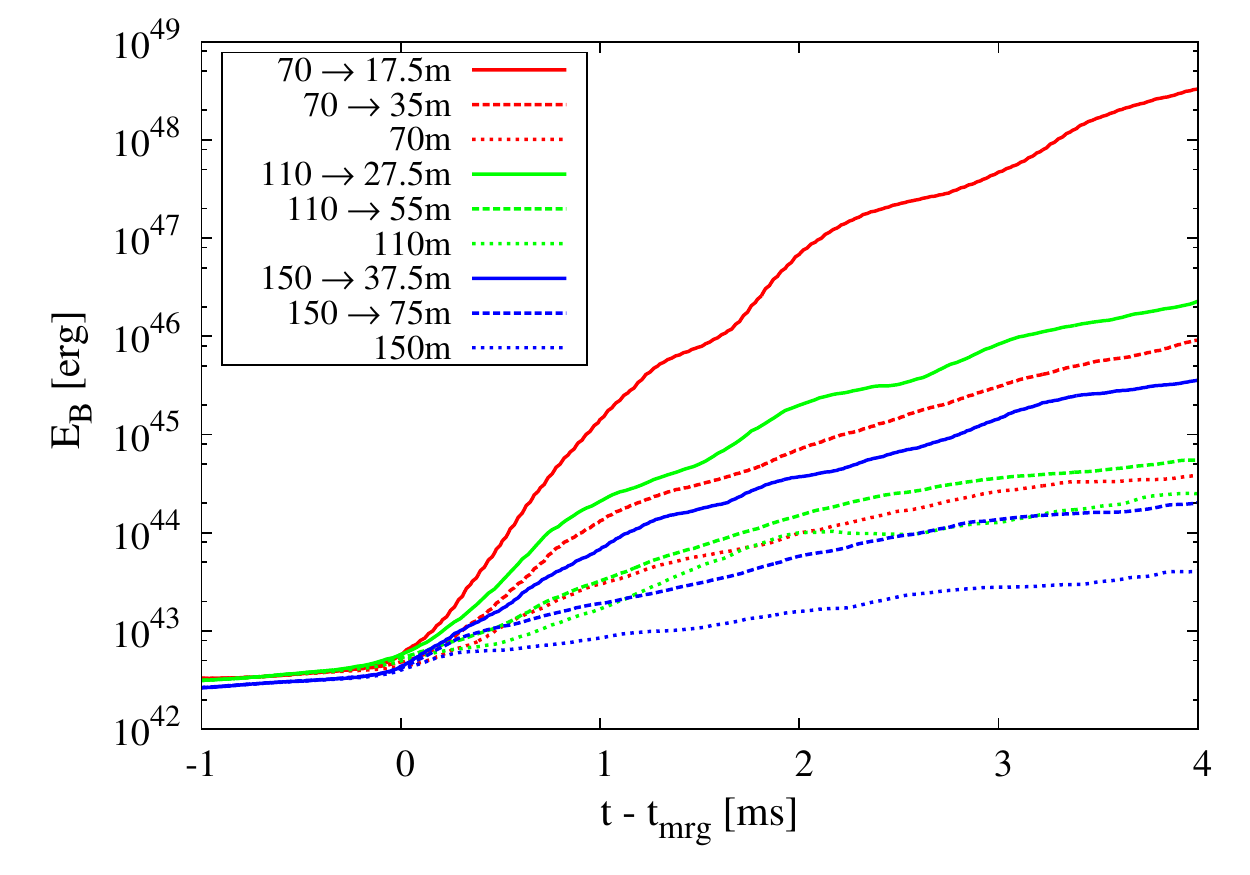}
\end{center}
\end{minipage}
\hspace{35mm}
\begin{minipage}{0.27\hsize}
\begin{center}
\includegraphics[width=9.0cm,angle=0]{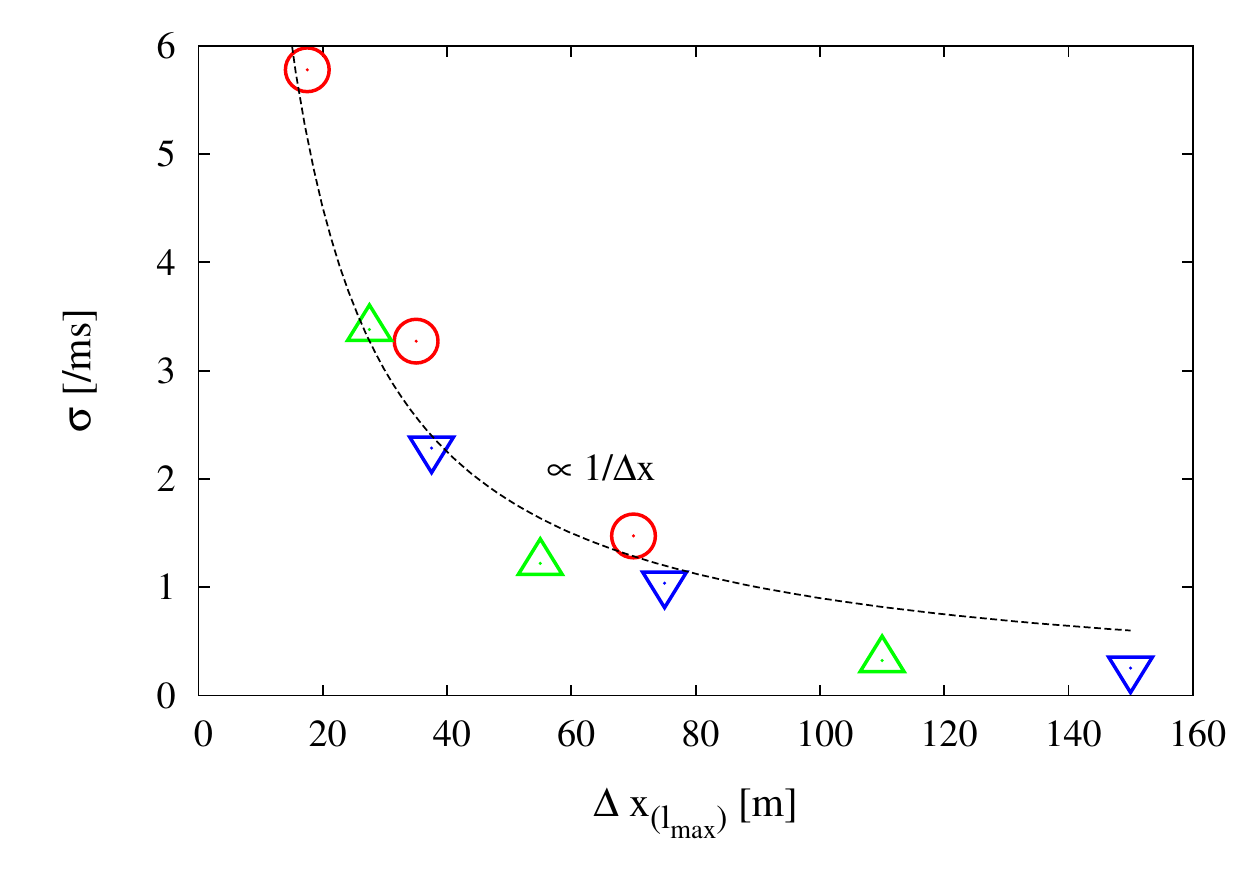}
\end{center}
\end{minipage}
\caption{\label{fig4}
(Left) Magnetic-field energy as a function of the time for the B13 run. (Right) The growth rate 
of the magnetic-field energy as a function of the final grid resolution. 
}
\end{figure*}

Figure~\ref{fig4} plots the time evolution of the magnetic-field energy. The initial magnetic-field strength is $10^{13}$ G for 
all the runs. The magnetic-field energy steeply increases for $t-t_{\rm mrg} \gtrsim 0$ ms and the amplification 
is more prominent in the higher-resolution runs. In the highest-resolution run with $\Delta x_{(l_{\rm max})} = 17.5$ m, 
the energy is amplified by a factor of $\approx 10^6$ at $t-t_{\rm mrg} \approx 4$ ms and this implies that the averaged 
magnetic-field strength increases up to $\sim 10^{16}$ G. 
The right hand-side panel of Fig.~\ref{fig4} plots the dependence of the growth rate of the magnetic-field energy 
on the finest grid resolution $\Delta x_{(l_{\rm max})}$. We fit the magnetic-field energy for 
$0~{\rm ms}~\lesssim t - t_{\rm mrg} \lesssim 1$ ms by an exponentially growing 
function $\propto \exp(\sigma t)$. The figure shows a divergent feature of the growth rate 
with respect to $1/{\Delta x}_{(l_{\rm max})}$. 
Even with $\Delta x_{(l_{\rm max})}=17.5$ m run, the saturation is not likely to be achieved. 
This behavior can be understood in terms of the property of the KH instability. As discussed in the previous subsection, 
the cut-off wavenumber $k_{\rm cutoff}$ of the KH instability increases with increasing the grid resolution. 
Therefore, the dispersion relation of the KH instability should be described as that of the shear layer with the infinitesimal thickness; 
$\sigma \propto k \propto 1/\Delta x_{(l_{\rm max})}$.

\begin{figure*}[t]
\begin{minipage}{0.27\hsize}
\begin{center}
\includegraphics[width=8.0cm,angle=0]{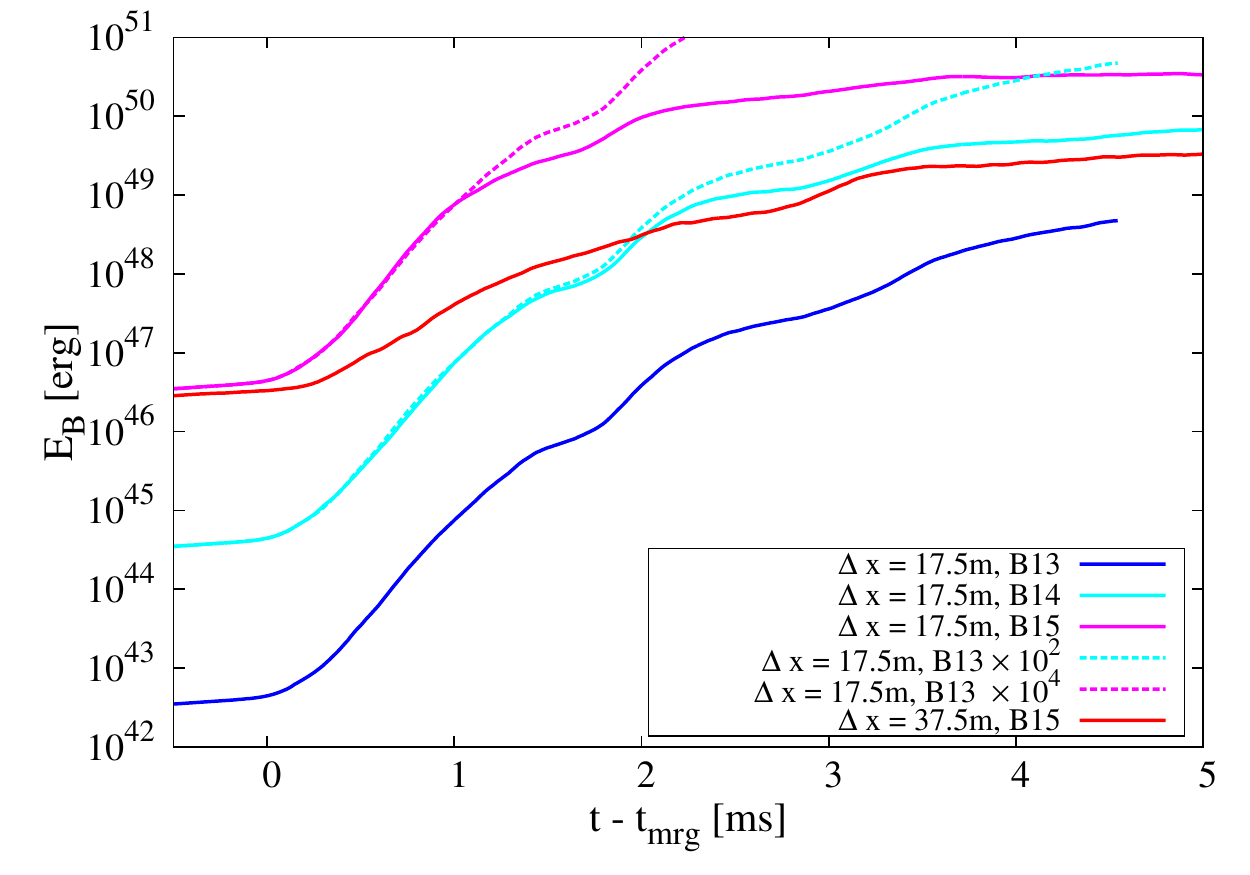}
\end{center}
\end{minipage}
\caption{\label{fig5}
Time evolution of the magnetic-field energy for the B13, B14, and B15 runs. The cyan- and magenta-dashed curves 
show the evolution of the B13 run magnified by $10^2$ and $10^4$, respectively.
}
\end{figure*}

To explore the saturation energy of the magnetic field at the end of the kinematic amplification phase, 
we also perform simulations for an initial 
maximum magnetic-field strength varied from $10^{13}$ G to $10^{15}$ G. 
In Fig.~\ref{fig5}, we plot $E_B$ as a function of time. 
We overplot the results of the B13 run magnified by $10^2$ and $10^4$. 
Comparing the curve for the B14 run with $\Delta x_{(l_{\rm max})}=17.5$ m to the magnified curve of the B13 run with the same resolution, 
we find a good overlap up to $t-t_{\rm mrg}\approx 2$ ms. This suggests that back reaction due to the amplified magnetic field is negligible. 
On the other hand for $t-t_{\rm mrg} \gtrsim 2$ ms, two curves do not overlap. This indicates that 
the back reaction probably due to magnetic braking turns on.  
At $t-t_{\rm mrg}\approx 5$ ms, the energy increases up to $\approx 10^{50}$ erg for the B14 run. 
For the B15 run with $\Delta x_{(l_{\rm max})}=17.5$ m, the back 
reaction is turned on at $t-t_{\rm mrg} \approx 1$ ms and subsequently the energy reaches $\approx 4\times 10^{50}$ erg 
at $t-t_{\rm mrg} \approx 3$ ms. The energy does not increase significantly after that. 
To assess the numerical effect on the saturation energy, we repeat the simulation 
with $\Delta x_{(l_{\rm max})}=37.5$ m for the B15 run. Figure~\ref{fig5} shows that 
the energy is amplified only up to $\approx 5 \times 10^{49}$ erg at $t-t_{\rm mrg} \approx 5$ ms. 
Therefore, the saturation energy of the magnetic-field is likely to be $\gtrsim 4 \times 10^{50}$ erg and the averaged magnetic-field 
strength could be $\gtrsim 10^{16}$ G.

\begin{figure*}[t]
\hspace{-40mm}
\begin{minipage}{0.27\hsize}
\begin{center}
\includegraphics[width=10.0cm,angle=0]{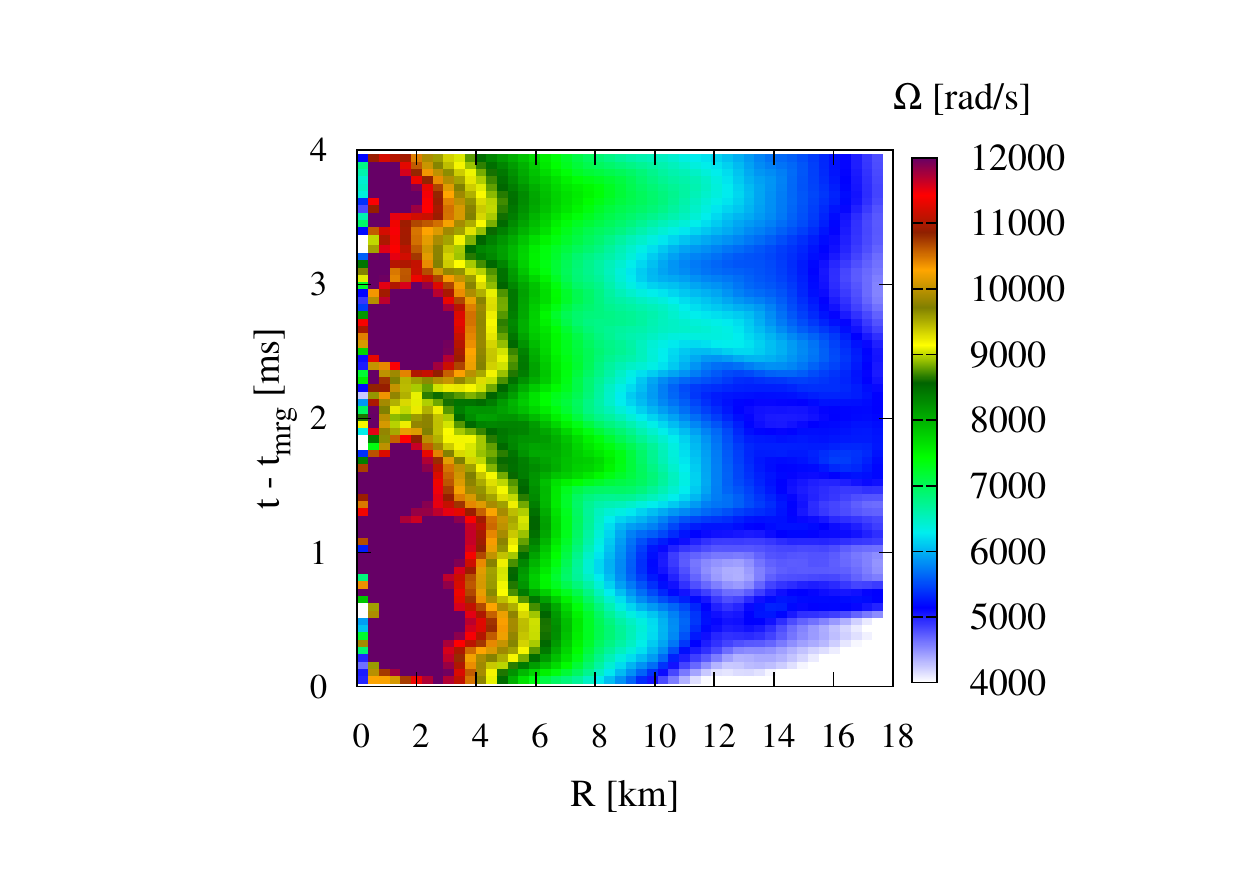}
\end{center}
\end{minipage}
\hspace{20mm}
\begin{minipage}{0.27\hsize}
\begin{center}
\includegraphics[width=10.0cm,angle=0]{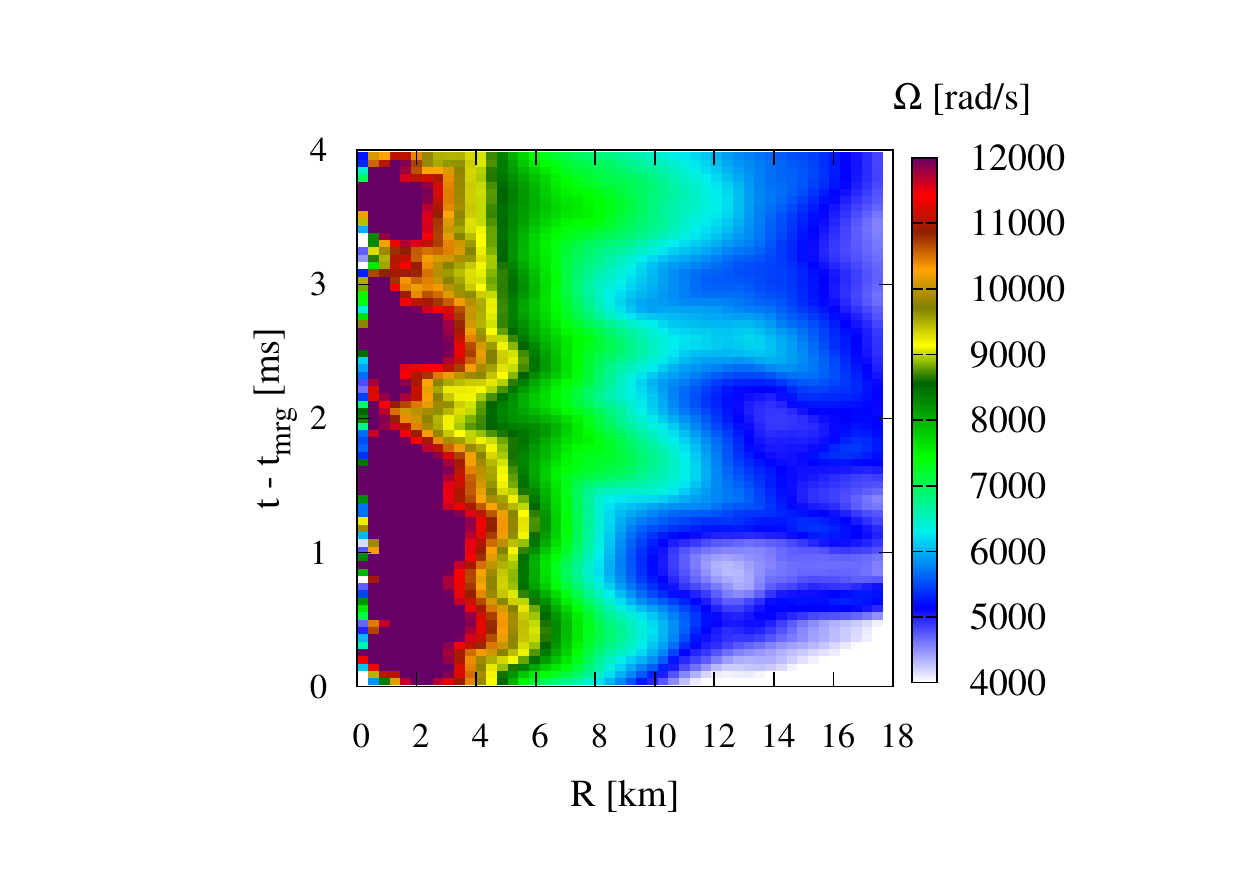}
\end{center}
\end{minipage}
\caption{\label{fig8}
Spacetime diagram of the angular velocity for the B13 run with $\Delta x_{(l_{\rm max})}=17.5$ m (left) 
and the B15 run with $\Delta x_{(l_{\rm max})}=17.5$ m (right). 
The horizontal axis is the radial coordinate and the angular velocity on the 
orbital plane is averaged in the azimuthal direction. 
}
\end{figure*}

Figure~\ref{fig8} plots a spacetime diagram of the averaged angular velocity on the orbital plane 
\begin{align}
\bar{\Omega}(t,R) = \frac{1}{2\pi} \int^{2\pi}_0 \Omega(t,R,z=0,\varphi) d\varphi,
\end{align}
where we employ cylindrical coordinates and $\Omega$ is an azimuthal component of the three velocity ($v^\varphi$). 
Because of the non-axisymetric structure of the HMNS, 
the angular momentum is transported outward. 
Figure~\ref{fig8} shows that the fluid elements moves gradually in the radial direction. 
The magnetic braking should be active in the Alfv\'{e}n timescale 
\begin{align}
t_A &= \frac{R_{\rm HMNS}}{v_A} \approx 2 \times 10^{-2} {\rm s} \left(\frac{\rho}{10^{15}{\rm g/cm^3}}\right)^{-1/2} \nonumber\\
&\times \left(\frac{R_{\rm HMNS}}{20{\rm km}}\right)\left(\frac{B}{10^{16}{\rm G}}\right)^{-1},
\end{align}
where $R_{\rm HMNS}$ and $v_A$ are the HMNS radius and the Alfv\'{e}n velocity, respectively. 
Although clear difference does not appear between the B13 and B15 runs in Fig.~\ref{fig8}, 
we find that the magnetic braking works efficiently in the late phase of the HMNS evolution. 

\section{Summary}

We have performed high-resolution GRMHD simulations of the BNS mergers. 
Focusing on the shear layer emerging at the merger, the simulations were performed by assigning a 
finer grid resolution than the resolution in the previous simulations. 
With this resolution, we have revealed that 
the small-scale turbulence-like motion is developed due to the KH instability and 
the magnetic field is amplified efficiently. 
Starting from initial maximum magnetic-field strength of $10^{13}$ G, we have found 
that the magnetic-field energy is amplified at least by a factor of $\approx 10^6$ at $\approx 4$ ms after the onset of the merger. 
Saturation energy of the magnetic-field is likely to be $\gtrsim 4 \times 10^{50}$ erg, which is $\gtrsim 0.1\%$ of the bulk kinetic energy. 
We should explore whether the physical saturation of the magnetic-field energy occurs below the equipartition value as a future work.

Our result shows that the efficient magnetic-field amplification during the BNS merger is realized in reality 
as pointed out in Refs.~\cite{Rasio,Price:2006fi}. This implies that it is always necessary to take into account the 
effects of high magnetic fields for modeling the post merger evolution of BNS. 

\acknowledgments
K. Kiuchi thanks to Andreas Bauswein and Sebastiano Bernuzzi for a discussion in University of Washington. 
Numerical computations were performed on K computer at AICS, XC30 at
CfCA of NAOJ, FX10 at Information Technology Center of the University of Tokyo, and SR16000 at YITP of Kyoto University. 
This work was supported by Grant-in-Aid for Scientific Research
(24244028, 15H00783,15H00836,15K05077,15H06857), for Scientific Research on Innovative
Area (24103001), by HPCI Strategic Program of Japanese MEXT/JSPS (Project numbers hpci130025, 140211, and 150225), 
by the RIKEN iTHES Research Project, by the European Research
Council (grant CAMAP-259276) and by local Spanish funds (grants AYA2013-40979-P and PROMETEOII/2014-069).

\appendix*
\section{Landau quantization effect}
Because the magnetic-field strength exceeds the QED limit of $4.414 \times 10^{13}$ G, 
the so-called Landau quantization would become important. The threshold density is characterized by 
the critical density
\begin{align}
\rho_B = 7.04\times 10^{10} \left(\frac{Y_e}{0.1}\right)^{-1}\left(\frac{B}{10^{16}{\rm G}}\right)^{3/2} {\rm g/cm^3},
\end{align}
where $Y_e$ is the electron fraction per baryon~\cite{Harding:2006qn}. Below $\rho_B$, the ground Landau level is populated 
and the Landau quantization effect becomes important. On the other hand, for $\rho_B \ll \rho$, many Landau levels 
are populated and the magnetic field is not quantized. As shown in Fig.~\ref{fig1}, the strong magnetic field 
exceeding the QED limit appears only in the high-density region. Therefore, the Landau quantization effect is 
irrelevant for the dynamics in the HMNS phase. Note that the range in which $\rho < \rho_B$ and $B^2/8\pi<\rho c^2$ 
are simultaneously satisfied is relatively narrow when the magnetic field is strong.

In this BNS model, the HMNS will collapse to a black hole (BH) at $t-t_{\rm mrg} \approx 10$ ms~\cite{Kiuchi:2014}. 
A massive accretion torus is formed after the BH formation. The density of the accretion torus is in the 
range of $10^{10-11}~{\rm g/cm^3}$. Because of the efficient amplification of the magnetic field in the HMNS phase, 
the accretion torus would be strongly magnetized at its birth and 
the typical magnetic-field strength would be $10^{15}$ G at about 25 ms after the BH formation~\cite{Kiuchi:2014}. 
The Landau quantization might play an important role in the accretion torus phase because the density becomes 
the same order of $\rho_B$. With the energy difference between two Landau levels $\Delta E_{\rm Landau}$, the magnetic temperature 
is defined by
\begin{align}
T_B \equiv \frac{\Delta E_{\rm Landau}}{k_B} =\left(\sqrt{1+\frac{2B}{B_{\rm QED}}}-1\right)\frac{m_e c^2}{k_B},
\end{align}
where $k_B$, $B_{\rm QED}$, and $m_e$ are the Boltzmann constant, the QED limit of the magnetic-field strength, 
and the electron rest mass, respectively~\cite{Harding:2006qn}. 
The second equality holds for the density below $\rho_B$. If the temperature exceeds the magnetic temperature, 
the Landau quantization effect is diminished. With $B=10^{15}$ G, $T_B\approx 3$--$4~{\rm MeV}(B/10^{15}{\rm G})^{1/2}$. 
Because the typical temperature inside the accretion torus is $\sim 10$ MeV~\cite{T-evaluation}, the Landau quantization effect 
is irrelevant for the early accretion torus phase. However, in the neutrino cooling timescale, the temperature will decrease 
and this effect would become important. 



\end{document}